\documentclass[longauth,traditabstract]{aa}

\usepackage{graphicx}
\usepackage{txfonts}
\usepackage{comment}
\usepackage{lscape}
\usepackage{longtable}

\usepackage{hyperref}
\hypersetup{breaklinks=true,colorlinks=true,linkcolor=blue,citecolor=blue,urlcolor=blue}

\usepackage{natbib,twoopt}
\bibpunct{(}{)}{;}{a}{}{,}             

\usepackage[usenames]{color}                                   

\newcommand\pycasso{{\sc p}y{\sc casso}}          	
\newcommand\starlight{{\sc starlight}}          	

\begin{document}
\defcitealias{cortijo17}{CF17}
\defcitealias{gonzalezdelgado2015}{GD15}
\defcitealias{gonzalezdelgado2016}{GD16}

\title{The spatially resolved star formation history of mergers: } 
\subtitle{A comparative study of the LIRGs IC1623, NGC6090, NGC2623, and Mice}

\authorrunning{IAA et al.}
\titlerunning{Star formation history of mergers}

\author{
C. Cortijo-Ferrero\inst{1},
R. M. Gonz\'alez Delgado\inst{1},
E. P\'erez\inst{1},
R. Cid Fernandes\inst{2},
R. Garc\'{\i}a-Benito\inst{1}, 
P. Di Matteo\inst{3},
S. F. S\'anchez\inst{4},
A. L.\ de Amorim\inst{2},
E. A. D. Lacerda\inst{2},
R. L\'opez Fern\'andez\inst{1},
C.Tadhunter\inst{5}
}

\institute{Instituto de Astrof\'isica de Andaluc\'ia (CSIC), PO 
Box 3004 18080 Granada, Spain. (\email{clara@iaa.es})
\and
Departamento de F\'{\i}sica, Universidade Federal de Santa Catarina, P.O. Box 476, 88040-900, Florian\'opolis, SC, Brazil
\and
GEPI, Observatoire de Paris, PSL Research University, CNRS, Place Jules Janssen, 92190 Meudon, France
\and
Instituto de Astronom\'\i a,Universidad Nacional Auton\'oma de M\'exico, A.P. 70-264, 04510 M\'exico D.F., Mexico
\and
Department of Physics and Astronomy, University of Sheffield, Sheffield S3 7RH, UK
}

\date{Feb 2017}


\abstract{This paper presents the spatially resolved 
star formation history (2D-SFH) 
of a small sample of four local mergers: the early-stage 
mergers IC1623, NGC6090, and the Mice, and the more advanced 
merger NGC2623, by analyzing IFS data from the CALIFA survey 
and PMAS in LArr mode. Full spectral fitting techniques are applied 
to the datacubes to obtain the spatially resolved mass growth histories, 
the time evolution of the star formation rate intensity ($\Sigma_{SFR}$), 
and the local specific star formation rate (sSFR), over three different 
timescales (30 Myr, 300 Myr, and 1 Gyr). The results are compared with 
non-interacting Sbc--Sc galaxies, to quantify if there is an enhancement 
of the star formation and to trace its time scale and spatial extent.
Our results for the three LIRGs (IC1623W, NGC6090, 
and NGC2623) show that a major phase of star 
formation is occurring in time scales of 10$^{7}$ yr to few 10$^{8}$ yr, 
with global SFR enhancements of $\sim$2--6 with respect to main-sequence 
star forming (MSSF) galaxies.
In the two early-stage mergers IC1623W and NGC6090, 
which are between first pericenter passage and 
coalescence, the most remarkable increase of the SFR with 
respect to non-interacting spirals occurred in the last 30 Myr, 
and it is spatially extended, with enhancements of factors 2--7 both 
in the centres ($r <$ 0.5 half light radius, HLR), and in the 
disks ($r >$ 1 HLR).
In the more advanced merger NGC 2623 an extended phase of 
star formation occurred on a longer 
time-scale of $\sim$1 Gyr, with a SFR enhancement of a 
factor $\sim$2--3 larger 
than the one in Sbc--Sc MSSF galaxies over the same period, 
probably relic of the first pericenter passage epoch.  
A SFR enhancement in the last 30 Myr is also present, but only in 
NGC2623 centre, by a factor 3. In general, the spatially resolved 
SFHs of the LIRG-mergers are consistent with the predictions 
from high spatial resolution simulations.
In contrast, the star formation in the Mice, 
specially in Mice B, is not enhanced but 
inhibited with respect to Sbc--Sc MSSF galaxies. 
The fact that the gas fraction of Mice B 
is smaller than in most non-interacting spirals, 
and that 
the Mice is close to a prograde orbit, represents a new 
challenge for the models, which must cover a larger 
space of parameters in terms of the the availability of gas 
and the orbital characteristics.}



\keywords{Techniques: Integral Field Spectroscopy -- galaxies: evolution -- galaxies: stellar content -- galaxies: interactions -- galaxies: star formation}
\maketitle

\section{Introduction}
\label{sec:Introduction}

Galaxy interactions, and in particular mergers, enhance the star formation, 
leading to temporarily increased star formation 
rates (SFRs), sometimes by large amounts.
However, even if triggering occurs, the observability time-scale 
of the burst may be 
shorter than the duration of the interaction. In other words, 
a galaxy encounter may not be recognized if observed long 
after the peak of activity.
Mergers of gas rich disks, such as those seen in luminous and 
ultraluminous infrared galaxies, LIRGs and 
ULIRGs \citep{surace1998,surace2000,veilleux2002,kim2013}, are 
extreme systems, where we are directly observing 
the merger-induced star formation \citep{sanders1996}. 
Given their IR luminosities, 
they could reach SFRs as high as $\sim$ 170 M$_{\odot}$ yr$^{-1}$, 
according to the relation between L$_{FIR}$ and SFR 
provided by \cite{kennicutt1998a}. 
However, on average, the SFR increase in mergers 
(not necessarily reaching U/LIRG luminosities) is more modest: a factor 
of a few with respect to control samples of non-interacting 
galaxies \citep{barrera-ballesteros2015,knapen2015}.
This limited SFR enhancement is also predicted by numerical 
modelling \citep{dimatteo2007,dimatteo2008,moreno2015}.  
\cite{dimatteo2007} found that the starbursts induced by major mergers 
have a moderate intensity, with SFRs rarely being enhanced by factors 
larger than 5 compared 
to isolated galaxies, even at the peak of the starbursts. 
However, about 15$\%$ of galaxy interactions models 
have a merger-triggered starburst with a 
relative enhancement higher 
than 5, in agreement with the vast majority of LIRGs and ULIRGs in 
the local universe.
Recent work (both theoretical and observational) 
has shown that minor mergers also
produce SFR enhancements of a few, that are similar to those 
produced by major mergers, and even a factor 10 or more in the most 
extreme cases \citep{kaviraj14,starkenburg16,hernquist89,mihos94}.

The spatial scale of the star formation is another important 
factor that must be taken into account when comparing 
the star formation in mergers with non 
interacting control galaxies. 
Low resolution merger simulations that parameterize the SFR as a function 
of the local gas density (Kennicutt-Schmidt law), and that do not treat 
gas fragmentation, predict that most star formation is strongly concentrated 
toward the central region of the merger remnant, and even 
suppressed in the outskirts \citep{mihos&hernquist1996,moreno2015}.
These models underestimate the spatially 
extended star formation observed in many galaxy mergers from early-stage merger 
or separated progenitors \citep{wang2004,elmegreen2006} to advanced or 
post coalesced systems \citep{evans2008,wilson2006}. 
Recently, \cite{smith2016}, compared the SFRs of 700 star forming complexes 
in 46 nearby interacting galaxy pairs (only early-stage mergers) with those 
in 39 normal spiral galaxies. They found that the regions with highest SFRs
in the interacting systems lie at the intersection of spiral/tidal 
structures. There exists an excess of high SFR clumps in interacting galaxies 
which is not observed in normal spirals.
 
\cite{barnes04} shows that SF is extended, when, instead of adopting 
a purely density dependent SF prescription, energy dissipation in 
shocks is also taken into account.
New high resolution models in the literature that 
determine the SFR efficiency in mergers by resolving parsec scale physical 
processes \citep{teyssier2010,hopkins2013,renaud2015,renaud16}, 
find that extended starbursts 
arise spontaneously (without any SFR prescription) 
due to fragmentation of the gas clouds produced by the increase of the 
supersonic turbulence of the interstellar medium (ISM) 
as a consequence of the tidal interaction itself.  
In these models, a merger-induced nuclear starburst is also present, 
but it occurs later in the merger sequence.
They conclude that the extended star formation is 
important in the early stages of the merger, while nuclear 
starbursts will occur in the advanced stages.
\cite{hopkins2013} note the importance of properly 
treating the stellar feedback in numerical models 
resolving the formation of dense clouds, as predictions critically 
depend on the physics that may (or may not) destroy/disperse those clouds. 

Most stars in the local Universe and up to 
redshift z $\sim$ 2, are not formed in mergers, but in normal 
star forming galaxies along the so-called 
main sequence, as revealed both by 
observations \citep{lofthouse17,rodighiero2011,wolf2005}, 
and simulations \citep{fensch17,dekel09}.
However, mergers are still a crucial 
phase towards the quenching of star formation and morphological 
transformation in galaxies. 
In that sense, local U/LIRGs are unique laboratories 
offering insight into the physical processes triggered by galaxy mergers.
Moreover, the observational characterization of star formation 
in mergers at different stages is necessary to test the validity 
and to constrain merger simulations. 
Current Integral Field Spectroscopy (IFS) 
surveys are very promising for studying merger evolution, 
as they offer spatially resolved information. 

The goal of this paper is to obtain the spatially resolved 
star formation history of  
mergers to determine if there is enhanced star 
formation, its time scale, and 
the spatial extent of these bursts of star formation. 
Here the study is limited to a small sample 
of local mergers: three early stage mergers (Mice, IC1623, and NGC6090) 
and a more advanced merger, NGC2623.
Except for the Mice, the other three systems are LIRGs. 
The four systems are representative 
cases of local mergers at distances of $\leq$127 Mpc, 
which correspond to a spatial 
sampling of $\leq$615 pc/arcsec. 

 This paper is organized as follows: 
Section \ref{sec:Sample} describes the main properties of the galaxies 
analyzed here. Section \ref{sec:Observations} describes briefly the 
observations and data reduction. Section \ref{sec:Results} summarizes 
our method for extracting the SFH, and the spatially resolved results. 
Section \ref{sec:RadialProfiles} deals with the radial profiles of the 
intensity of the star formation ($\Sigma_{SFR}$) and local specific star 
formation rate. In Section \ref{sec:Discussion}, we discuss the results 
in relation to the enhancement of star formation, its spatial extent 
and the time scale of the star formation in comparison with spiral galaxies. 
Section \ref{sec:Summary} reviews of our main findings.

\renewcommand{\arraystretch}{1.3}

\begin{table*} 
\label{tab:sample}
\centering
\begin{tabular}{ccccc}
\hline \hline
\multicolumn{1}{l}{Property}         & Mice	       & IC 1623     &  NGC 6090     & NGC 2623   \\
\hline
\multicolumn{1}{l}{CALIFA ID}	   & 577 (A); 939(B)  &   -		 &  2945	 & 213        \\
\multicolumn{1}{l}{RA}		   & 12 46 10.7       & 01 07 46.3  &  16 11 40.8   & 08 38 23.8  \\
\multicolumn{1}{l}{DEC}			    & +30 43 38        & -17 30 32   &  +52 27 27    & +25 45 17   \\
\multicolumn{1}{l}{Interaction stage}	    & IIIa	   & IIIb	 &  IIIb	 & IV	      \\
\multicolumn{1}{l}{z}			    & 0.022049         & 0.020067    &  0.029304     & 0.018509    \\
\multicolumn{1}{l}{Scale (kpc/$\tt{''}$)}	    & 0.47	   & 0.42	 &  0.61	 & 0.39        \\
\multicolumn{1}{l}{HLR (kpc)}		    & 4.6 (A); 3.8 (B) & 2.8		 &  4.2 	 & 3.3         \\
\multicolumn{1}{l}{Stellar mass (M$_{\odot}$)}  & 1.2$\times$10$^{11}$(A), 1.5$\times$10$^{11}$(B)&3.9$\times$10$^{10}$ &6.8$\times$10$^{10}$&5.4$\times$10$^{10}$ \\
\multicolumn{1}{l}{log(L$_{IR}$[L$_{\odot}$])}  & 10.62	       & 11.65       &  11.51	     & 11.54	  \\
\multicolumn{1}{l}{SFR$_{30 \ Myr}$ (M$_{\odot}$/yr)} &  3(A), 2(B) &     20	 &   51 	 &  8	       \\

\hline \hline
\end{tabular}
\caption{Sample properties. Rows from top to bottom are: (1) CALIFA identification number, 
(2) and (3) Right Ascension and Declination from Equinox J2000 extracted 
from \cite{sanders2003}, (4) Interaction stage
according to \cite{veilleux2002} classification based on the optical morphologies, 
(5) Redshift extracted from NED webpage, (6) Spatial scale, (7) Half light radius, 
(8) Stellar masses extracted from \citet{wild14,cortijo17,cortijo17b}, 
(9) Infrared luminosities 
L$_{IR}$=L(8-1000)$\mu$m from IRAS fluxes extracted from \cite{sanders2003} for the LIRGs. 
They are in logarithm and in units of solar bolometric luminosity, 
L$_{\odot}$=3.83 $\times$ 10$^{33}$ erg s$^{-1}$. Mice L$_{IR}$ 
is extracted from \citet{wild14} and (10) SFR calculated in the last 30 Myr, 
from the spectral synthesis in this paper.}

\end{table*}

\begin{figure*}
\centering
\includegraphics[width=0.45\textwidth]{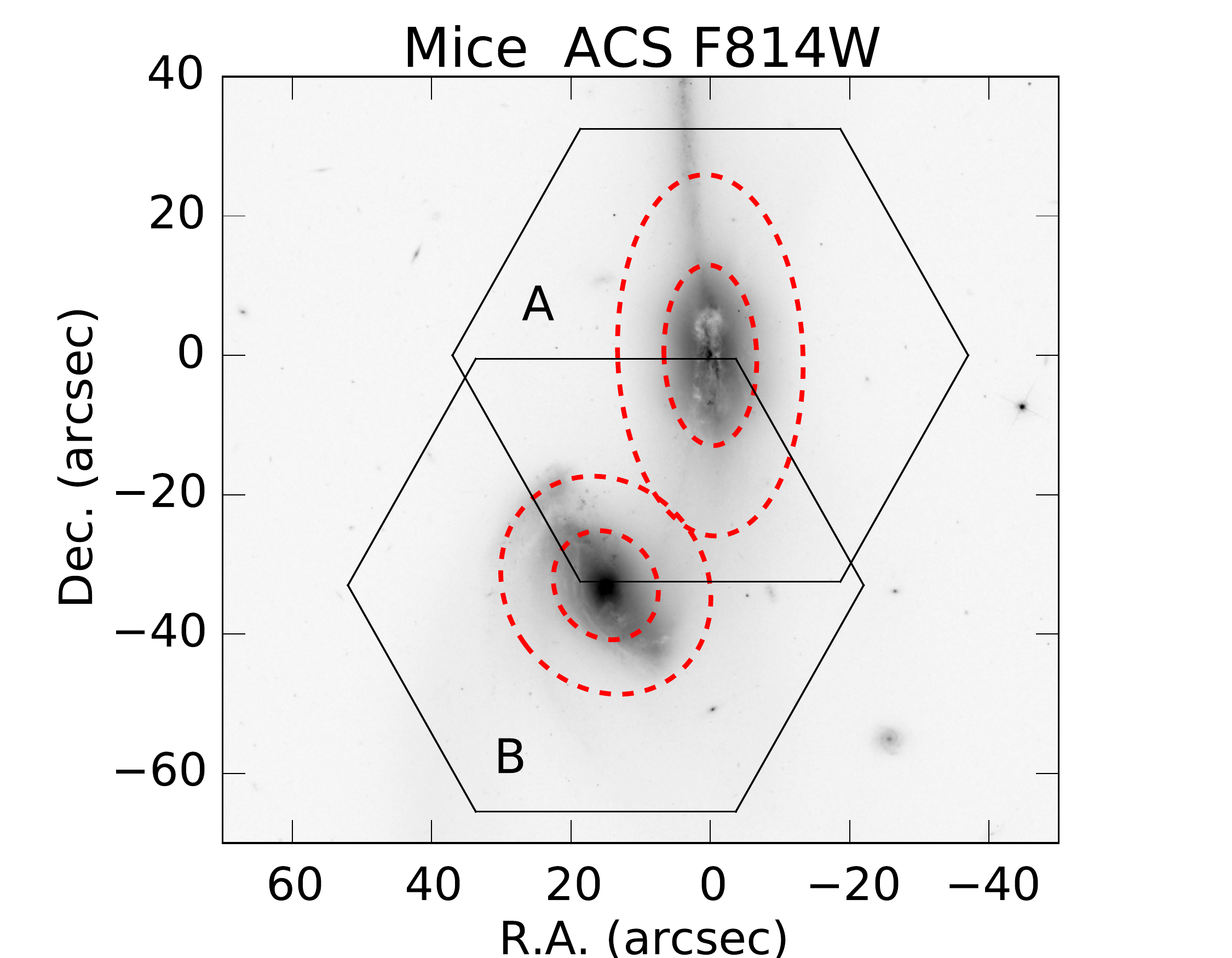}
\includegraphics[width=0.45\textwidth]{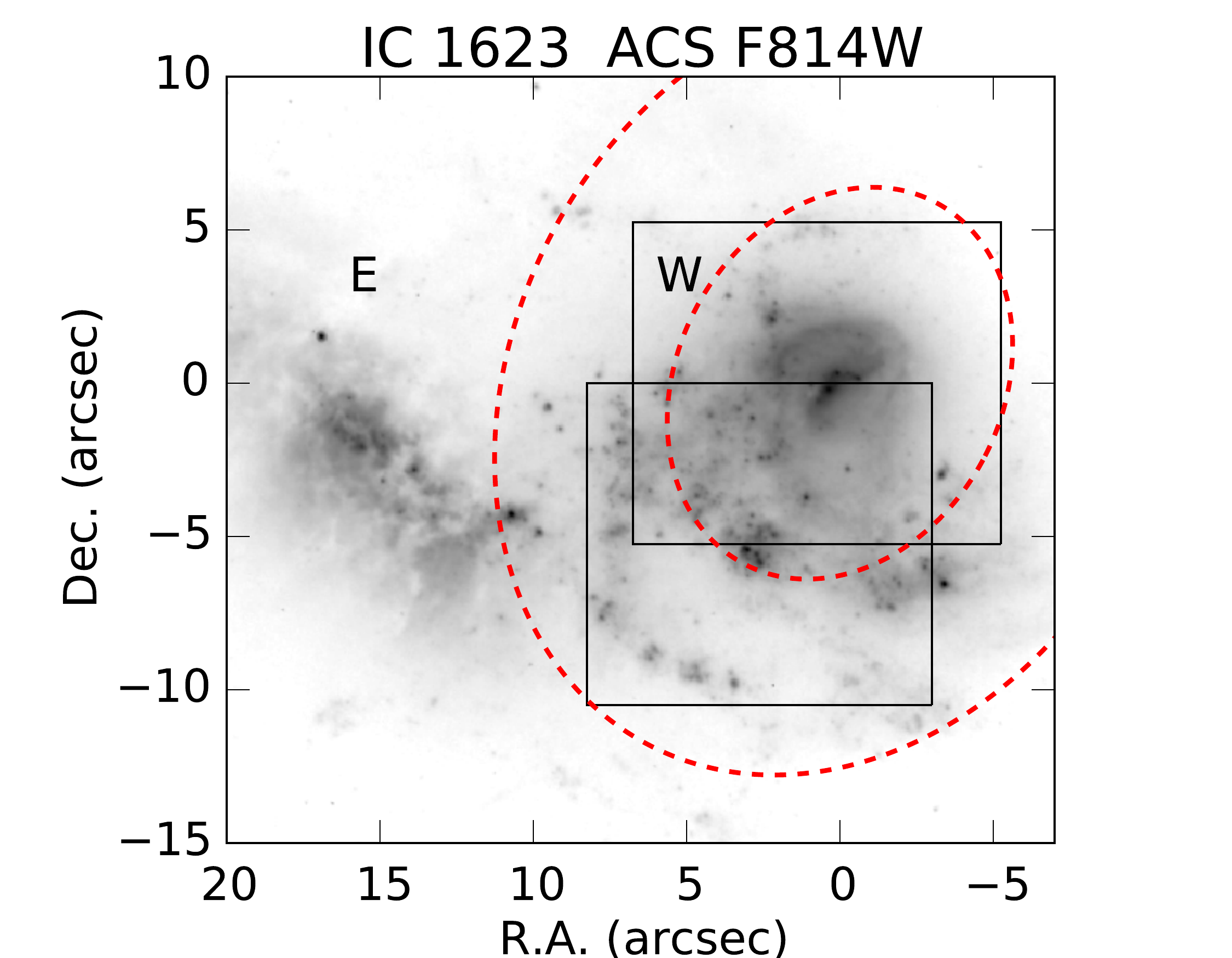}
\includegraphics[width=0.45\textwidth]{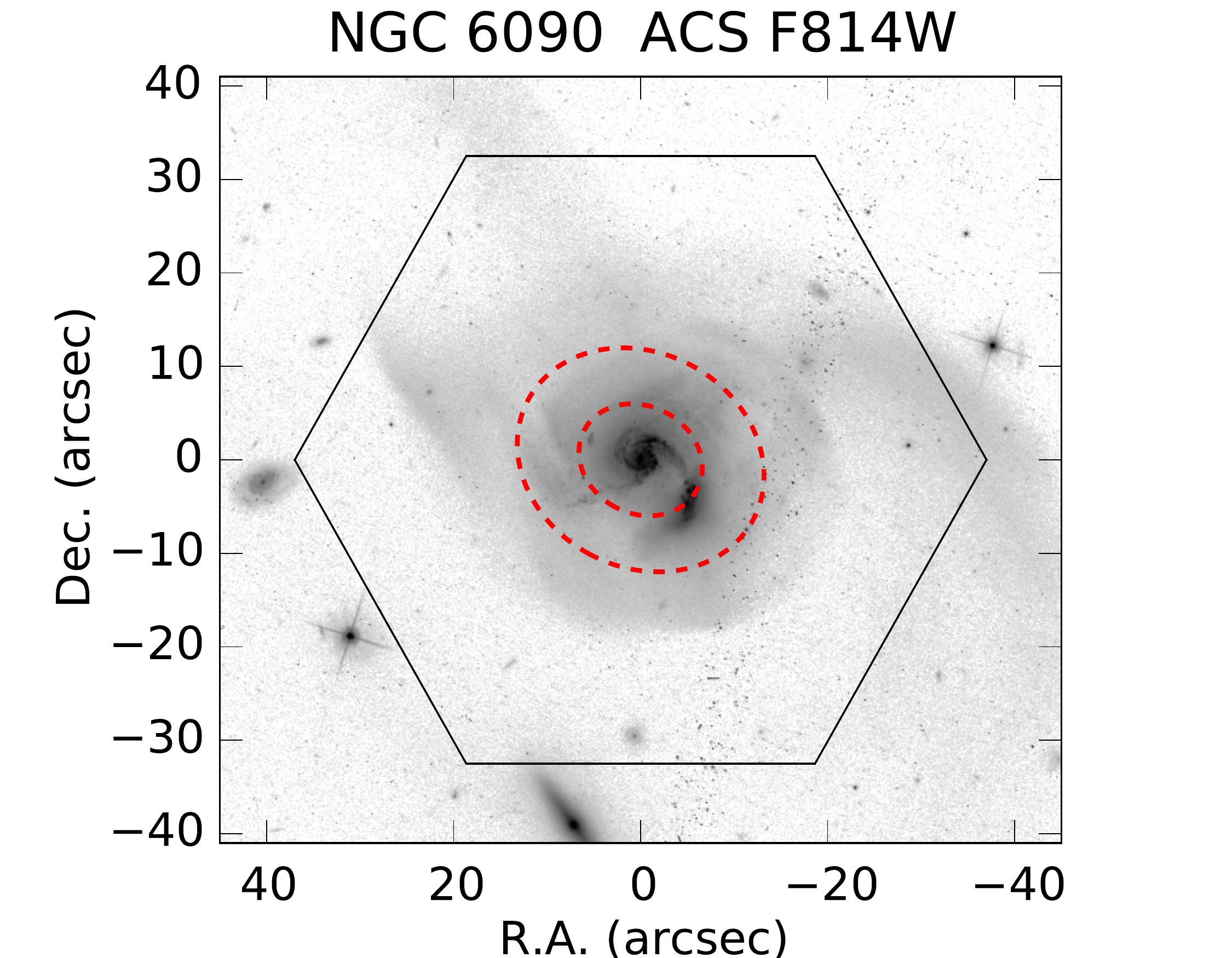}
\includegraphics[width=0.45\textwidth]{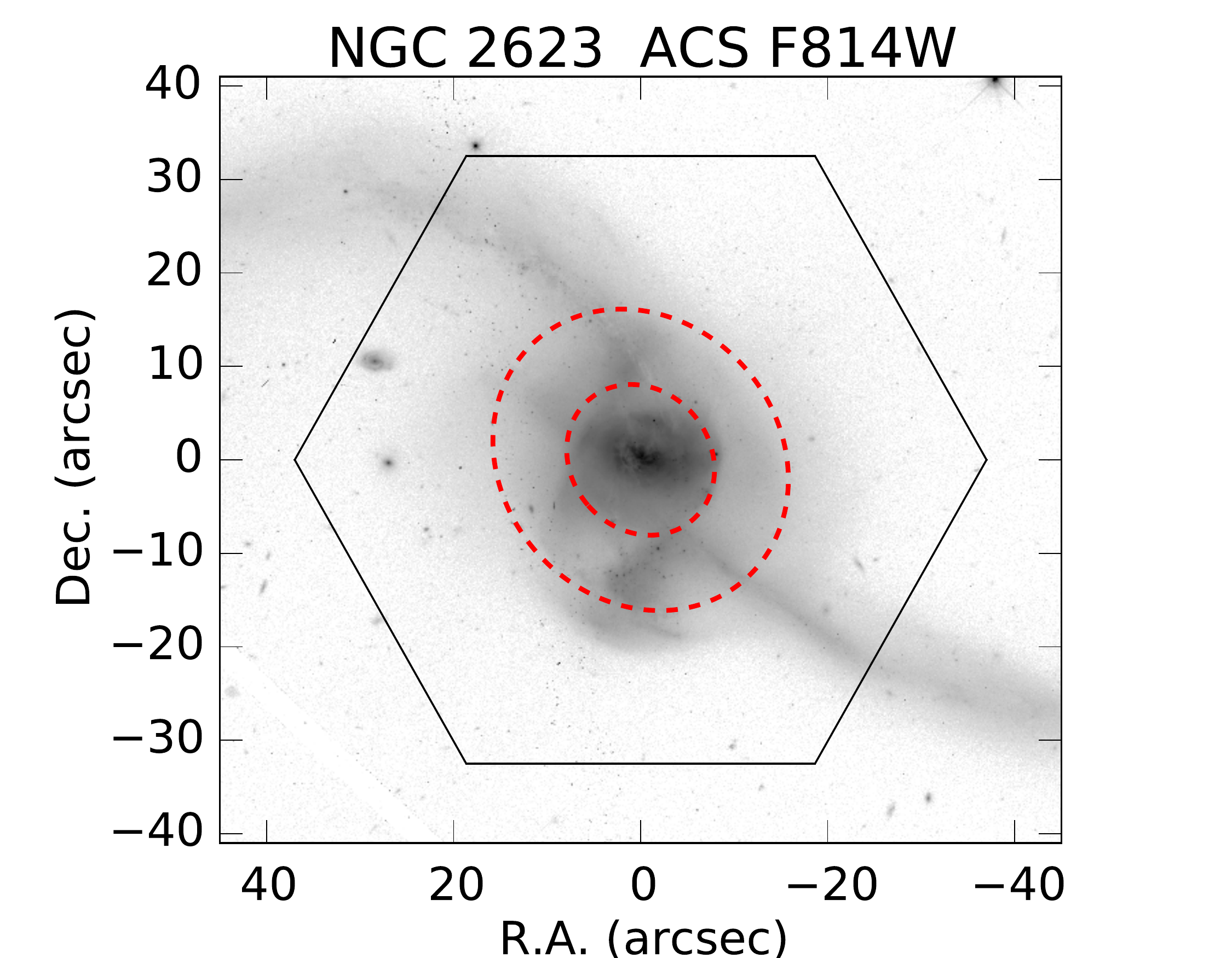}
\caption{HST/ACS (F814W) images of the four galaxies in our sample. 
The fields observed with IFS are shown by hexagons (PPaK) or squares (LArr). 
The red dashed circles indicate the position 
of 1 and 2 HLR centered at the nuclei of the main galaxy.} 
\label{fig:lumden_hlr}
\end{figure*}

\begin{figure*}
\centering
\includegraphics[width=0.48\textwidth]{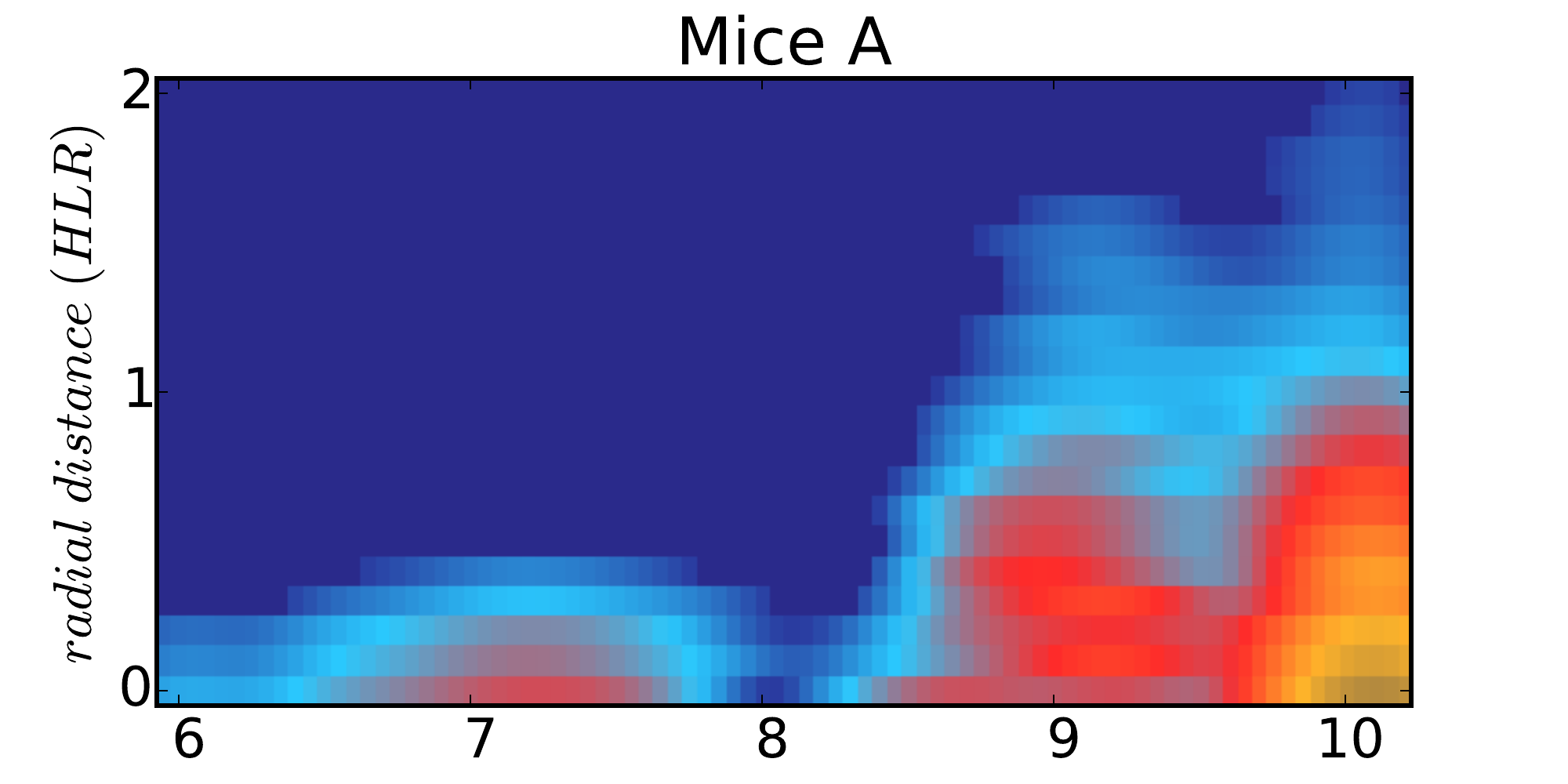}
\includegraphics[width=0.48\textwidth]{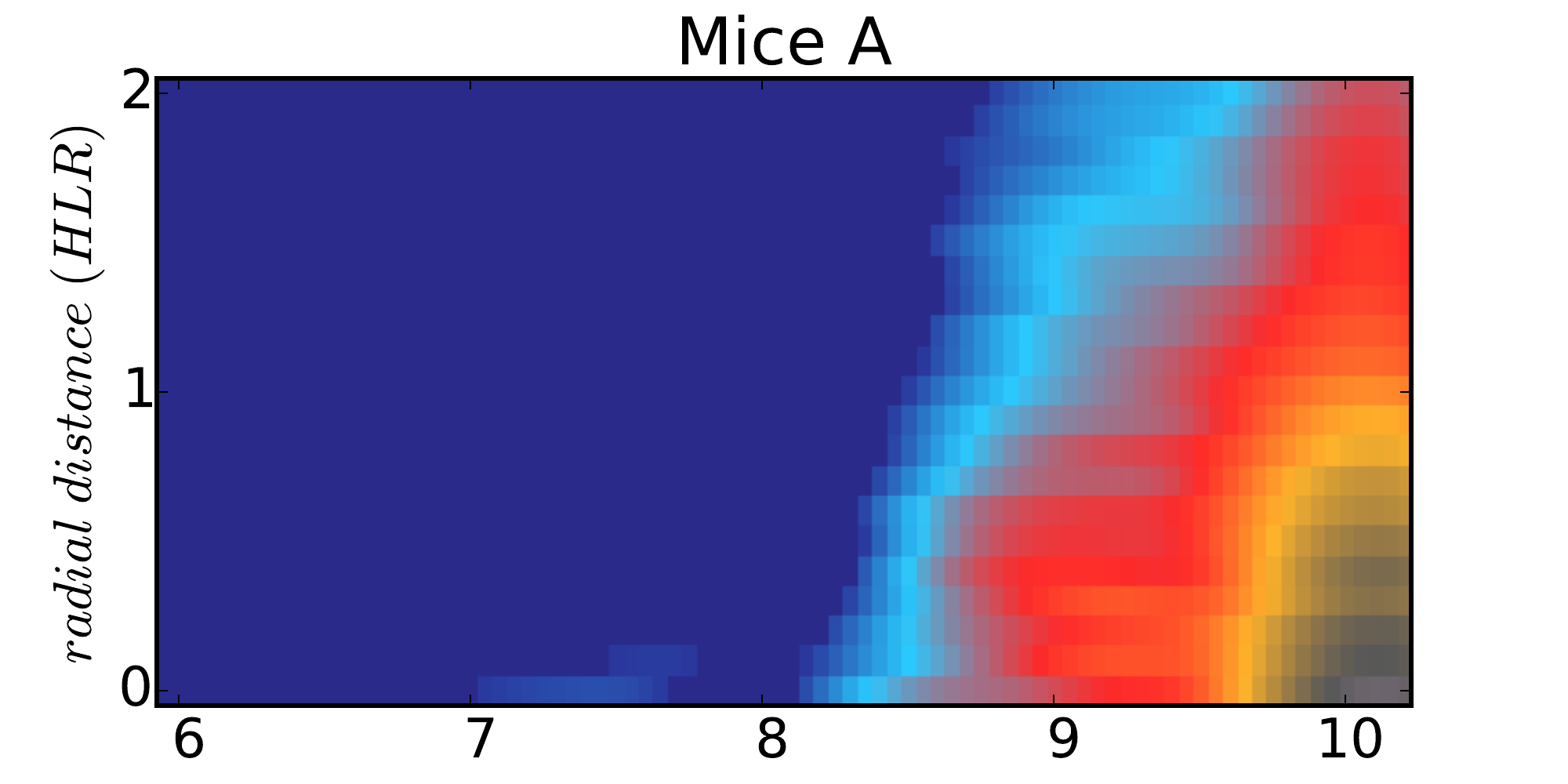}
\includegraphics[width=0.48\textwidth]{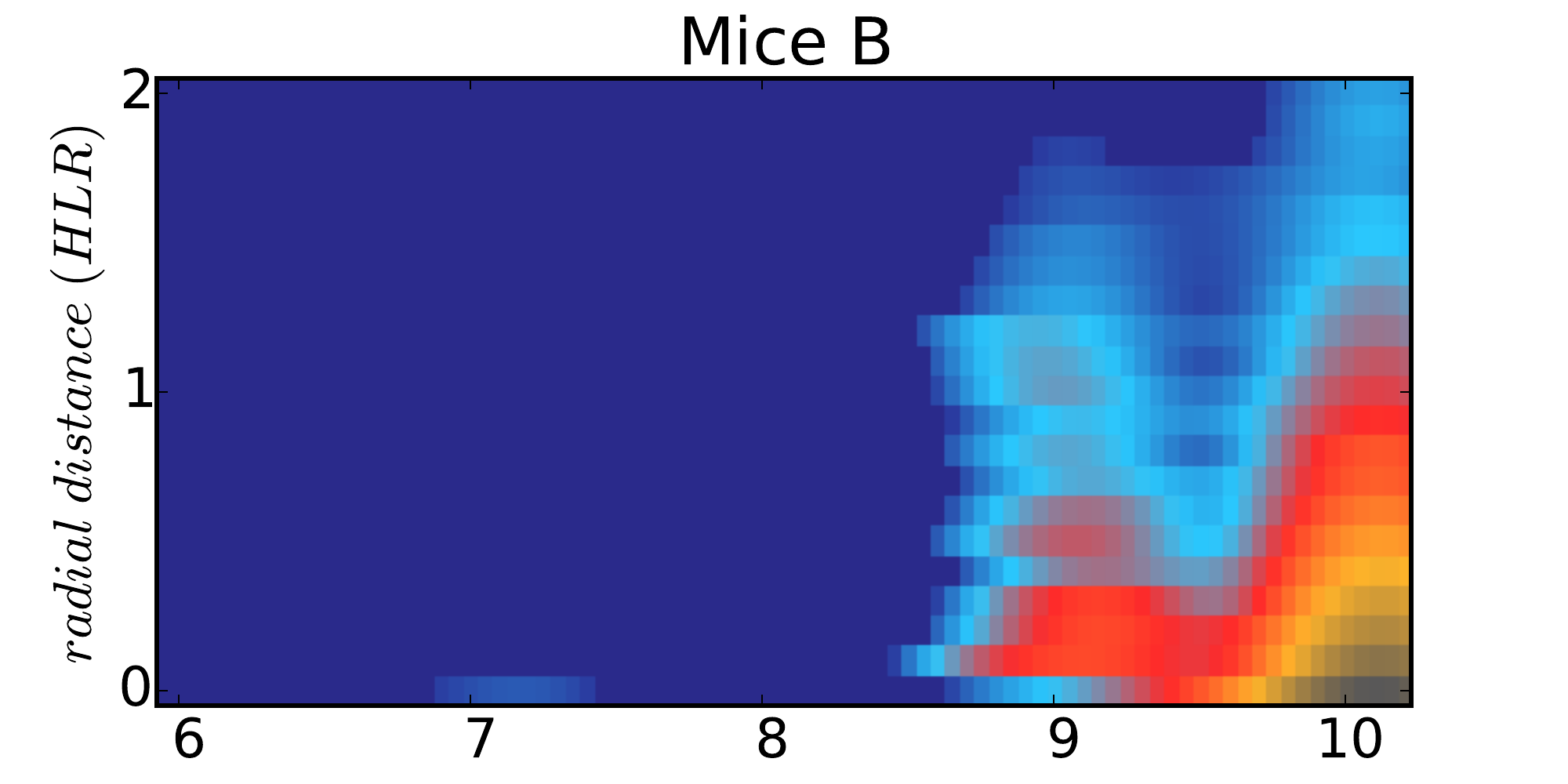}
\includegraphics[width=0.48\textwidth]{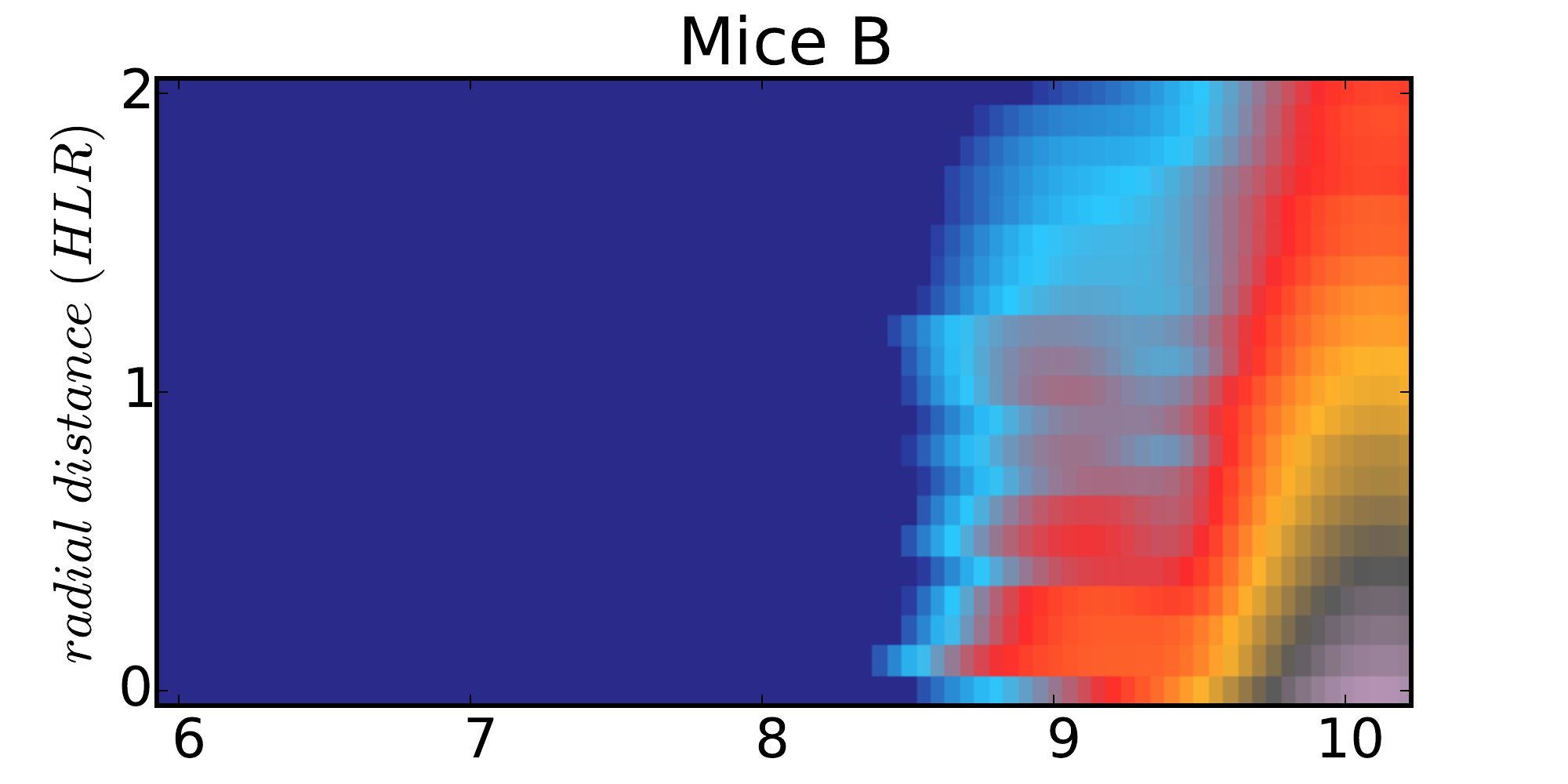}
\includegraphics[width=0.48\textwidth]{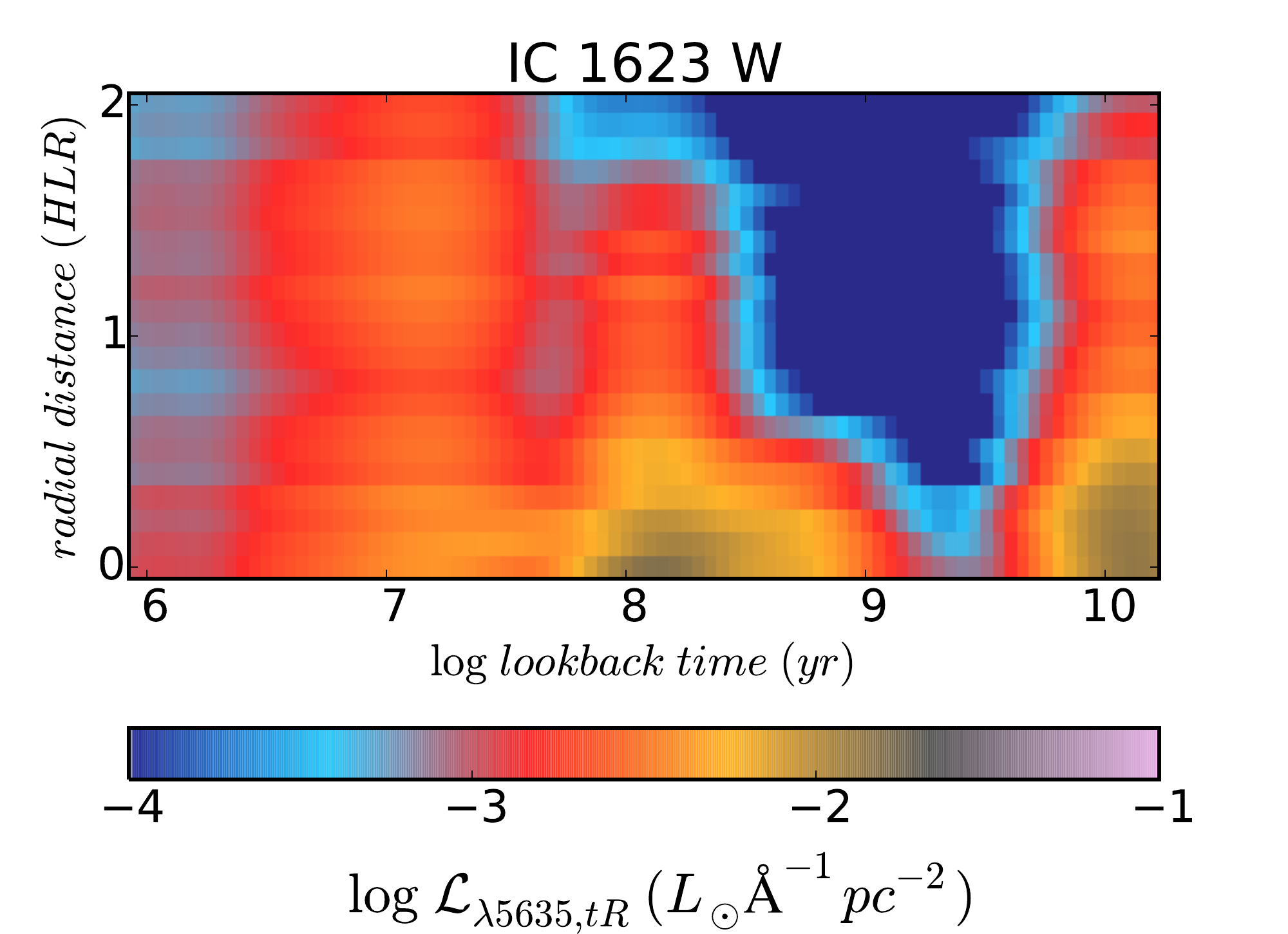}
\includegraphics[width=0.48\textwidth]{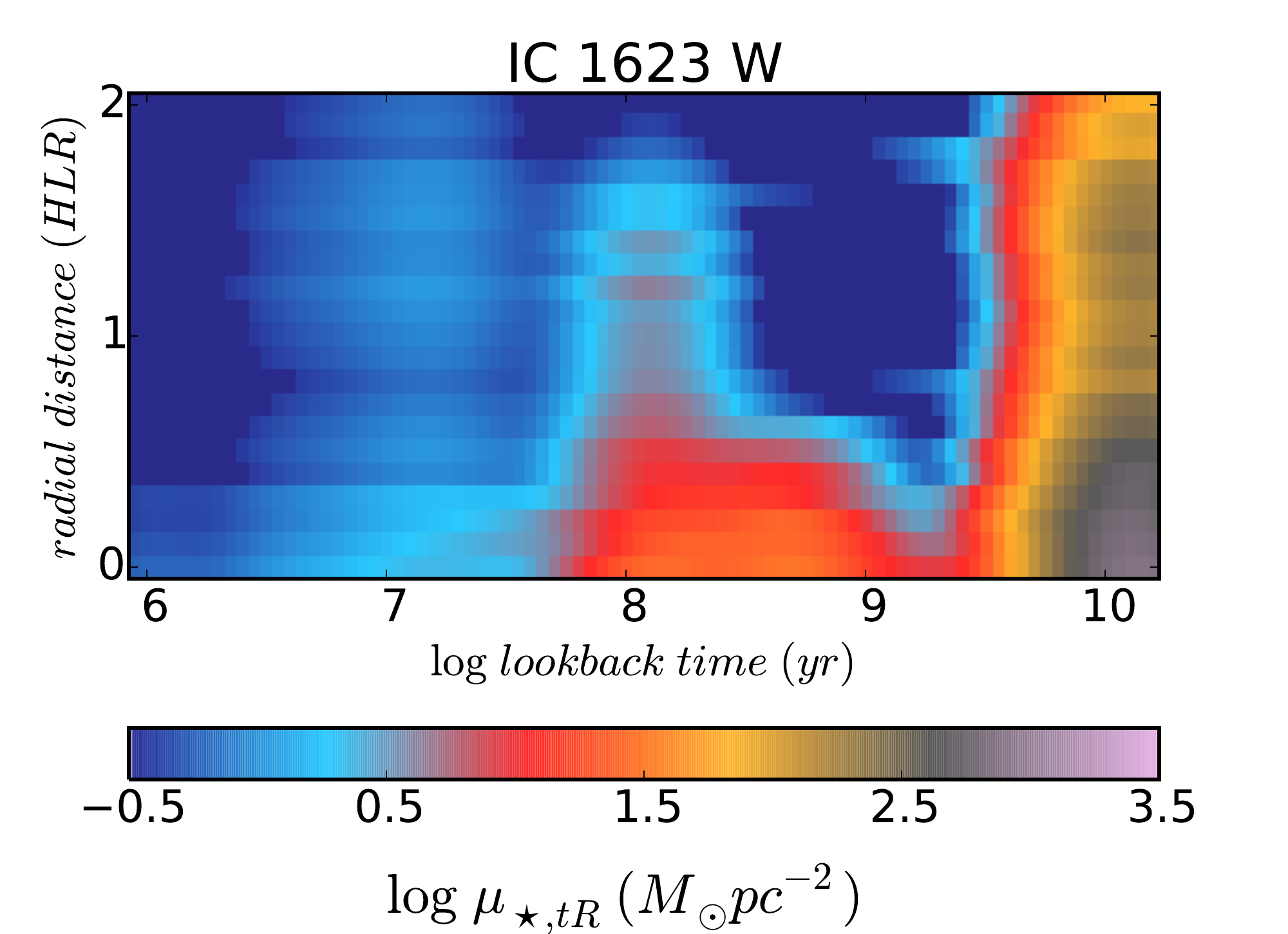}
\caption{{\em $R\times$t } diagrams showing the radially averaged distribution of the 
mass and light as a function of the distance (in HLR units) from the 
nucleus and lookback time (in log units). The intensity of the 
map shows: Stellar mass formed per unit area ($M_\odot \ pc^{-2}$) (right panels), 
and luminosity at $5635 \AA$ per unit area ($L_\odot  \ \AA^{-1} \ pc^{-2}$) (left panels). 
From top to bottom: Mice A, Mice B, IC1623W. The figure continues in next page.} 
\label{fig:2DSFH}
\end{figure*}
\addtocounter{figure}{-1}
\begin{figure*}
\centering
\includegraphics[width=0.48\textwidth]{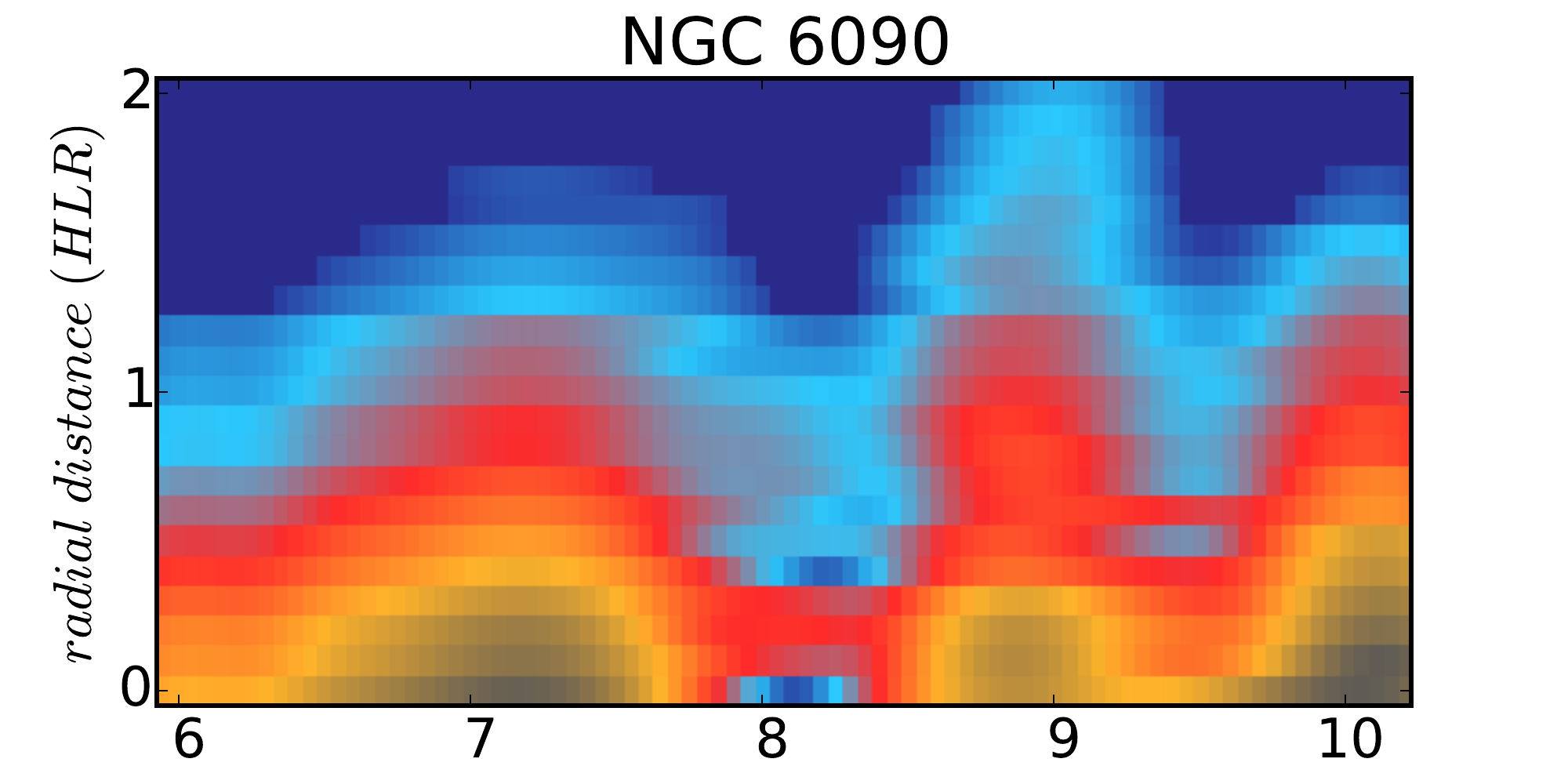}
\includegraphics[width=0.48\textwidth]{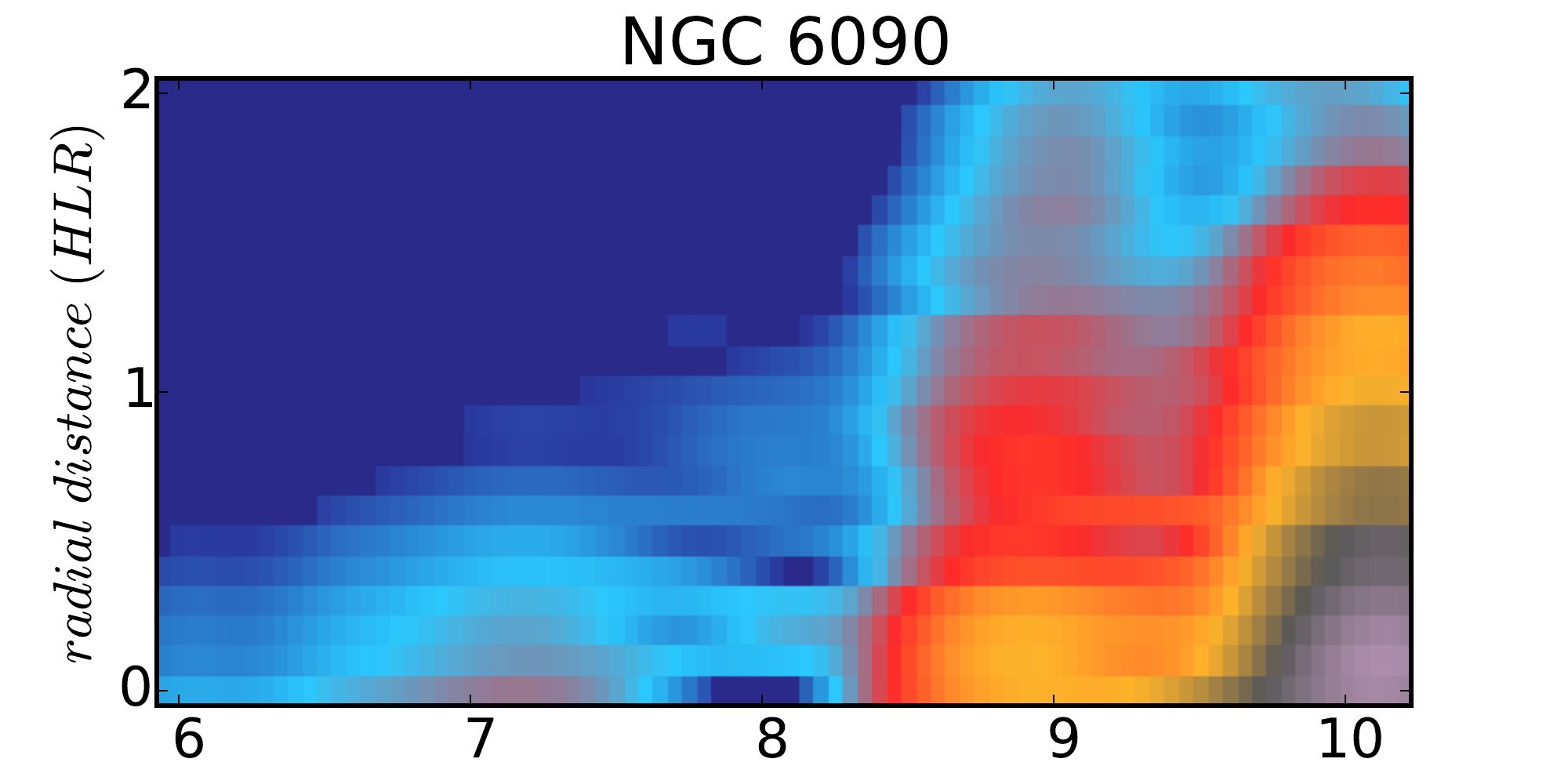}
\includegraphics[width=0.48\textwidth]{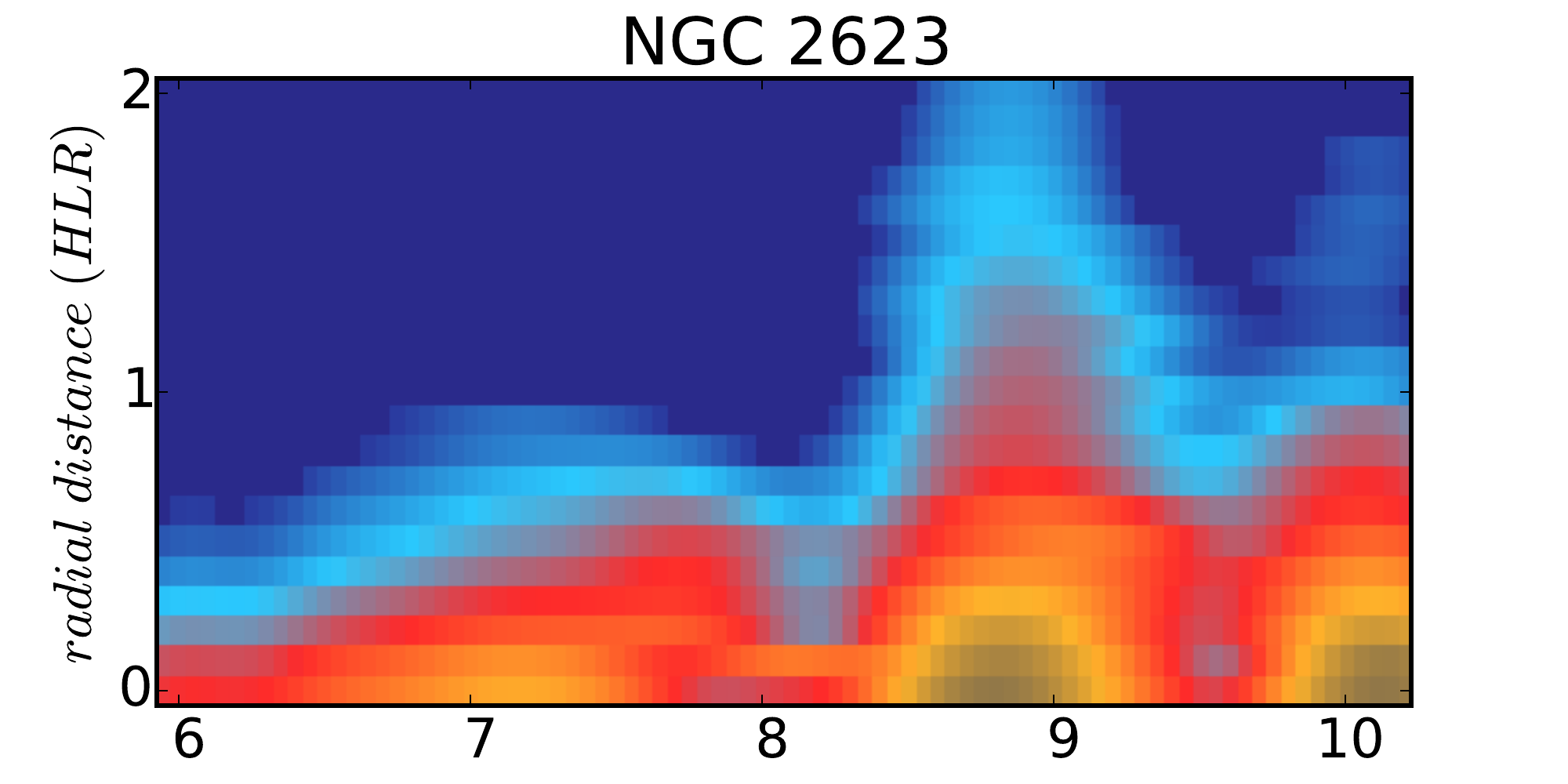}
\includegraphics[width=0.48\textwidth]{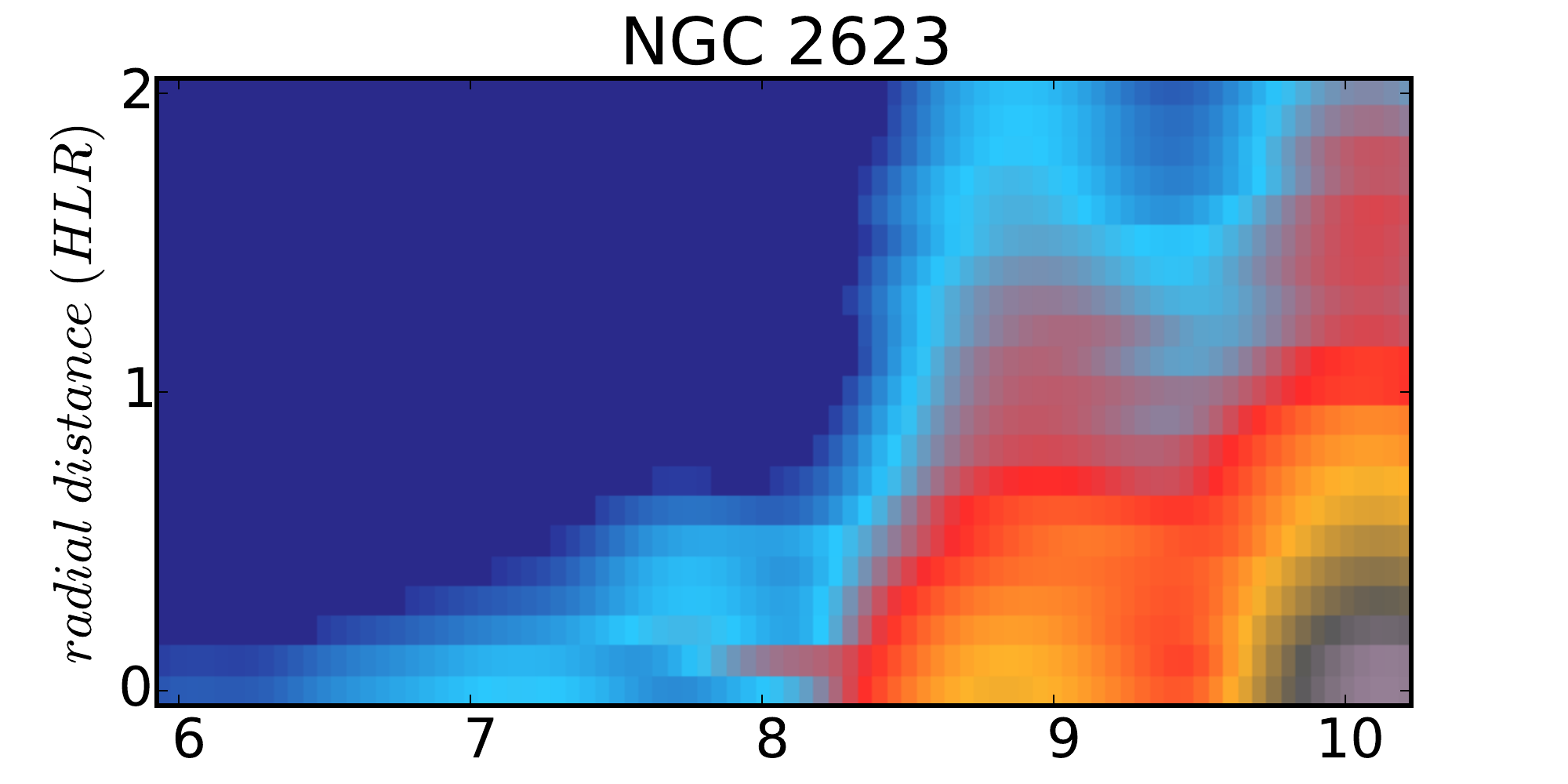}
\includegraphics[width=0.48\textwidth]{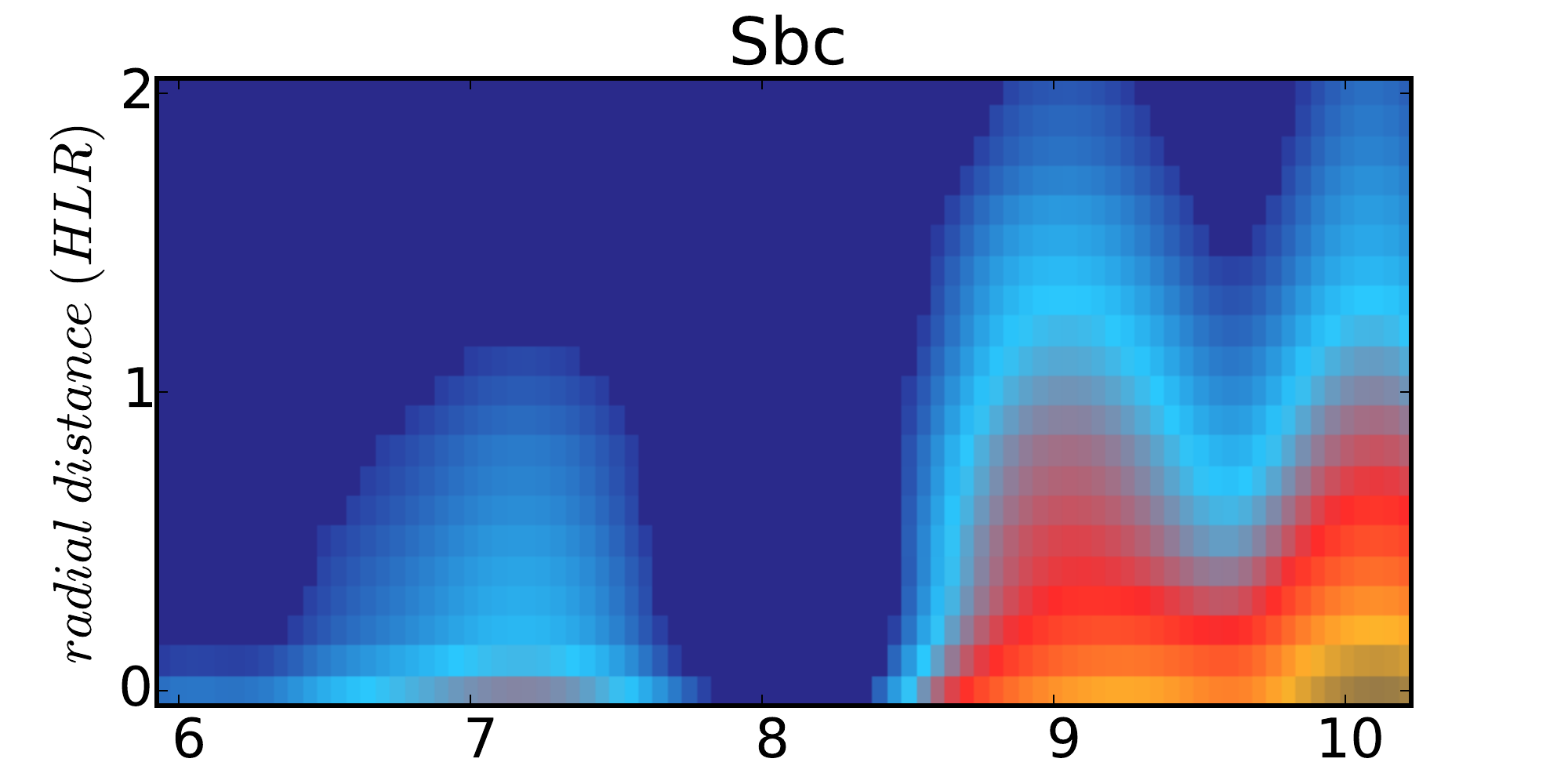}
\includegraphics[width=0.48\textwidth]{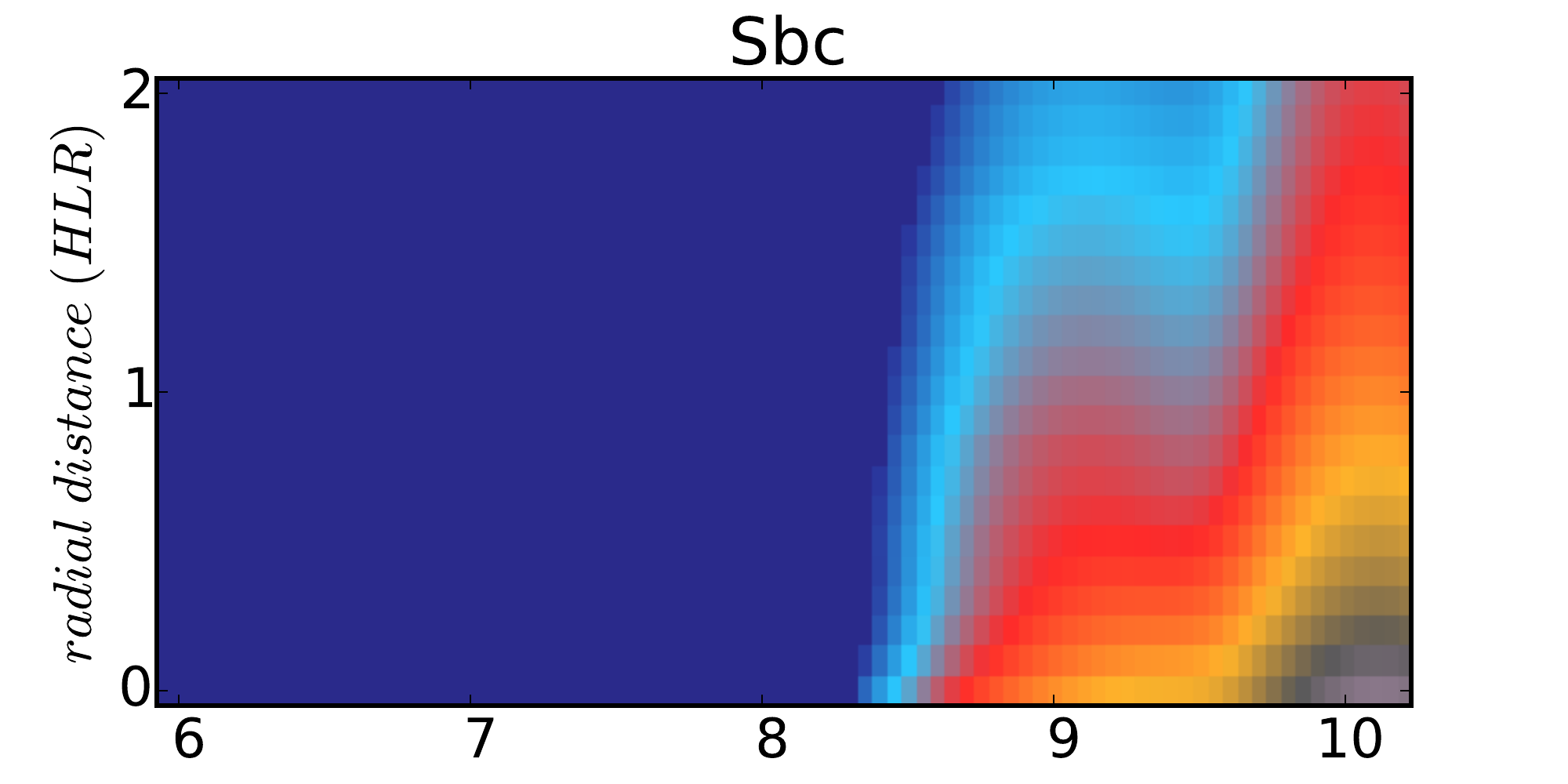}
\includegraphics[width=0.48\textwidth]{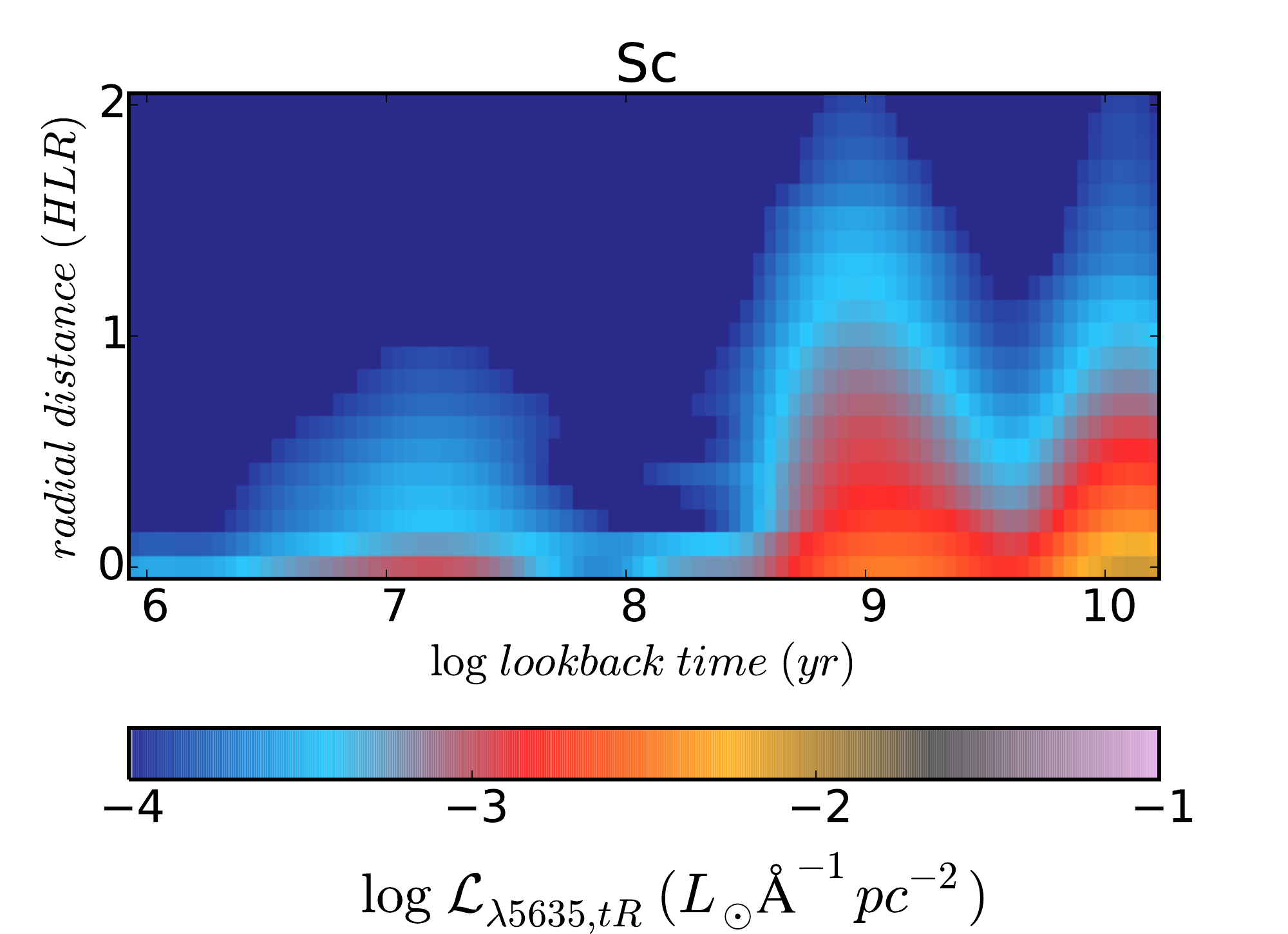}
\includegraphics[width=0.48\textwidth]{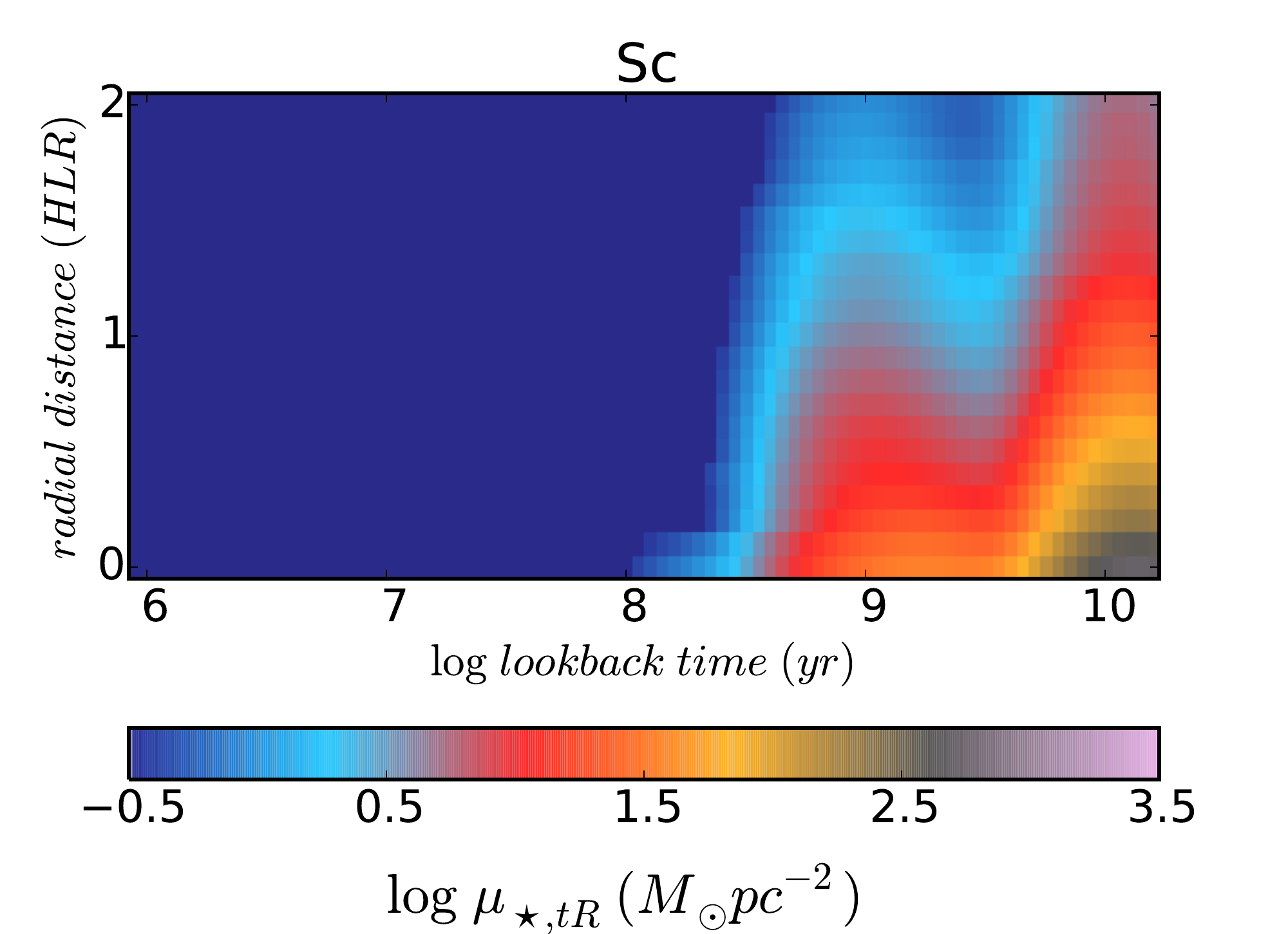}

\caption{{\it Cont.} From top to bottom: 
{\em $R\times$t} diagrams for NGC6090, NGC2623, Sbc, and Sc control spirals.} 
\label{fig:2DSFHbis}
\end{figure*}


\section{Sample}
\label{sec:Sample}

Our study concentrates on four mergers: the Mice, IC1623, NGC6090, 
and NGC2623. Except for IC1623, the other systems are part of the 
CALIFA survey \citep{sanchez12, sanchez16}. These galaxies do not make 
a complete sample of 
interacting/merger systems of CALIFA, 
but they are already well characterized systems in terms 
of the spatially resolved properties,
and they are proto-type of early-stage 
mergers (Mice, IC1623, NGC6090) and more advanced mergers (NGC2623). 
Thus, they constitute
a suitable sample for this pilot study that will be extended 
in the future for a larger CALIFA sub-sample. 
Based on IFS data, a detailed analysis on the spatially resolved stellar 
population properties for IC1623 and NGC6090 are presented 
in \citet{cortijo17}, for NGC2623 in \citet{cortijo17b}, and for the Mice 
in \citet{wild14}. The main properties of the sample are summarized 
in Table \ref{tab:sample}. Also, 
a summary of the spatially resolved 
stellar population properties is:

\begin{itemize}

\item Mice: The stellar populations of both galaxies are predominantly older 
than several Gyrs. Only the northern tidal tail and the NE arm 
of NGC 4676B have distinctly younger stellar populations, 
with light-weighted ages of $\sim$0.6 Gyr. 
The fraction of stellar mass contributed 
by stars younger than 140 Myr is less that 5$\%$ in all regions, 
although $\sim$30$\%$ of optical light in the nuclear regions of NGC 4676A 
and tidal arm of NGC 4676B arises from stars younger than 140 Myr. 
There is an excess of light from intermediate age stars 
seen in the western half of NGC 4676A, inter-galaxy region, and NE bar 
of NGC 4676B, compared to the disks of these galaxies. This is consistent 
with the triggering of low level star formation in the gas flung out 
from the galaxy disks at first passage $\sim$170 Myr ago, 
perhaps through dynamic instabilities or shocks. 
However, the ongoing merger has had little significant 
effect on the stellar populations in terms of total stellar mass or 
global star formation rate. In agreement with the observations, 
a standard hydrodynamical merger simulation of the Mice also shows little 
enhancement in star formation during and 
following first passage \citep{wild14}.

\item IC 1623: Merger-induced star formation is extended and recent, 
as revealed by the widespread star clusters. The W progenitor has 
an average light-weighted age of $\sim$50 Myr. IC 1623 W centre ($<$0.5 HLR) 
is older, $\sim$300 Myr, 
than the surrounding area, $\sim$30 Myr, due to the presence 
of a significant contribution of intermediate-age stellar 
populations, probably induced in a previous stage 
of the merger. However, the light is everywhere dominated 
by young stellar populations, with contributions  
much higher than the typical 10--20\% in Sbc/Sc control galaxies. 
The extinction in the W progenitor 
is $\sim$0.2 mag, and has a flat radial profile in comparison with 
Sbc/Sc control spirals. However, we note that IC 1623 E is affected by much 
higher levels of extinction (up to 2--6 mag) according to the star 
cluster photometry. IC 1623 W has a mean 
stellar metallicity of $\sim$0.6 Z$_{\odot}$, with a flat or 
slightly positive radial profile. 
The average SFR($t <$ 30Myr) of 
IC 1623 W is a factor 6 above the SFR of main-sequence Sbc-Sc galaxies 
of the same mass \citep{cortijo17}.

\item NGC 6090: Merger-induced widespread clusters are also present in this 
early-stage merger, with an average light-weigthed age of the stellar populations 
$\sim$50--100 Myr. 
The age profiles are significantly flatter 
than in Sbc/Sc galaxies, in agreement with a general rejuvenation 
of the progenitor galaxies due to merger-induced star formation.
There exists a difference of $\sim$ 1 mag between the 
stellar extinctions of the two progenitors, with an average 
of 1.3 mag in NGC 6090 NE, 
and 0.5 mag in NGC 6090 SW, and whose 
radial profiles are flat in comparison 
with Sbc/Sc control spirals. NGC 6090 
has a mean stellar metallicity of $\sim$0.6 Z$_{\odot}$, 
with a flat or slightly positive radial profile.
The average SFR($t <$ 30Myr) is increased by a factor 9 with 
respect to main-sequence Sbc-Sc galaxies. In this study we 
will better quantify the spatial extension and timescale of 
the SFR enhancement \citep{cortijo17}.

\item NGC 2623: There were two periods of 
merger-induced star formation in NGC 2623: 
a recent episode, 
traced by young stellar populations ($<$ 140 Myr) located 
in the innermost ($<$ 0.5 HLR $\sim$ 1.4 kpc) 
central regions, and in some isolated clusters 
in the northern tidal tail and in the star forming knots south 
of the nucleus; and a previous and 
widespread ($\sim$ 2 HLR $\sim$ 5.6 kpc) episode, traced 
by the spatially extended intermediate-age stellar 
populations with ages between 140 Myr--1.4 Gyr. 
Due to this distribution of the stellar populations, the centre of 
NGC 2623 ($<$0.2 HLR) is younger, 500 Myr on average,
than at 1 HLR, $\sim$ 900 Myr, and the age gradient 
is positive ($\sim$ 400 Myr), in constrast with the negative 
gradients in Sbc--Sc galaxies.
The stellar extinction is high in the inner 0.2 HLR ($\sim$ 1.4 mag), 
and shows a negative gradient, which is much 
steeper ($-0.9$ mag) than in Sbc--Sc galaxies, $-0.3$ mag, 
indicating that in the last stages of a merger most of the 
gas and dust content of both progenitors is concentrated in the 
mass centre of the system \citep{cortijo17b}.

\end{itemize}

\renewcommand{\arraystretch}{1.3}

\begin{table*} 
\caption{Light and mass contributions of SSPs over three 
timescales ($t \leq$ 30 Myr, 300 Myr, and 1 Gyr) and for three 
spatial regions: from 0 -- 0.5 HLR (top), 
1 -- 1.5 HLR (middle), and 0 -- 2 HLR (bottom).}
\label{tab:natbib}
\centering
\begin{tabular}{cccccccc}
\hline \hline
\multicolumn{6}{ c }{0.0 -- 0.5 HLR} \\
\hline \hline
Light ($\%$)      &  Sbc   &  Sc       & Mice A   &   Mice B	 & IC 1623 W   & NGC 6090  & NGC 2623   \\   
\hline
\multicolumn{1}{l}{SSPs $\leq$ 30 Myr}   &  8.0(0.2)     &  13.3(0.2)  & 13.5(0.7) &   
1.4(0.1)   &  20.5(2.7)         &  49.2(0.8)	   & 21.7(0.6)           \\ 
\multicolumn{1}{l}{SSPs $\leq$ 300 Myr}  &  8.4(0.2)     &  17.7(0.1)   & 17.1(0.7) &   
1.4(0.1)   &  57.1(0.9)         &  52.20(0.02)	   & 29.9(0.3)           \\ 
\multicolumn{1}{l}{SSPs $\leq$ 1 Gyr}  &  19.0(2.0)  &  32.4(2.9)      & 27.0(0.9)&   
4.2(0.4)	 &  70.9(0.9)         &  67.5(0.9)	   & 64.7(1.1)           \\ 
\hline
Mass ($\%$)       &  Sbc   &  Sc       & Mice A   &   Mice B	 & IC 1623 W   & NGC 6090  & NGC 2623   \\   
\hline
\multicolumn{1}{l}{SSPs $\leq$ 30 Myr}   &  0.061(0.004)   &  0.133(0.006)    & 0.13(0.02)	&  
0.007(0.002) &  0.6(0.1)        &  0.66(0.02)	   &  0.4(0.1)        \\  
\multicolumn{1}{l}{SSPs $\leq$ 300 Myr}  &  0.09(0.01)   &  0.27(0.01)    &0.3(0.1)	  &  
0.007(0.002) &  4.2(0.3)        &  0.92(0.01)	   &  1.1(0.1)         \\ 
\multicolumn{1}{l}{SSPs $\leq$ 1 Gyr}  &  2.2(0.5)   &  4.3(0.9)      & 2.5(0.4)     &  
0.5(0.1) &  7.4(0.3)       &  6.5(0.4)	   &  10.6(0.7)        \\ 
\hline \hline
\\
\hline \hline
\multicolumn{6}{ c }{1.0 -- 1.5 HLR} \\
\hline \hline
Light ($\%$)      &  Sbc   &  Sc       &  Mice A  &  Mice B	 & IC 1623 W   & NGC 6090  & NGC 2623	\\
\hline
\multicolumn{1}{l}{SSPs $\leq$ 30 Myr}   &  16.7(0.2)    &  15.2(0.4)     & 2.9(0.1)	 &  
8.4(0.3)	 &  57.7(0.5)         &  33.1(1.7)	   &  10.2(0.5)     	 \\
\multicolumn{1}{l}{SSPs $\leq$ 300 Myr}  &  22.7(0.4)    &  24.2(0.4)       & 5.1(0.4)	  &  
9.5(0.4)	 &  79.7(0.3)         &  43.9(1.0)	   &  17.8(1.1)      \\
\multicolumn{1}{l}{SSPs $\leq$ 1 Gyr}  &  42.5(3.1)    &  52.8(3.6)       & 17.8(0.6)	  &  
28.1(3.7) 	 &  79.7(0.3)         &  66.2(1.6)	   &  60.9(2.1)      \\ 
\hline
Mass ($\%$)       &  Sbc   &  Sc       &  Mice A  &  Mice B	 & IC 1623 W   & NGC 6090  & NGC 2623	\\
\hline
\multicolumn{1}{l}{SSPs $\leq$ 30 Myr}    &  0.23(0.01)   &  0.30(0.02)    & 0.026(0.001)	  & 
0.072(0.004) 	 &  1.37(0.01)        &  0.7(0.1)	   &  0.28(0.02)    	\\ 
\multicolumn{1}{l}{SSPs $\leq$ 300 Myr}  &  0.7(0.1)   &  1.1(0.1)      & 0.2(0.1)	  & 
0.15(0.02) 	 &  3.8(0.1)       &  1.8(0.1)	   &  1.5(0.3)    	\\ 
\multicolumn{1}{l}{SSPs $\leq$ 1 Gyr}  &  7.6(1.3)   &  13.2(1.8)       & 2.6(0.2)	  & 
3.9(0.8) 	 &  3.8(0.1)        &  11.0(0.9)	   &  23.1(1.7)     	\\ 
\hline \hline
\\
\hline \hline
\multicolumn{6}{ c }{0.0 -- 2.0 HLR} \\
\hline \hline
Light ($\%$)      &  Sbc   &  Sc       &  Mice A  &  Mice B	& IC 1623 W   & NGC 6090   & NGC 2623	\\    
\hline
\multicolumn{1}{l}{SSPs $\leq$ 30 Myr}  &  10.5(0.2)    &  14.0(0.2)     &  10.2(0.4)	  & 
2.9(0.1) 	&  35.5(2.1)	      &  46.8(0.8)	   &  19.3(0.6)	 \\   
\multicolumn{1}{l}{SSPs $\leq$ 300 Myr} &  12.2(0.3)  & 19.4(0.2)   &  13.5(0.4)       &  
3.2(0.2)	&  65.6(1.6)	      &  51.1(0.1)	   &  28.3(0.4) 	 \\   
\multicolumn{1}{l}{SSPs $\leq$ 1 Gyr}   &  25.3(2.3)    &  37.9(3.2)       &  25.6(1.1)	  & 
8.1(1.0) 	&  73.6(0.4)	      &  66.9(1.0)	   &  62.6(1.2)	\\    
\hline
Mass ($\%$)       &  Sbc   &  Sc       &  Mice A  &  Mice B	& IC 1623 W   & NGC 6090   & NGC 2623	\\	   
\hline
\multicolumn{1}{l}{SSPs $\leq$ 30 Myr}   &  0.144(0.007)   &  0.22(0.01)      & 0.062(0.005)	& 0.034(0.003) 	&  
1.0(0.1)        &  0.7(0.1)	   &  0.3(0.1)	    \\  
\multicolumn{1}{l}{SSPs $\leq$ 300 Myr}  &  0.36(0.03)   &  0.6(0.1)      &  0.2(0.1)  & 0.08(0.02) 	&  
3.7(0.1)        &  1.2(0.1)	   &  1.3(0.2)	    \\  
\multicolumn{1}{l}{SSPs $\leq$ 1 Gyr}    &  4.8(0.9)   &  8.3(1.4)      &  2.9(0.3)	 &1.5(0.3)     &  
4.9(0.2)  &   8.1(0.7)	   &  13.4(0.9)	  \\  
\hline \hline\hline
\end{tabular}
\\
{\small
\small }
\end{table*}


\section{Observations and data reduction}
\label{sec:Observations}

The observations were carried out at Calar Alto observatory (CAHA) with 
the 3.5m telescope and the Potsdam Multi-Aperture 
Spectrometer PMAS \citep{roth05}. For IC1623 we used data with the Lens 
Array (LArr) configuration with a spatial magnification of $0.75{\tt''}$/lens, 
covering a $12{\tt''} \times 12{\tt''}$ field of view (FoV). 
Two pointings were taken to cover the whole western galaxy 
(see \citet{cortijo17} for details). For the other systems, 
we use the PPaK mode \citep{verheijen04} with
a FoV of $74{\tt''} \times 64{\tt''}$, and 382 fibers 
of $2.7{\tt''}$ diameter each \citep{kelz06}. IC1623 was observed with 
the V300 grating and the other galaxies using a combination of the 
gratings V500 and V1200, named in CALIFA as COMBO mode. These different 
set ups, however, cover a similar wavelength range $\sim 3700-7100 \AA$ and 
with a spectral resolution of FWHM $\sim 7 \AA$ for V300 and $\sim 6 \AA$ for 
COMBO data. While in the LArr data the spatial sampling is $0.75{\tt''}$/spaxel, 
for PPaK the final sampling is 1 arcsec/spaxel. 

Data of IC1623 was calibrated as explained in \citet{cortijo17} and the data 
for the other galaxies were calibrated with version V1.5 of the reduction 
pipeline \citep{garciabenito15}. We refer to \citet{sanchez12}, 
\citet{husemann13}, \citet{garciabenito15}, and \citet{sanchez16} for 
details on the observational strategy and data processing of these data cubes.
 
In Fig.\ \ref{fig:lumden_hlr} we show 
the HTS/ACS (F814W) maps of the four 
galaxies in our sample. The red dashed lines indicate the position 
of 1 and 2 half light radius (HLR). Our radial distances will be 
expressed in units of HLR, this is defined as the semi-major axis 
length of the elliptical aperture which 
contains half of the total light of the galaxy at the 
rest-frame wavelength 5635 $\AA$ (see \citealt{cidfernandes13} and 
\citealt{gonzalezdelgado14} for details). In these galaxies, 
1 HLR is $\sim$ 4.6, 3.8, 2.8, 4.2, and 3.3 kpc for the Mice A, 
Mice B, IC 1623 W, NGC 6090, and NGC 2623, respectively, on average, while 
for the Sbc and Sc control galaxies, 1 HLR $\sim$ 5.0 kpc, 
and 4.1 kpc, respectively.


\section{Spatially resolved star formation histories}
\label{sec:Results}

\subsection{Method of analysis}
\label{sec:Base}

LArr and PPaK datacubes are analyzed using the full spectral synthesis 
technique applied in a similar way as we have done in 
\citet{perez13} and \citet{cidfernandes13, cidfernandes14}. The individual 
spectra are extracted from the CALIFA data by coadding spaxels 
into Voronoi zones 
\citet{cappellari03} to get a $S/N > 20$; from the LArr we use the individual 
spaxels if the $S/N > 5$.
To extract the star formation histories we use the \starlight\ code 
\citep{cidfernandes05}, which fits the observed 
spectra through a $\chi^{2}$ minimization, using a linear combination of 
simple stellar populations (SSPs) 
from a base of models spanning different ages and metallicities. 
For the SSPs we use a combination of the \citet{vazdekis10} for ages older 
than 63 Myr and \citet{gonzalezdelgado05} for younger ages, formed by 156 spectra 
of 39 different ages from 1 Myr to 14 Gyr 
\footnote{SSP ages are: 1 Myr, 3, 4, 5.6, 8.9, 10, 12.6, 14.1, 
17.8, 20, 25.1, 31.6, 39.8, 56.2, 63, 63.1, 70.8, 100, 112, 126 
159, 200, 282, 355, 501, 708, 891 Myr, and 1.1 Gyr, 1.3, 1.4, 2.0, 
2.5, 3.5, 4.5, 6.3, 7.9, 10, 12.3, and 14.1 Gyr}, and four 
metallicities ($Z =$ 0.2, 0.4, 1, and 1.6 $Z_\odot$). 
As in \citet{cortijo17}, dust effects are modeled as a 
foreground screen with a \citet{calzetti00} reddening law with R$_{V}$ = 4.5, 
and assuming the same reddening for all the SSP components. The output is then 
processed through \pycasso\ (the Python CALIFA \starlight\ Synthesis Organizer) 
to produce a suite of the spatially resolved stellar population properties. 

As most U/LIRGs, and in particular, 
the systems in our paper, present a strong 
contribution of the stellar populations younger 
than 1 Gyr \citep{cortijo17,cortijo17b,alonsoherrero10,pereira15}, our 
non-parametric approach is the best to unveil their complex SFH. 
Other methods of analysis commonly used to predict the SFR of 
normal, non-interacting galaxies, like the parametric 
exponentially declining SFH fitting 
or $\tau$ models \citep{maraston13,maraston10}, 
will not be able to predict correct and simultaneously the 
contribution of the young 
and/or the old stellar population.

Radial profiles of the age, metallicity, stellar mass surface density, 
and extinction were presented in \citet{cortijo17} for IC1623 and NGC6090, 
and in \citet{cortijo17b} for NGC2623. In the next sections, the results for 
spatially resolved SFH and radial profiles of the specific SFR and the intensity 
of the SFR (SFI) are presented and compared with similar results obtained for two 
subsamples of Sbc/Sc galaxies from the CALIFA mother sample that were analyzed 
in \citet{gonzalezdelgado16} and \citet{gonzalezdelgado17} (submitted). 
They were selected in the same mass 
range as the mergers, and are formed by 70 Sbc and 14 Sc galaxies.
These late spirals are chosen for this comparison study because 
the progenitors of the galaxy mergers in our sample 
are gas-rich spirals without significant bulges.
This allows us to find out if there is an enhancement of the star 
formation with respect to non-interacting systems.


\subsection{SFH: The 2D $R \times t$ maps}

As explained in \citet{cidfernandes13} an interesting way to visualize 
the spatially resolved star formation history of a galaxy is the 
$R\times$t diagrams. There, the y-axis is the radial distance ($R$) and 
the x-axis is the lookback time ($t$) and the intensity of the map is the 
luminosity density ${\cal{L}}_{\lambda 5635,tR}$ (corrected by extinction) or 
the mass surface density $\mu_{\star,tR}$. 
These maps represent the mass in stars per pc$^2$ at each radial distance 
formed at each cosmic time epoch and the luminosity emitted by these 
stars. These $R\times$t diagrams are obtained as follows. For each 
spectrum that corresponds to an {xy} location in the galaxy, \starlight\ provides 
the light fraction ($x_{xytZ}$) at the normalized 
wavelength ($\lambda$ = 5635 \AA\, as in previous studies) due to the base 
population with age $t$ and metallicity $Z$. Taking into account the 
stellar extinction and $M/L$ spatial variation, \starlight\  also 
provides the mass fraction ($m_{xytZ}$). Then, $x_{xytZ}$ and 
$m_{xytZ}$ are collapsed in the $Z$ domain, and the {xy} locations 
are compressed in an azimuthally average radial distance, to get the 
luminosity density, 
${\cal{L}}_{\lambda 5635,tR}$ (L$_\odot$ $\AA^{-1}$ pc$^{-2}$) and mass 
density $\mu_{\star,tR}$ at each radial position $R$ and for each 
age (equivalent to time evolution, expressed in lookback time). 

The $R\times$t diagrams of the galaxies are presented 
in Fig.\ \ref{fig:2DSFH}, where the luminosity 
density ${\cal{L}}_{\lambda 5635,tR}$ (L$_\odot$ $\AA^{-1}$ pc$^{-2}$) and mass 
density ${\mu}_{\star,tR}$ (M$_\odot$ pc$^{-2}$) are on the left 
and right panels, respectively. 
For the purpose of visualization, 
the original ${\cal{L}}_{\lambda 5635,tR}$ 
and $\mu_{\star,tR}$ maps 
\footnote{Due to the lack of temporal resolution, for ages $<$30 Myr 
we assign an average value of the mass and luminosity densities using 
all the SSPs younger than that age.} 
are smoothed in $\log t$, applying a Gaussian filter with FWHM=0.5 dex. 
For comparison, 
the ${\cal{L}}_{\lambda 5635,tR}$ and $\mu_{\star,tR}$ 
maps of the Sbc/Sc control non-interacting spirals are also shown.
We note how complex the luminosity and mass densities 
of LIRG-mergers are in comparison with Sbc/Sc ones. They 
present prominent spatial and temporal distorsions, 
specially for ages younger than 1 Gyr, whose contribution to the 
luminosity density is significantly higher than in Sbc/Sc galaxies.

A first way to summarize the SFH is condensing the age distribution 
in several age ranges. This strategy originated in methods based on fitting 
equivalent width spectral indices \citep{bica88, cidfernandes03}, but it was 
also applied  to full spectral fits \citep{gonzalezdelgado04}. Here we chose 
four age scales that typically trace the periods of enhanced star formation 
in these mergers. They are: 1) $\leq$ 30 Myr, which represents 
the youngest ionizing stellar populations, and that is comparable to the SFR 
derived from the H$_\alpha$ \citep{valeasari07, gonzalezdelgado16}; 
2) $\sim$ 300 Myr which represents young populations emitting 
in the UV including up to early type A stars; 3) $\sim$ 1 Gyr 
which are the intermediate age populations dominated by A and F stars; 
and 4) populations older than 1 Gyr. 
In order to quantify the dependence of the results with the spatial 
information, we extract the SFH in three different spatial scales: 
1) the central 2 HLR that for each galaxy corresponds approximately 
to the whole galaxy; 2) the central 0.5 HLR, that in early 
spirals is usually dominated by the bulge component; 
3) and between 1 and 1.5 HLR, that usually represents well the 
average properties of regions located in the disk of spirals. 
In Table \ref{tab:natbib} we summarize the contributions in 
light and mass over the three timescales ($t \leq$ 30 Myr, 300 Myr, and 1 Gyr) 
and for three spatial regions: from 0 -- 0.5 HLR, 
1 -- 1.5 HLR, and 0 -- 2 HLR. The uncertainties are 
shown in brackets, and are the result of propagating the dispersions 
due to slight age ($\pm$ 1 SSP) and radial variations ($\pm$ 0.1 HLR 
in the central and outer regions, 
and $\pm$ 0.5 HLR in the integrated 0--2 HLR range.).
The uncertainties calculated in this way are quite small. 
Indeed, the major uncertainties associated to this method 
come from the models base choice, and the extinction law. 
For example, using different spectral fits from NGC 2623, 
we have found: a) when maintaining the same set of SSPs, 
but changing to Cardelli extinction law \citep{cardelli89}, 
the fraction of light in ages younger than 32 Myr, x$_{< 32 Myr}$, 
presents a maximum increase of up to $+8 \%$ in the 0--0.5 HLR bin, 
and a maximum decrease of up to $-6\%$ for x$_{< 300 Myr}$, 
but the differences in the mass fractions are 
significantly smaller.
However, the variations in light and mass fractions 
for the age bin $<$ 1000 Myr are larger. In terms of light, 
they present a maximum decrease of $-4\%$ in the 0--0.5 HLR region, 
and an increase of $+5\%$ in the 1--1.5 HLR, with mass variations, 
m$_{< 1 Gyr}$, of $\sim - 2\%$ and $\sim + 2\%$, respectively.
b) Analogously, when maintaining 
the Calzetti extinction law but changing the model base to one 
built from a 2007 update of the \cite{bruzual&charlot2003} models,
BC07 models give x$_{< 1000 Myr}$ that are $11 \%$ smaller than with 
GSD base, but still of $2-3\%$ in mass fraction. In summary, 
although these variations are significantly larger than those 
obtained by the propagation of uncertainties in SSP ages 
and space intervals, they are still of $\lesssim 2-3\%$. 

A brief description of the results based on these 
quantifications are:

\subsubsection{SFH: Mice A and B}

The optical light of the Mice is dominated by the old stellar populations. 
Populations younger than 1 Gyr contribute, on average, to the luminosity 
density, with 25.6$\pm$1.1 $\%$\footnote{dispersions are 
only based on those tabulated in Table 2.} (Mice A) 
and 8.1$\pm$1.0 $\%$$\%$ (Mice B). In the Mice B these 
populations are more relevant in the 1--1.5 HLR 
regions (28.1$\pm$3.7$\%$) than in 
the central 0.5 HLR regions (4.2$\pm$0.4$\%$), 
while in contrast, 
in Mice A they are less relevant in the 1--1.5 HLR 
regions (17.8$\pm$0.6$\%$) 
than in the central 0.5 HLR (27.0$\pm$0.9$\%$). 
Old populations contribute with 74$\%$ (92$\%$) 
to the optical light of 
the Mice A (B), and dominate the stellar mass density, accounting 
for more than 97$\%$ of the total mass. 

\subsubsection{SFH: IC1623}

Stellar populations younger than 300 Myr account for 
more than 65.6$\pm$1.6$\%$
of the luminosity density of this galaxy, being approximately 
similar in the 
inner 0.5 HLR (57.1$\pm$0.9$\%$), and significant 
higher (79.7$\pm$0.3$\%$) at regions 
between 1--1.5 HLR. Populations younger than 30 Myr are also relevant 
in this galaxy, contributing with 20.5$\pm$2.7$\%$ in 
the central 0.5 HLR, 
and up to 57.7$\pm$0.5$\%$ in 1--1.5 HLR regions. 
Old populations contribute 
with less than 26$\%$, but these old components account for more 
than 95$\%$ of the mass.

\subsubsection{SFH: NGC6090}

The optical light of this galaxy is dominated by very young stars ($t <$ 30 Myr). 
They contribute to the luminosity density with 46.8$\pm$0.8$\%$,  
and with similar fractions in the central 0.5 HLR 
regions (49.2$\pm$0.8$\%$), and slighly 
lower in the 1--1.5 HLR regions (33.1$\pm$1.7$\%$). 
Stellar populations with ages 
between 30 Myr and 1 Gyr, account in this galaxy by only $\sim$20$\%$, although 
is higher (33$\%$) in outer regions. Old populations contribute 
with $\sim 34\%$ to the optical light, and dominate 
the stellar mass density, 
as in IC1623 by 91$\%$ of the total mass.

\subsubsection{SFH: NGC2623}

Young ($t <$ 30 Myr) and  intermediate (30 Myr $< t <$ 1 Gyr) stars are 
relevant in this galaxy, they contribute 
with $\sim$ 19.3$\pm$0.6$\%$ 
and 43.3$\pm$1.3$\%$
each one to the total light density, and with similar 
fractions, $\sim$ 21.7$\pm$0.6$\%$ and 
$\sim$ 43.0$\pm$1.3$\%$ for the young and intermediate 
ages, respectively, in the 
central 0.5 HLR, and in 1--1.5 HLR these fractions 
are $\sim$ 10.2$\pm$0.5$\%$ 
and $\sim$ 50.7$\pm$1.8 $\%$. 
Populations older than 1 Gyr contribute 
by $\sim$ 37$\%$ of 
the light density, and with 86$\%$ of the 
mass density, on average in 0--2 HLR region, being this contribution similar in the 
central region (89$\%$), and slighly lower in the disk (77$\%$).


\begin{figure*}
\includegraphics[width=0.32\textwidth]{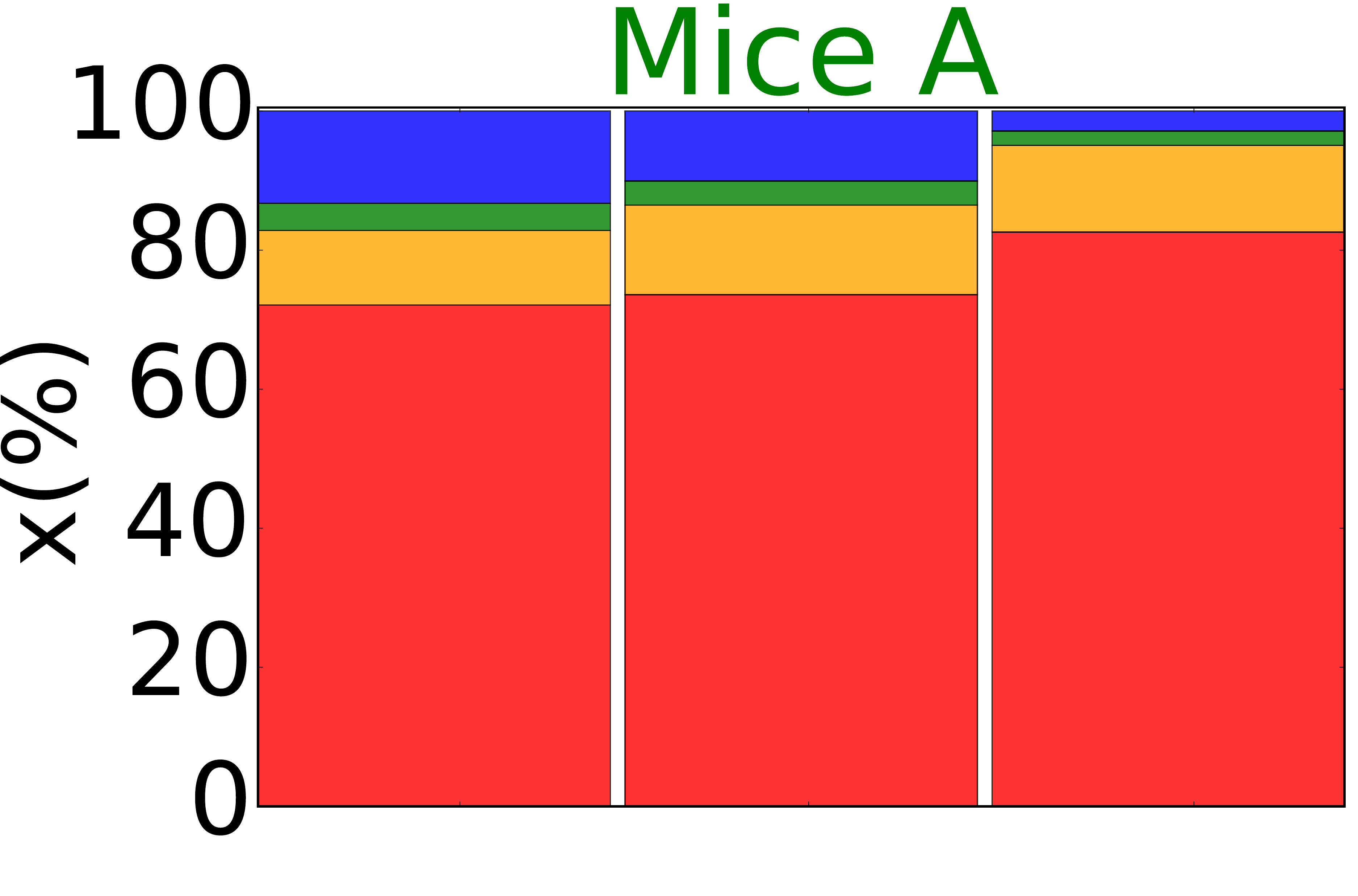}  
\includegraphics[width=0.32\textwidth]{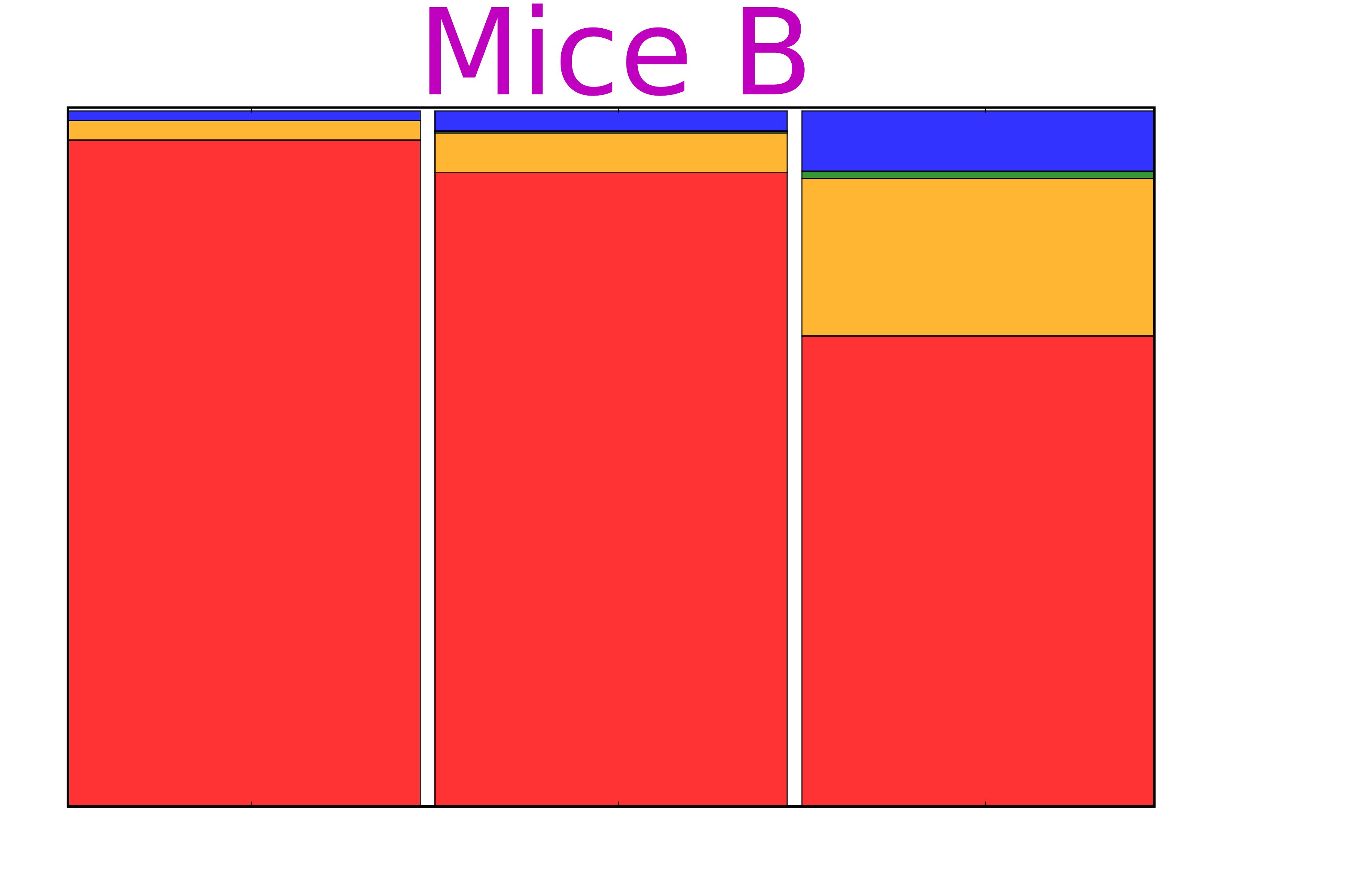}  
\includegraphics[width=0.32\textwidth]{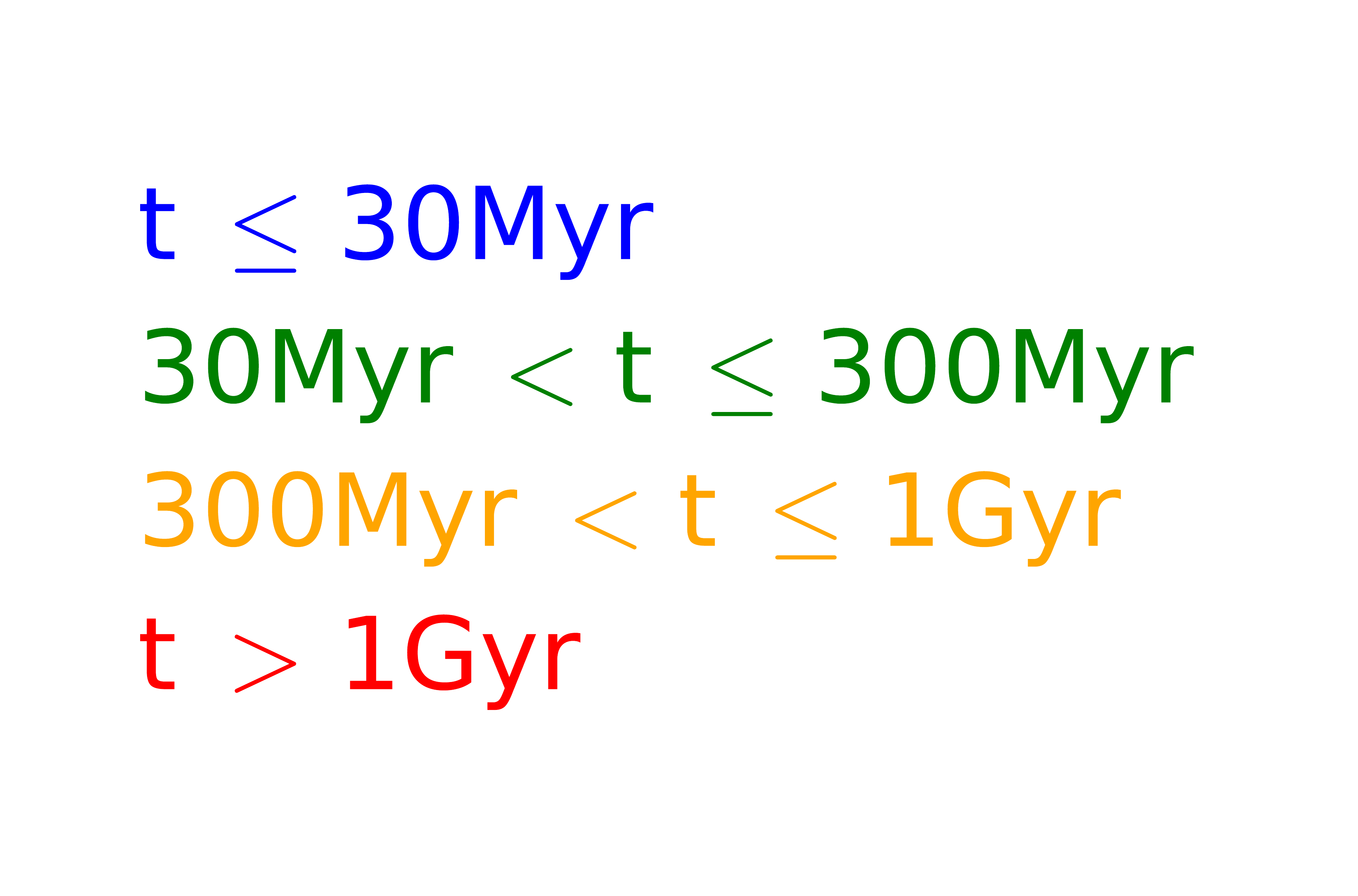}
\includegraphics[width=0.32\textwidth]{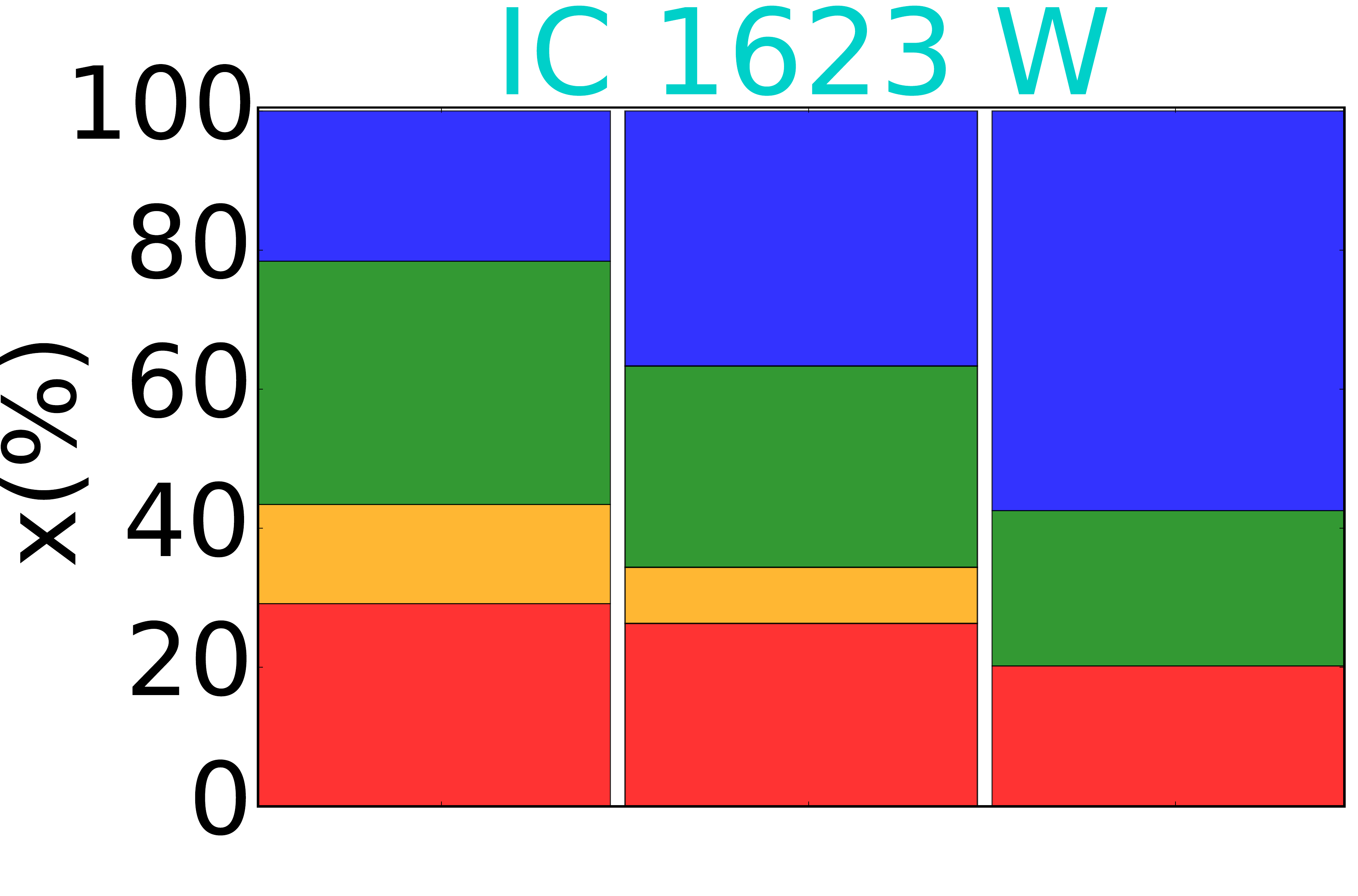} 
\includegraphics[width=0.32\textwidth]{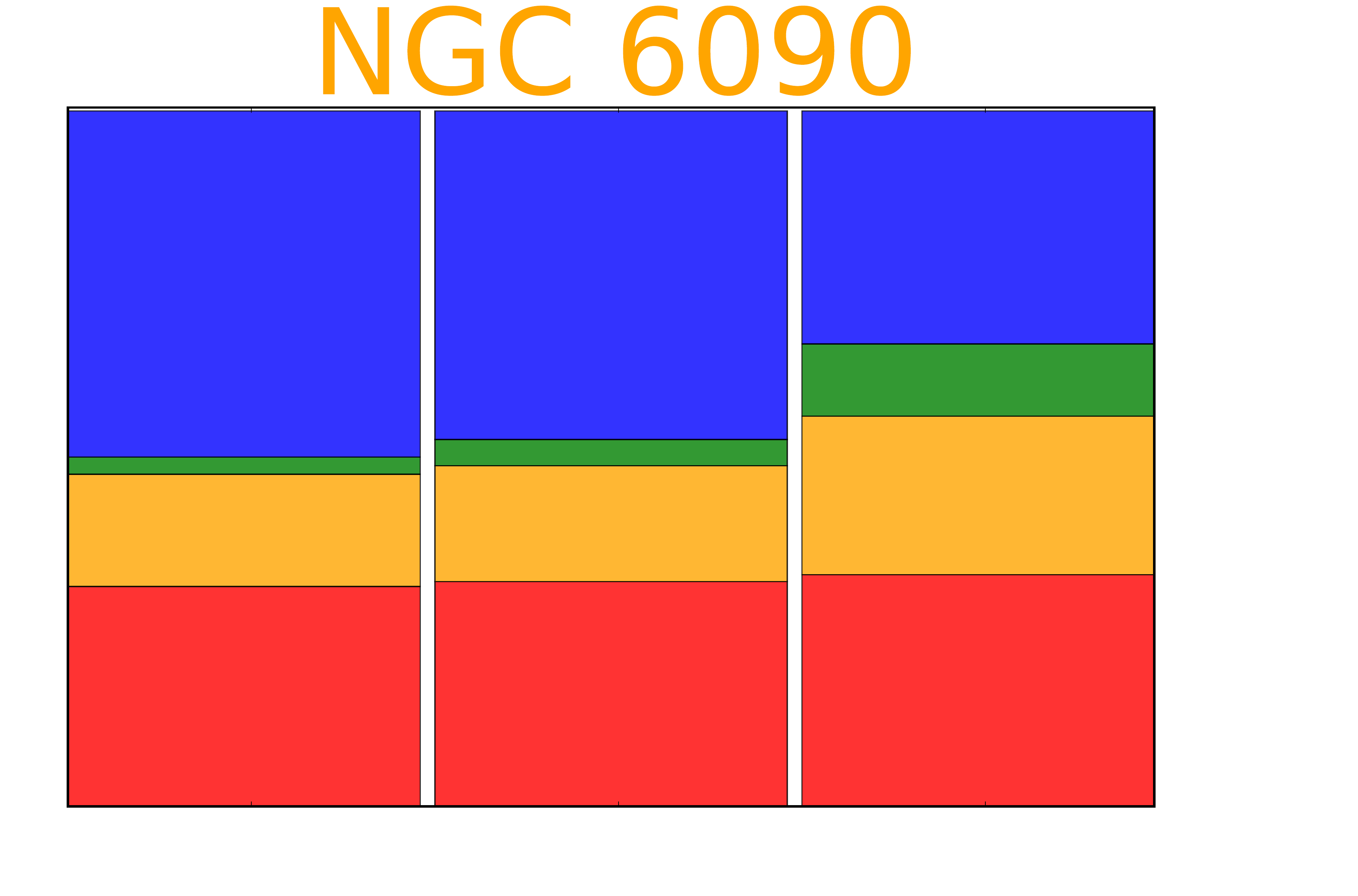}  
\includegraphics[width=0.32\textwidth]{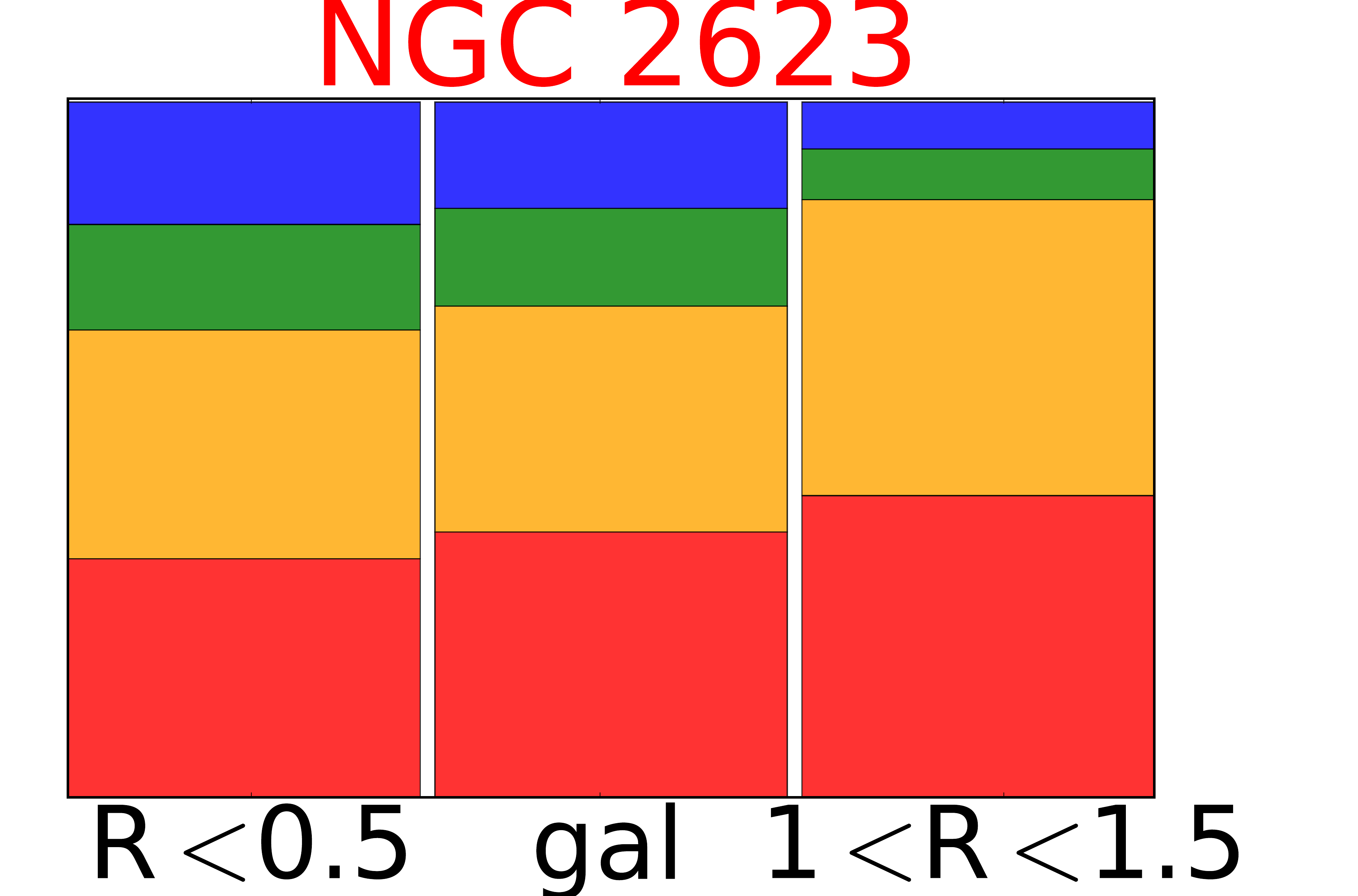} 
\includegraphics[width=0.32\textwidth]{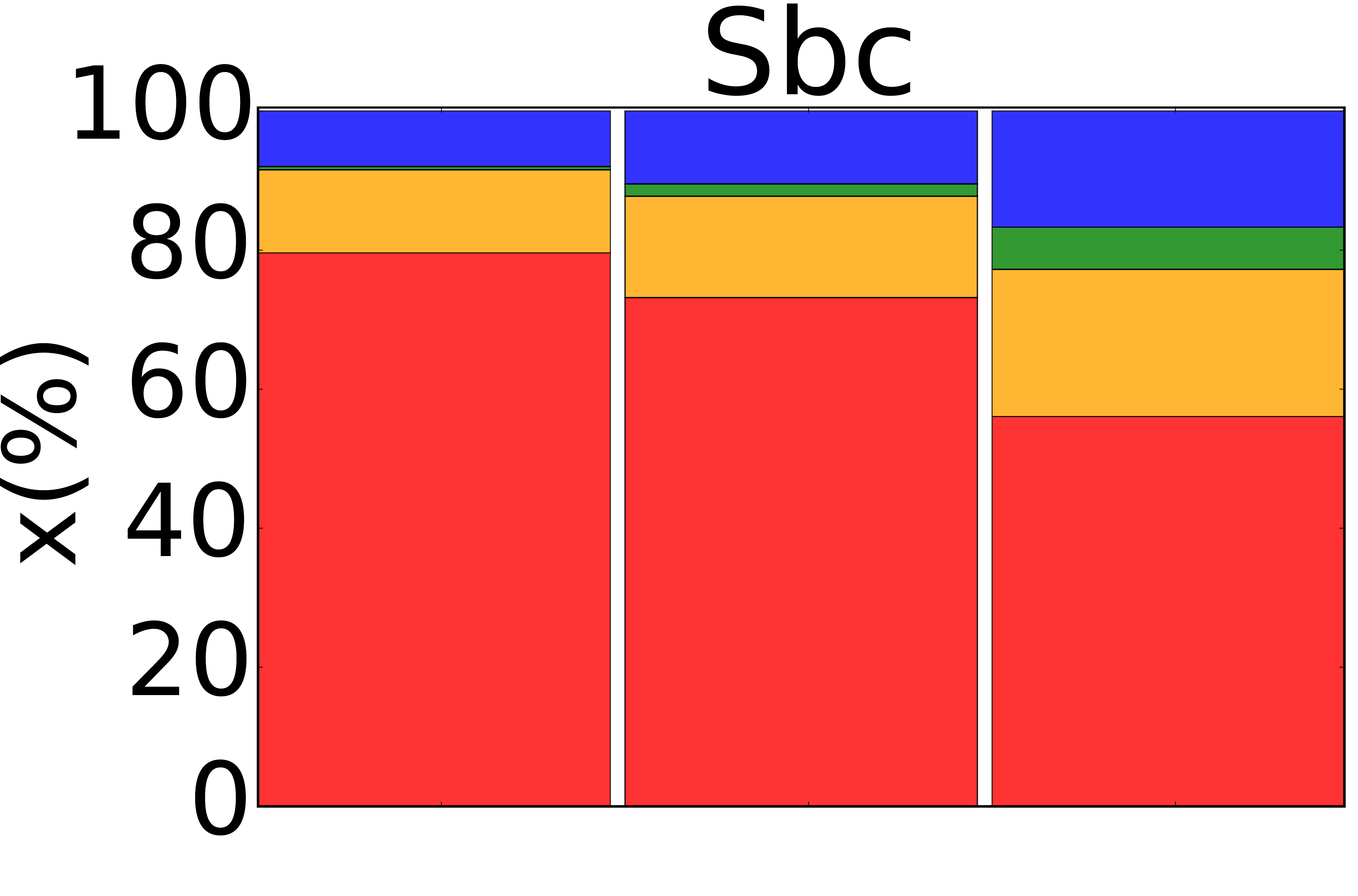} 
\includegraphics[width=0.32\textwidth]{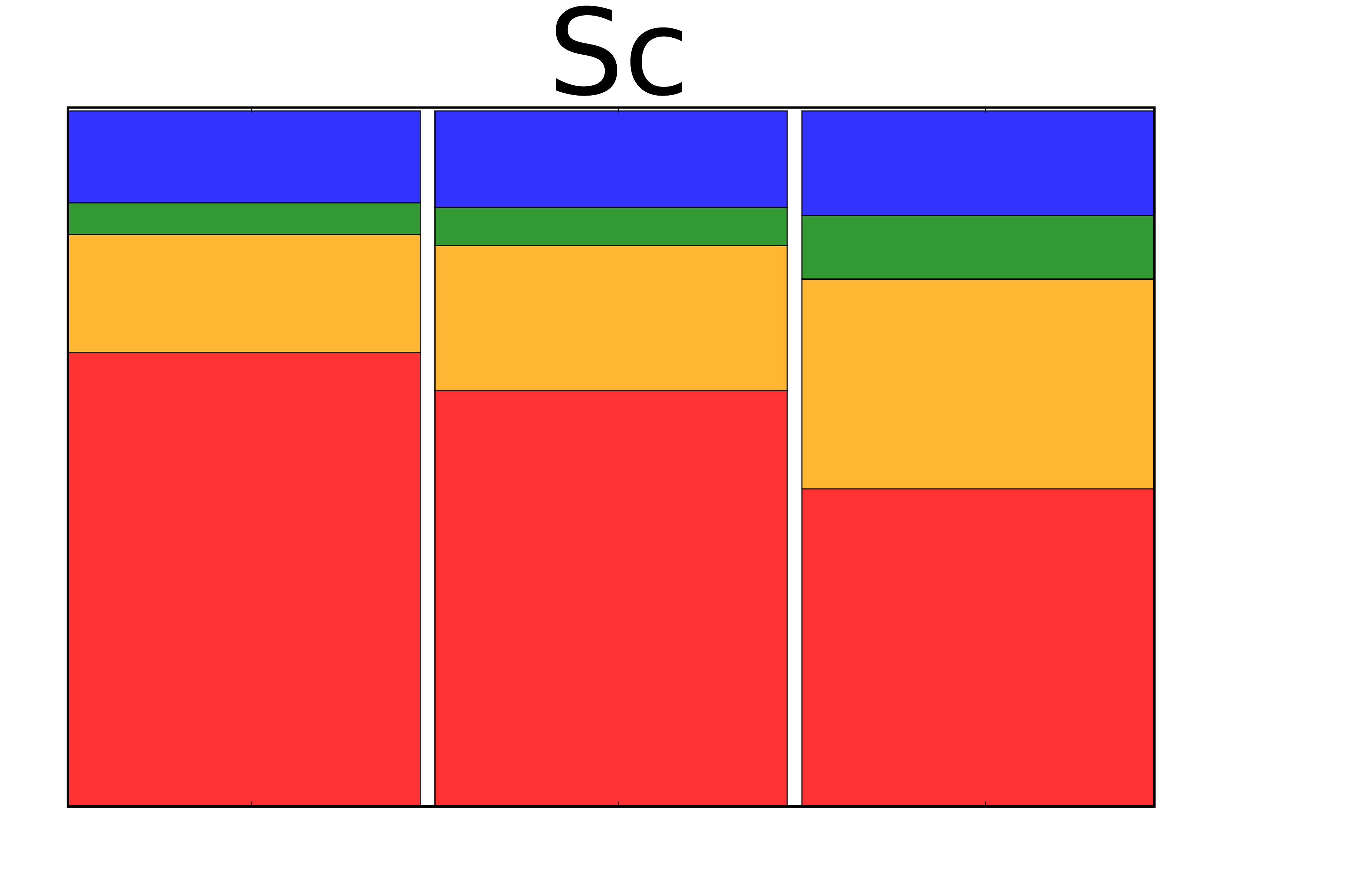}
\caption{Average light fractions due to stars in different age and
radial ranges. Each panel corresponds to a galaxy merger (first and second
rows), and the comparison Sbc/Sc galaxies are shown in the third row. 
The three bar chart histograms in each panel correspond to different 
galaxy regions: the inner region R $<$ 0.5 HLR (left bar); 
the whole galaxy R $<$ 2 HLR (central bar); and outer regions 
1 $<$ R $<$ 1.5 HLR (right bar). 
Colors represent the different age ranges.} 
\label{fig:1DSFHLight}
\end{figure*}



\subsection{SFH: Light  fraction}
In order to compare the similarities and differences 
between the SFH of these galaxies, 
Fig.\ \ref{fig:1DSFHLight} presents an alternative way of visualizing the 
results by discretizing the light fraction ($x$) in a few relevant 
$R$ and $t$ ranges. Each panel shows three rectangular bar charts, 
each corresponding to a spatial region: 
$R <$ 0.5 HLR (left), 0 $< R <$ 2 HLR (middle), 
and 1 $< R <$ 1.5 HLR (right). The middle charts are meant to represent the 
galaxy as a whole. The age information is compressed into four 
color-coded representative ranges: $t \leq$ 30 Myr (blue), 
30 Myr $< t \leq$ 300 Myr (green), 300 Myr $< t \leq$ 1 Gyr (orange), 
and $t >$ 1 Gyr (red), tracing the main epochs where merger-induced 
star formation is triggered.

The most important difference that emerges between the 
LIRG-mergers (IC1623, NGC6090, and NGC2623) and disk galaxies 
is that in the former there is a significantly larger light 
fraction in $\lesssim$1 Gyr components than in Sbc and Sc galaxies. 
This is clearly evident in the inner ($<$ 0.5 HLR) for 
the 3 LIRG-mergers and also in 
the outer regions (1--1.5 HLR) for the pre-merger LIRGs 
IC1623 and NGC6090. 
This is expected because of their FIR luminosity, these galaxies 
had to experience a burst of star formation in the last 1 Gyr, 
and stellar populations of ages younger than 1 Gyr 
dominate the light at optical wavelengths, as we have seen 
in the descriptions above. 

The SFH of the Mice, however, are more similar to spirals than to the 
LIRG-mergers, in the sense that most of the optical light is dominated 
by stellar populations of 1 Gyr and older. 
In particular, the outer regions of Mice A and Mice B seem 
to have suppressed the star formation in the last $<$1 Gyr 
in comparison with the control spirals.

Using Table \ref{tab:natbib}, we have quantified this in more detail:

\begin{itemize}
\item{ Inner 0.5 HLR: We found that the contributions to 
light of the youngest 
SSPs (< 30 Myr) in IC 1623 W, NGC 6090, and  NGC 2623 are respectively a 
factor of 3 (2), 6 (4), and 3 (2)  times higher than in Sbc (Sc) galaxies. 
In contrast, in Mice A are similar to Sbc (Sc) galaxies, by 2 (1), and 
significantly suppressed in Mice B, by factors 
of 6 (13) lower than 
in Sbc (Sc) galaxies.
Also, when integrating over the last 300 Myr, the light 
fractions we measure in IC 1623 W, NGC 6090, and NGC 2623 
are a factor of 7 (3), 6 (3), and 4 (2) higher than the light 
fraction in Sbc (Sc) galaxies. 
In Mice A is similar to Sbc (Sc) galaxies, and 
suppressed in Mice B, as in the youngest time scale.
Lastly, the central star formation in the last 1 Gyr is also enhanced in the 
three LIRGs with respect to the control spirals by a 
factor 4--3 (2), in terms 
of light, with respect to Sbc (Sc). This is approximately a factor $\sim$2 less 
than what we measured 
for the shorter 300 Myr timescale. This indicates that the star 
formation has not been continuous in the last 1 Gyr but rather in bursts. 
In Mice A, the star formation in the 1 Gyr timescale is similar to the control 
spirals, and suppressed in Mice B by the same amount as in the other
timescales.}

\item{ 1--1.5 HLR regions:
In the outer-regions of the two early-stage mergers IC 1623 W and NGC 6090, the star 
formation in the last 30 Myr is, respectively, a factor $\sim$4 and $\sim$2 higher 
in light fraction than the 
average of the control Sbc and Sc.
Otherwise, in the merger NGC 2623 the contribution is comparable or 
slightly lower 0.6 (0.7) than in Sbc (Sc) galaxies, suggesting that the 
star formation in the last 30 Myr has been significantly enhanced in the 
central region of NGC 2623 but not in its "disk".
Again, in the outer-regions of the early-stage merger LIRGs, the star formation 
in the last 300 Myr is a factor of 3 (for IC 1623 W) and 2 (for NGC 6090) 
higher than in the control spirals. The contribution to light of the 
SSPs $<$ 300 Myr in NGC 2623 is comparable to the one in the control spirals.
In terms of light, the star formation in the three LIRGs in the 
last 1 Gyr is not enhanced (or very slightly) with respect to the control spirals. 
In the disk of both Mice galaxies, the star formation in the three 
timescales is suppressed with respect to Sbc (Sc) galaxies 
by 0.3--0.4 in Mice A, 
and 0.3--0.5 in Mice B.}

\item{ Central 2.0 HLR, global average:
The global average results are intermediate between the ones 
described above.}

\end{itemize}

In conclusion, in the two early-stage merger LIRGs the star formation 
in the last 30 Myr, also in the last 300 Myr, is enhanced with respect to the 
control spirals both in the central regions and in the "disks". 
In the merger NGC 2623, the young star formation in the last 30 Myr and 300 Myr 
is significantly enhanced in the central region, while in the outer parts it 
is comparable to the Sbc-Sc spirals.
In contrast, in Mice A and B there is no evidence of enhanced SF 
with respect to isolated spirals; on the contrary, 
it is suppressed in the disk of Mice A and both in the disk and center of Mice B.


\subsection{SFH: Mass assembly}

\begin{figure}[!h]
\includegraphics[width=0.5\textwidth]{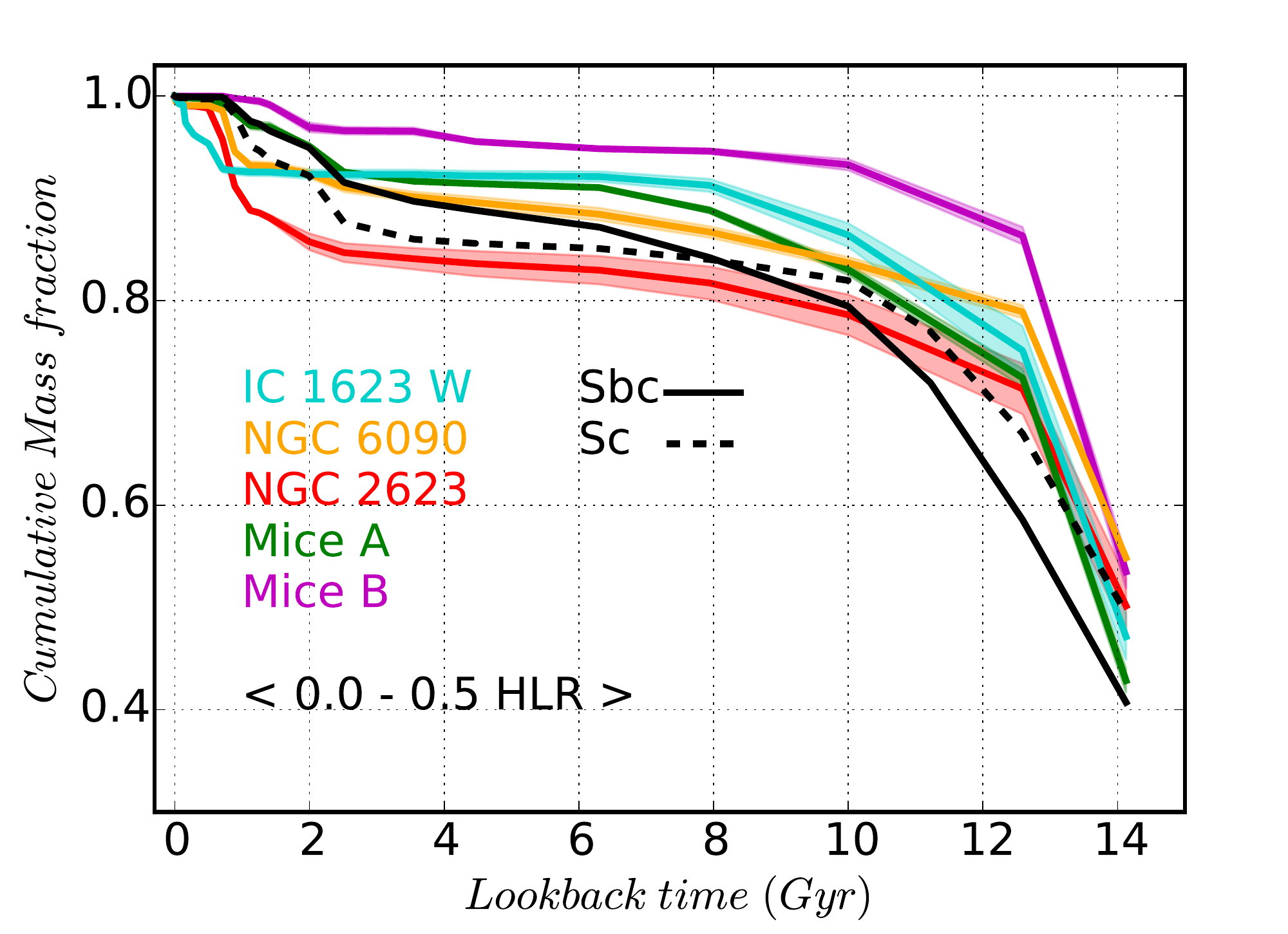}
\includegraphics[width=0.5\textwidth]{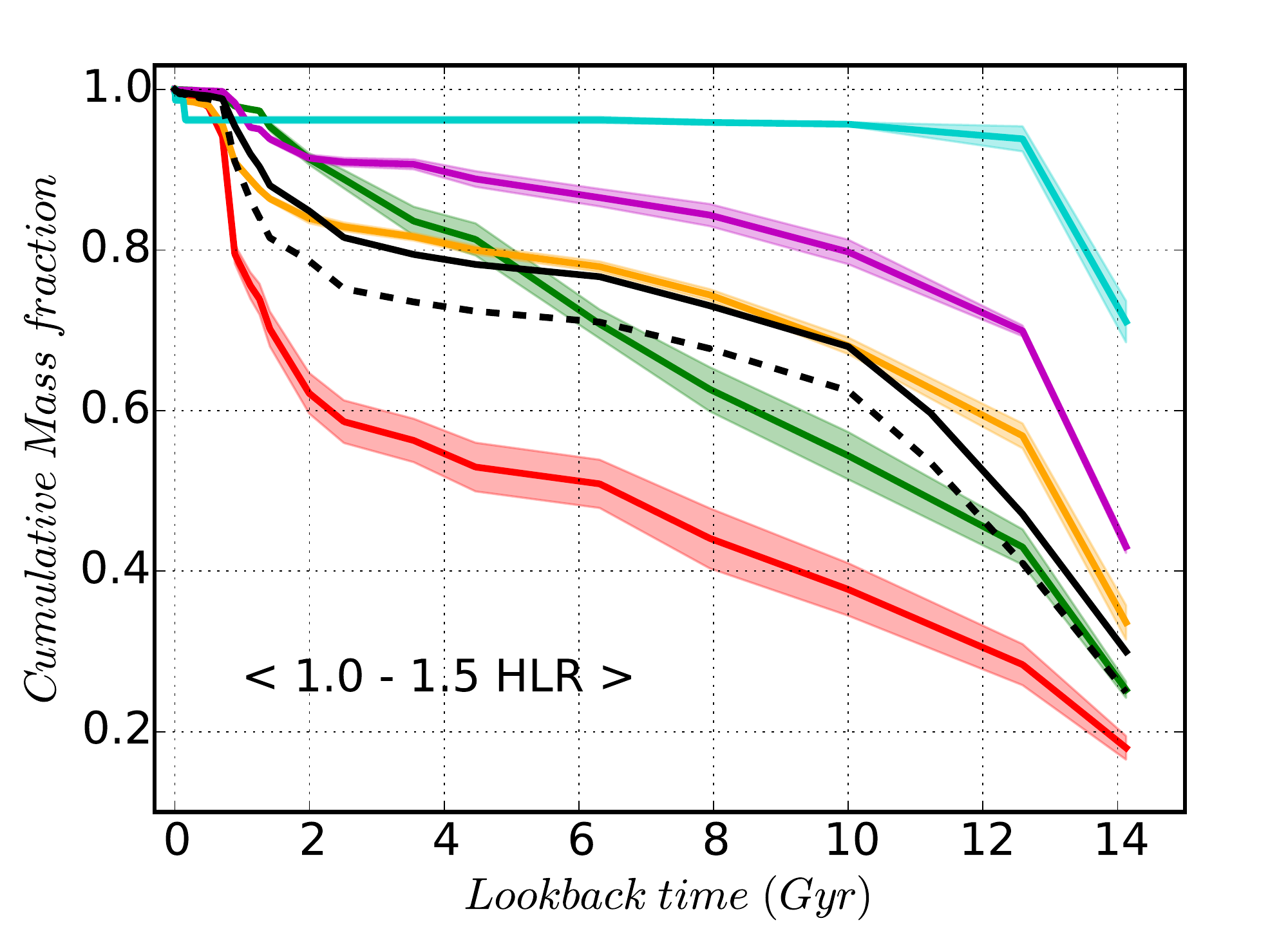}
\includegraphics[width=0.5\textwidth]{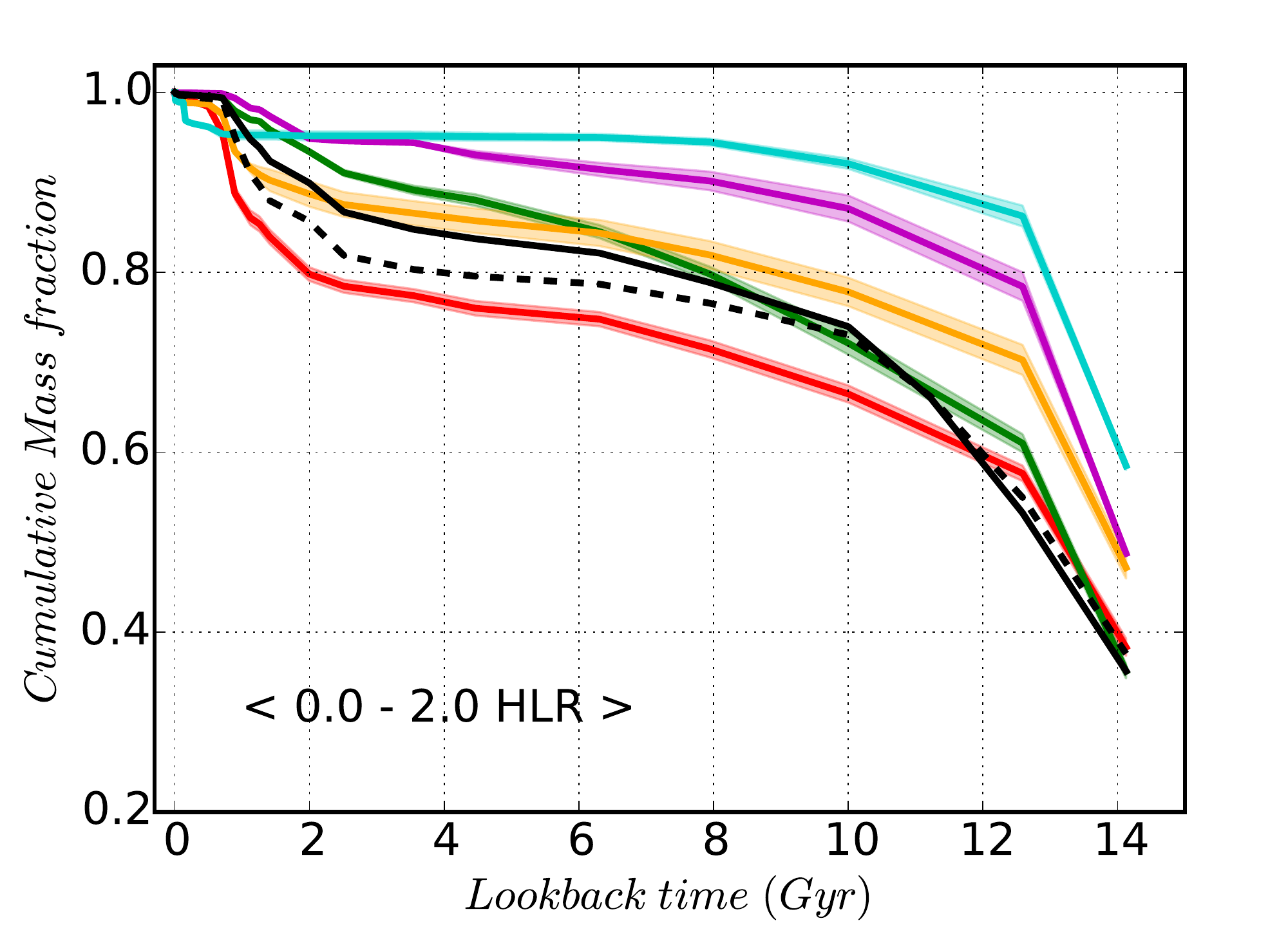}
\caption{The cumulative mass growth for the inner 0.5 HLR (upper panel), 
between 1 and 1.5 HLR (middle panel), and the central 2 HLR (bottom panel). 
The uncertainties are shaded in light colors, 
calculated as the dispersion in the profiles due to $\pm$0.1 HLR 
variations in the radial distance in the 0--0.5 HLR and 1--1.5 HLR 
regions, and $\pm$0.5 HLR for the global 0--2 HLR region.
The merger galaxies are color coded, and they are compared with the results 
obtained for Sbc (black line) and Sc (dashed-black line) galaxies.}
\label{fig:1DSFHMass}
\end{figure}

Fig.\ \ref{fig:1DSFHMass} presents the cumulative  functions of the SFH, 
showing the mass growth curve as a function of lookback time. This is obtained 
by adding all the mass formed in the system up to a given lookback time and 
dividing by the present day stellar mass. 
This cumulative mass function is derived 
by adding the contribution of regions located at: 
(i) the central 0.5 HLR, (ii) between 1 and 1.5 HLR, 
and (iii) adding almost the whole galaxy (between 
0 and 2 HLR). 
The uncertainties are shaded in light colors. 
They are calculated as the dispersion in the profiles due to $\pm$0.1 HLR 
variations in the radial distance in the 0--0.5 HLR and 1--1.5 HLR regions, 
and $\pm$0.5 HLR for the global 0--2 HLR region. 

As expected, the young stellar populations ($<$ 1 Gyr) that 
dominated the optical light contributes little to the mass; and populations 
older than 2 Gyr are the main contributors. This is also true for the Sbc and 
Sc galaxies, where the populations older than 10 Gyr contribute 
with $\sim$ 73$\%$ of the total mass. These old stellar populations 
also contribute significantly to the mass in these mergers, with a fraction 
than range between $\sim$ 66$\%$ for the NGC2623, to $\sim$ 92$\%$ for 
IC1623. However, in terms of mass assembly the SFH of 
mergers are diverse, and present interesting 
differences and similarities to the SFH of late-type spiral galaxies. The most 
relevant are: a) The mass growth in these mergers, as also in Sbc-Sc galaxies, 
was very fast, although the growth rate is different 
for each galaxy. Globally, mergers have grown their mass faster 
or at similar rate than Sbc and Sc galaxies, except in NGC 2623, 
where outer regions have grown more slowly.  
b) Except in IC 1623, the mass is assembled faster in the inner part 
than in the outskirts (compare middle and upper 
panels Fig.\ \ref{fig:1DSFHMass}). These galaxy mergers have grown 
most of their mass inside-out, as we have found in most of the CALIFA 
galaxies (\citealt{perez13}, \citealt{garciabenito17} submitted). 

To quantify the rate at which these galaxies have grown their mass, 
we calculate the epoch ($t_{80}$) at which roughly  $\sim$80$\%$ of the 
present stellar mass had been assembled. 

\begin{itemize}

\item{Inner 0.5 HLR:
The central regions of these mergers formed at similar epoch than 
in Sbc (Sc) galaxies.
In particular, $t_{80}$ was 12.9 Gyr ago in Mice B, 12.0 Gyr ago in NGC 6090, 
11.4 Gyr ago in IC 1623 W, 10.7 Gyr ago in Mice A, 9.7 Gyr (10.5 Gyr) ago 
in Sbc (Sc) galaxies, and 9 Gyr ago in NGC 2623.
We find that the mass contribution of SSPs < 1 Gyr in the central regions 
of the two early-stage mergers IC 1623 W and NGC 6090 is a factor $\sim$ 2 higher 
than in Sbc-Sc spirals, while it is a factor $\sim$3 higher for the merger NGC 2623. 
In contrast, in Mice A it is comparable to Sbc-Sc galaxies, and suppressed by 
a factor 8 in Mice B.}

\item{1--1.5 HLR:
The outer regions of all the early-stage 
mergers (IC 1623 W, NGC 6090, and Mice) 
formed earlier than Sbc (Sc) galaxies, and these earlier 
than the merger LIRG NGC 2623.  
Quantitatively, $t_{80}$ was 13.5 Gyr ago in IC 1623 W, 9.8 Gyr ago in Mice B, 
4.7 Gyr ago in Mice A, 4.4 Gyr ago in NGC 6090, 
3.3 Gyr ago (1.7 Gyr ago) in Sbc (Sc), and 0.9 Gyr ago in NGC 2623.
Also, beyond 1 HLR, NGC 2623 formed 2 times more mass fraction in 
the last 1 Gyr than Sbc and Sc galaxies. 
This increment is similar, but slighly lower, 
to the one found in the central part ($\sim$3). In the early-stage merger 
LIRG NGC 6090 the relative mass formed in 
the last 1 Gyr is comparable to Sbc-Sc spirals and in IC 1623 W it is a 
factor 3 less, as if the star formation would have been inhibited 
in this system. The same happens in Mice A and Mice B, that have formed 
a factor 5 and 2, respectively, less mass than the control spirals.}

\item{Central 2.0 HLR:
The global average results are something similar, and intermediate, 
between the ones already derived in the central and the outer regions.
Quantitatively, $t_{80}$ was there 12.9 Gyr ago in IC 1623 W, 
12.1 Gyr ago in the Mice B, 8.9 Gyr ago in NGC 6090, 7.8 Gyr ago in the Mice A,
7.3 Gyr ago (3.9 Gyr ago) in Sbc (Sc), and 1.9 Gyr ago in NGC 2623.
Again, from the global averages, only NGC 2623 has formed in 
the last 1 Gyr a factor 2 more mass fraction than the control spirals, 
while for the early-stage merger LIRGs it is comparable or even lower.}
\end{itemize}

In summary, we conclude that most of the mass was assembled in the 
main body of these 
early-stage merger LIRGs very early on, similarly to the same epoch of early type 
spirals and massive galaxies \citep{gonzalezdelgado17}; but in 
NGC 2623 it has happened later than in the early-stage 
mergers and Sbc--Sc. In terms of the mass formed in 
the last 1 Gyr, in the early-stage merger LIRGs it is enhanced by a factor 2 in 
the central regions compared to Sbc/Sc galaxies, but comparable to them in the 
disks (or even lower as in the external regions of IC 1623 W). In contrast, 
the merger 
NGC 2623 has formed more mass fraction than the control spirals 
in this period, by a factor 
3 in the central regions and 2 in the outer regions. 
Mice A (B) formed, on average, 
a factor 2 (4) less mass fraction in the 
last 1 Gyr than Sbc--Sc galaxies.


\section{Spatially resolved recent star formation rates}
\label{sec:RadialProfiles}

This section presents the radial structure of the intensity of the 
star formation rate ($\Sigma_{SFR}$) and the local specific star 
formation rate of the merger 
galaxies and the comparison 
with the control spiral galaxies Sbc-Sc. 
These two quantities 
are very suitable to obtain spatially resolved information 
about the rate at which stars form in a galaxy in recent 
epochs, and are independent 
of the galaxy size or stellar mass. These two properties 
have been well characterized for the CALIFA sample as a 
function of Hubble type by \citet{gonzalezdelgado16} 
and \citet{gonzalezdelgado17}, and constitute 
a very rich and a highly valuable source of comparison 
between mergers and non-interacting spiral galaxies. First, we need 
to briefly explain the choices of the recent star formation time scales.

\subsection{Choices for the star formation time scale}

From the SFH we can calculate the SFR as the 
ratio of the mass formed during a given time interval 
to the duration of this time interval.

This represents a time-averaged star formation during 
each time step. To compute the recent star formation rate, 
we need to specify what we mean by "recent past" by 
defining $t_{SF}$ as the age of the oldest stars to be included 
in the computation of our recent SFR. The mean rate of star 
formation is calculated in each {xy} position in the galaxy 
by summing over all the populations younger than $t_{SF}$ and 
dividing by $t_{SF}$.

Although the choice of $t_{SF}$ is arbitrary, we have already 
found that $t_{SF}$ = 30 Myr is a good choice because it shows 
the best correlation between the stellar and H$\alpha$ based 
estimates of the SFR \citep{gonzalezdelgado16} 
(see also \citet{valeasari07} that for their study of star forming 
galaxies with SDSS found $t_{SF}$ = 25 Myr). This choice also 
coincides with the life time of O and early type B stars, which 
are the major contributors to the ionized gas in galaxies.

Considering the relevance of the intermediate age stellar 
populations in these mergers, we also need to define two 
complementary star formation time scales: $t_{SF}$ = 300 Myr, 
which represents young populations emitting in the UV including 
up to early type A stars;  and $t_{SF}$= 1 Gyr,  which are 
the intermediate age populations dominated by A and F stars. 
These choices are similar to those proposed by \citet{pereira15} that 
found that the UV and FIR diagnostics trace well the SFR 
on scales of $\sim$100 Myr and 1 Gyr in a sample of local LIRGs.

Thus, in the next sections we will obtain $\Sigma_{SFR}$ 
and the local sSFR by using these three time scales for 
the star formation: $t_{SF}$ = 30 Myr, 300 Myr, and 1 Gyr.


\subsection{The intensity of the star formation rate, $\Sigma_{SFR}$}
\label{sec:IntensitySFR}

\begin{figure}
\includegraphics[width=0.5\textwidth]{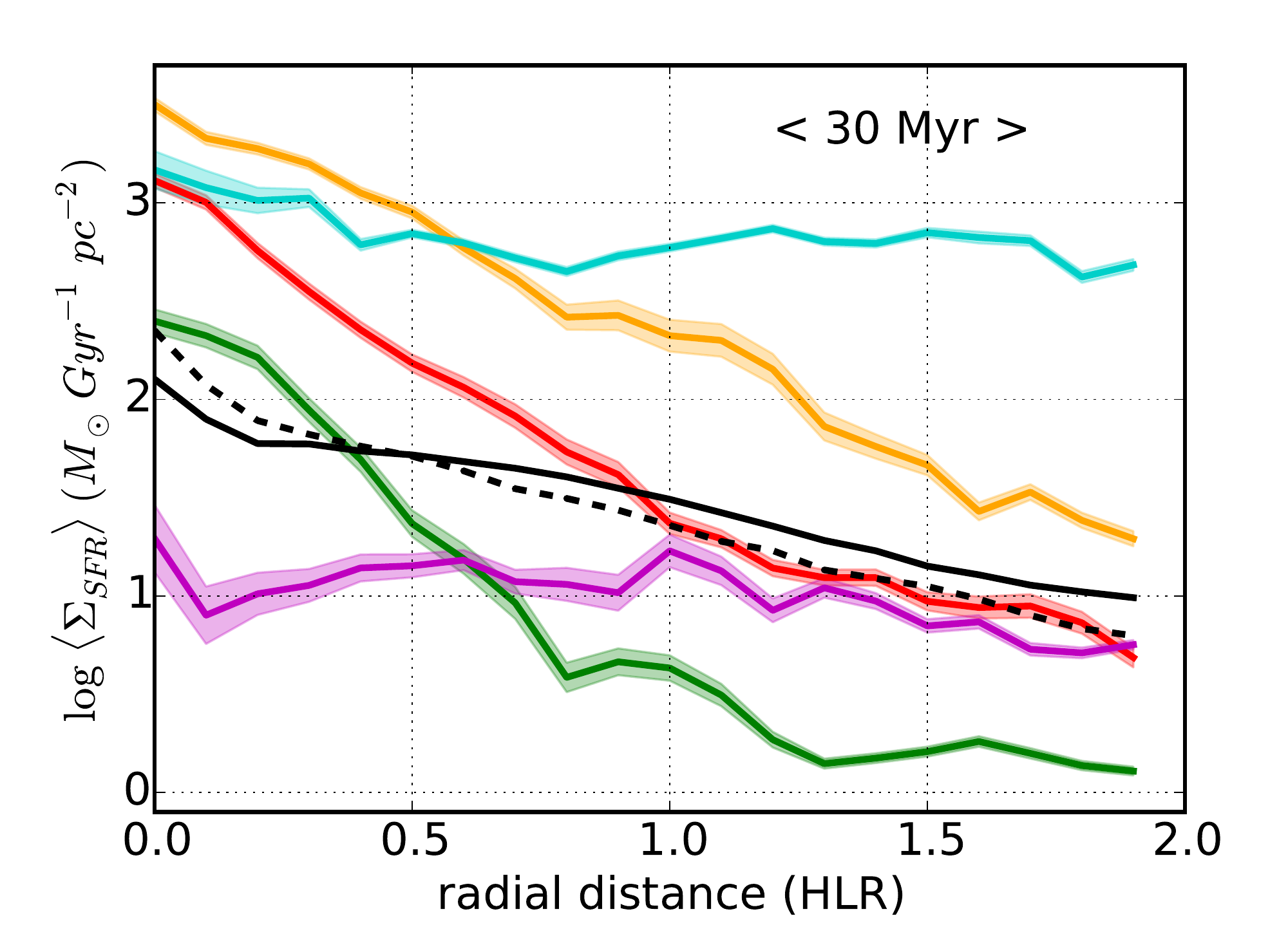}
\includegraphics[width=0.5\textwidth]{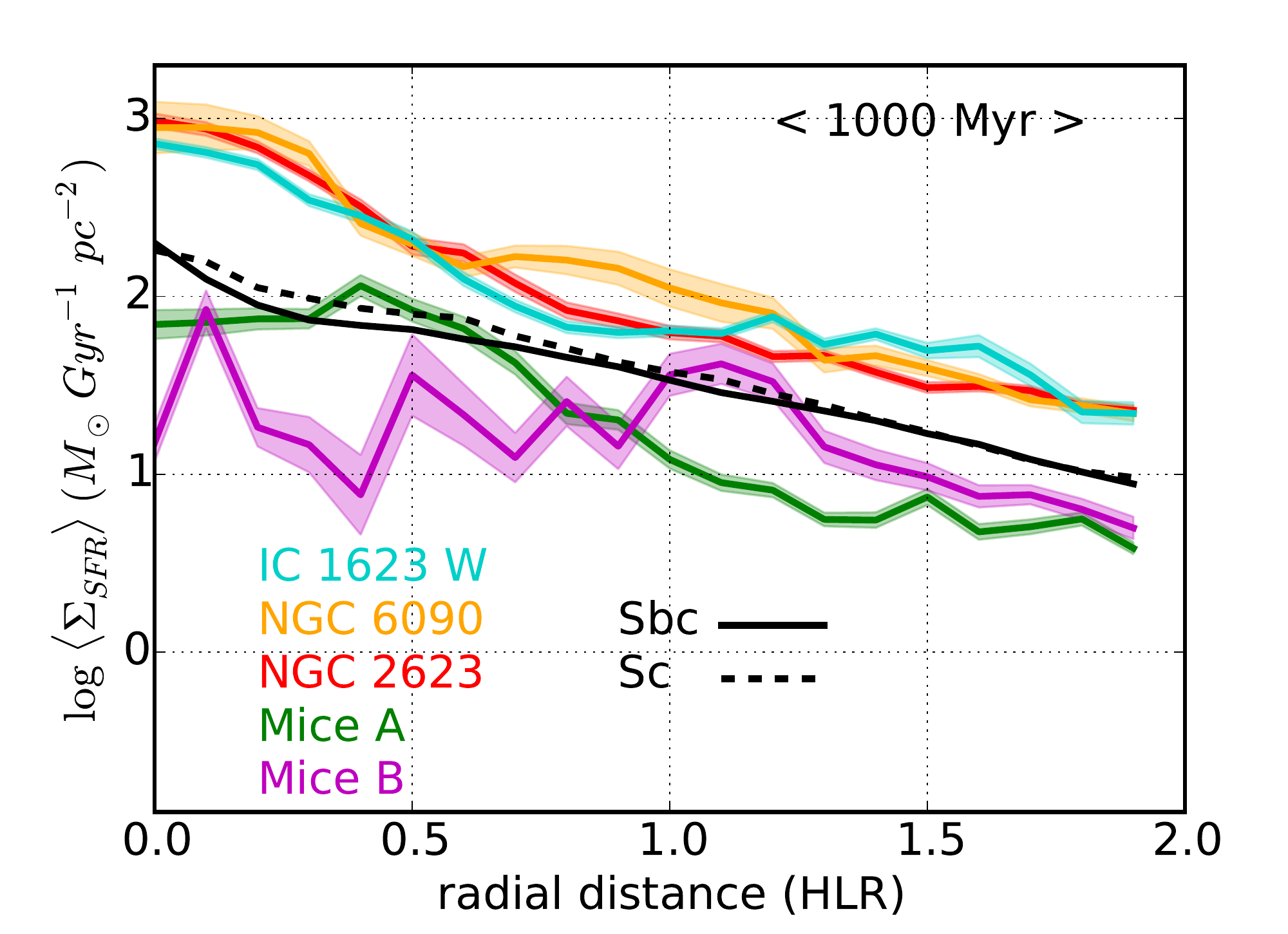}
\caption{The radial profiles of the intensity of the star formation 
rate ($\Sigma_{SFR}$) obtained for two different time scales: 
the last 30 Myr, and 1.0 Gyr. The uncertainties, 
calculated as the standard error of the mean 
in each $\pm$0.1 HLR bin, are shaded in light colors.} 
\label{fig:RadialProfiles_Intensity}
\end{figure}

The intensity of star formation rate is defined as the rate of star 
formation per area; i.e. the star formation rate 
surface density. We compute $\Sigma_{SFR} (t_{SF})$ from 
the ${\mu}_{\star,tR}$ 2D 
maps by adding the contribution 
of components with $t \leq t_{SF}$, and dividing by 
the $t_{SF}$ time scale. 

Previously, in the analysis of spiral galaxies, we found that 
$\Sigma_{SFR}$ at the present time shows declining radial profiles that 
exhibit very little differences between spirals; at a given radial 
distance, there is a small dispersion in $\Sigma_{SFR}$ that can 
be understood because galaxies in the main sequence of 
star formation (MSSF) have nearly constant SFI. 
A comparison of the $\Sigma_{SFR}$ radial 
profiles of the mergers obtained for different $t_{SF}$ 
and those of Sbc-Sc galaxies can tell us if there is 
an enhanced SFR in these mergers, where does it happen, and 
over which time scale. 

This comparison for $t_{SF} =$ 30 Myr and 1 Gyr 
is shown in Fig.\ \ref{fig:RadialProfiles_Intensity}. 
The uncertainties are shaded in light colors. 
They represent the standard 
error of the mean, calculated as the standard deviation divided 
by the square root of the number of points in each distance bin, 
with 0.2 HLR of width. 
As it is normalized by the square root of the number of points/spaxels 
in each HLR bin, the differences in the uncertainties between the inner 
and outer regions are visually not so strong, as the larger the radius, 
the larger is N. In terms of dispersion (=standard deviation) we find 
that with respect to the centre, the dispersion 
is, on average, a factor $\sim$2 larger at 0.5 HLR, 
$\sim$3 larger at 1 HLR, and $\sim$5 at 2 HLR. 

For the time scale $<$ 30 Myr, the $\Sigma_{SFR}$ in 
LIRG-merger galaxies is 
enhanced with respect to spirals. 
On the contrary, the Mice 
show a significant depression of the star formation with respect 
to the LIRG-mergers. Although at the inner 0.5 HLR of Mice A, 
$\Sigma_{SFR}$ is similar to Sc galaxies, in the outer 
regions, and in Mice B, 
$\Sigma_{SFR}$ is significantly below the control spirals.
In the time scale $t_{SF}$ = 1 Gyr, the LIRG-mergers also show 
some enhancement of the SFR with respect to Sbc-Sc galaxies, 
but the radial structure of $\Sigma_{SFR}$ of the three mergers 
is more similar and also similar to the declining profile 
in spirals. $\Sigma_{SFR}$ in Mice A also shows a declining profile, 
being in the central $\sim$0.5 HLR similar to the control spirals. 
In Mice B, $\Sigma_{SFR}$ is more variable, and significantly depressed 
with respect to Sbc galaxies in the central 1 HLR.

Quantifying the results in more detail:

\begin{itemize}
\item{{\em $t_{SF}$ = 30 Myr}: 
Taking as the reference the inner 0.5 HLR, we find that the 
$\Sigma_{SFR}$($t_{SF}$ = 30 Myr) of IC 1623 W, NGC 6090, and NGC 2623 are 
enhanced by 1.15 dex (1.05 dex), 1.38 dex (1.28 dex), and 0.83 dex (0.72 dex), 
respectively, with respect to the average of Sbc (Sc) galaxies, 
while it is much more similar for Mice A, 0.16 dex (0.06 dex), 
and significantly 
depressed in Mice B, by $-$0.74 dex ($-$0.84 dex).
Analogously, 
taking now as reference the outer parts from 1--1.5 HLR, we find differences 
between our mergers and Sbc (Sc) galaxies of 1.50 dex (1.63 dex) for IC 1623 W, 
0.69 dex (0.82 dex) for NGC 6090, $-$0.16 dex ($-$0.03 dex) for NGC 2623, 
$-$1.00 dex ($-$0.87 dex) for Mice A, and $-$0.30 dex ($-$0.17 dex) for Mice B.
These results indicate that in the two early-stage merger LIRGs the star 
formation in the last 30 Myr is enhanced with respect to the control spirals 
both in the inner and outer regions, while for the merger NGC 2623, 
$\Sigma_{SFR}$($t_{SF}$ = 30 Myr) is significantly enhanced in the central 
region, but at the outer parts it is similar or even slightly decreased 
in comparison to the control spirals. On the contrary, the only region 
of the Mice with $\Sigma_{SFR}$($t_{SF}$ = 30 Myr) similar to Sbc--Sc galaxies 
is the center of Mice A; in its disk and the whole Mice B is significantly decreased.
With respect to the gradients, in the inner 1 HLR 
\footnote{$\nabla_{in} log \Sigma_{SFR} = log \Sigma_{SFR} (1 HLR) - log \Sigma_{SFR} (0)$, 
as in \citet{gonzalezdelgado15}.}
we found that for $\Sigma_{SFR}$($t_{SF}$ = 30 Myr) the gradient 
is negative for the mergers and control spirals, with values of 
$-0$.40 dex/HLR in IC 1623 W, $-$1.18 dex/HLR in NGC 6090, 
$-$1.74 dex/HLR in NGC 2623, $-$1.77 dex/HLR in Mice A, 
$-$0.06 dex/HLR in Mice B, $-$0.61 dex/HLR in Sbc, 
and $-$0.99 dex/HLR in Sc galaxies.
The main differences are in the slope, 
while in NGC 6090 it is only a bit steeper than in Sc galaxies, 
in NGC 2623 and Mice A it is significantly steeper, 
and in IC 1623 W and Mice B the gradient is even 
flatter than in Sbc galaxies.
A negative gradient indicates a central triggering of the star 
formation in comparison with that at 1 HLR.

Although not shown, the results for 
the longer young-intermediate time-scale ($t_{SF} \lesssim$ 300 Myr) 
are qualitatively consistent with these for $t_{SF} =$ 30 Myr.}

\item{{\em $t_{SF}$ = 1 Gyr}: We found 
that in the inner 0.5 HLR the difference in 
$\Sigma_{SFR}$($t_{SF}$ = 1 Gyr) between our mergers and Sbc (Sc) 
galaxies is 0.64 dex (0.57 dex) in IC 1623 W, 
0.74 dex (0.67 dex) in NGC 6090, 0.73 dex (0.65 dex) in NGC 2623, 
$-$0.07 dex ($-$0.15 dex) in Mice A, 
and $-$0.65 dex ($-$0.73 dex) in Mice B.
Analogously, taking now as reference the outer parts from 1--1.5 HLR, 
$\Sigma_{SFR}$($t_{SF}$ = 1 Gyr) of IC 1623 W, NGC 6090, NGC 2623, Mice A, 
and Mice B, differ from the Sbc (Sc) galaxies by 0.40 dex (0.37 dex), 
0.42 dex (0.39 dex), 0.28 dex (0.24 dex), 
$-$0.50 dex ($-$0.53 dex), and $-$0.07 dex ($-$0.10 dex), respectively. 
Thus, in the central 0.5 HLR star formation in the last 1 Gyr is 
enhanced in IC 1623 W, NGC 6090, and NGC 2623 
with respect to control spirals ($\leq$5, significantly 
less than for 
the young 30 Myr timescale), and 
in the "disks" / outer parts (1--1.5 HLR) of the 
LIRGs $\Sigma_{SFR}$($t_{SF}$ = 1 Gyr) is only slightly 
enhanced with respect to the control spirals. 
At this timescale, the Mice show a decreased 
star formation, in comparison with Sbc-Sc galaxies, 
except maybe in the center of Mice A, 
where it is similar to Sbc galaxies.
With respect to the gradients, 
$\Sigma_{SFR}$($t_{SF}$ = 1 Gyr) gradient in the inner 1 HLR is 
$-$1.05 dex/HLR for IC 1623 W, $-$0.90 dex/HLR for NGC 6090, 
$-$1.19 dex/HLR for NGC 2623, $-$0.76 dex/HLR for Mice A, 
0.38 dex/HLR for Mice B, $-$0.77 dex/HLR for Sbc galaxies, 
and $-$0.68 dex/HLR for Sc.
Except for Mice B, that shows a flat gradient, 
the rest of them have a negative gradient, with 
the one in NGC 6090 and Mice A being similar to Sbc galaxies, 
and in IC 1623 W and NGC 2623 steeper than for 
the control spirals.}

\end{itemize}


\subsection{sSFR}
\label{sec:sSFR}

As in \cite{gonzalezdelgado16}, we obtain the local specific star formation 
rate as the ratio between the intensity of the recent star formation 
rate and the stellar mass surface density, 
sSFR(R) = $\Sigma_{SFR}(R)$/$\mu_\star(R)$. It measures the relative 
rate of ongoing star formation with respect to the past in each 
position in a galaxy. This local definition is equivalent to the global 
sSFR defined for a galaxy as the ratio between the total 
SFR and the galaxy stellar mass.  

For $t_{SF}= 30 Myr$ we have 
found for spiral galaxies that sSFR(R) increases outwards as a function 
of the radial distance, growing faster in the inner 1 HLR 
than outwards, probably signaling the bulge-disk 
transition, and the effect of the quenching that progresses 
inside-out of galaxies \citep{tacchella15, gonzalezdelgado16}.

\begin{figure}[!h]
\includegraphics[width=0.5\textwidth]{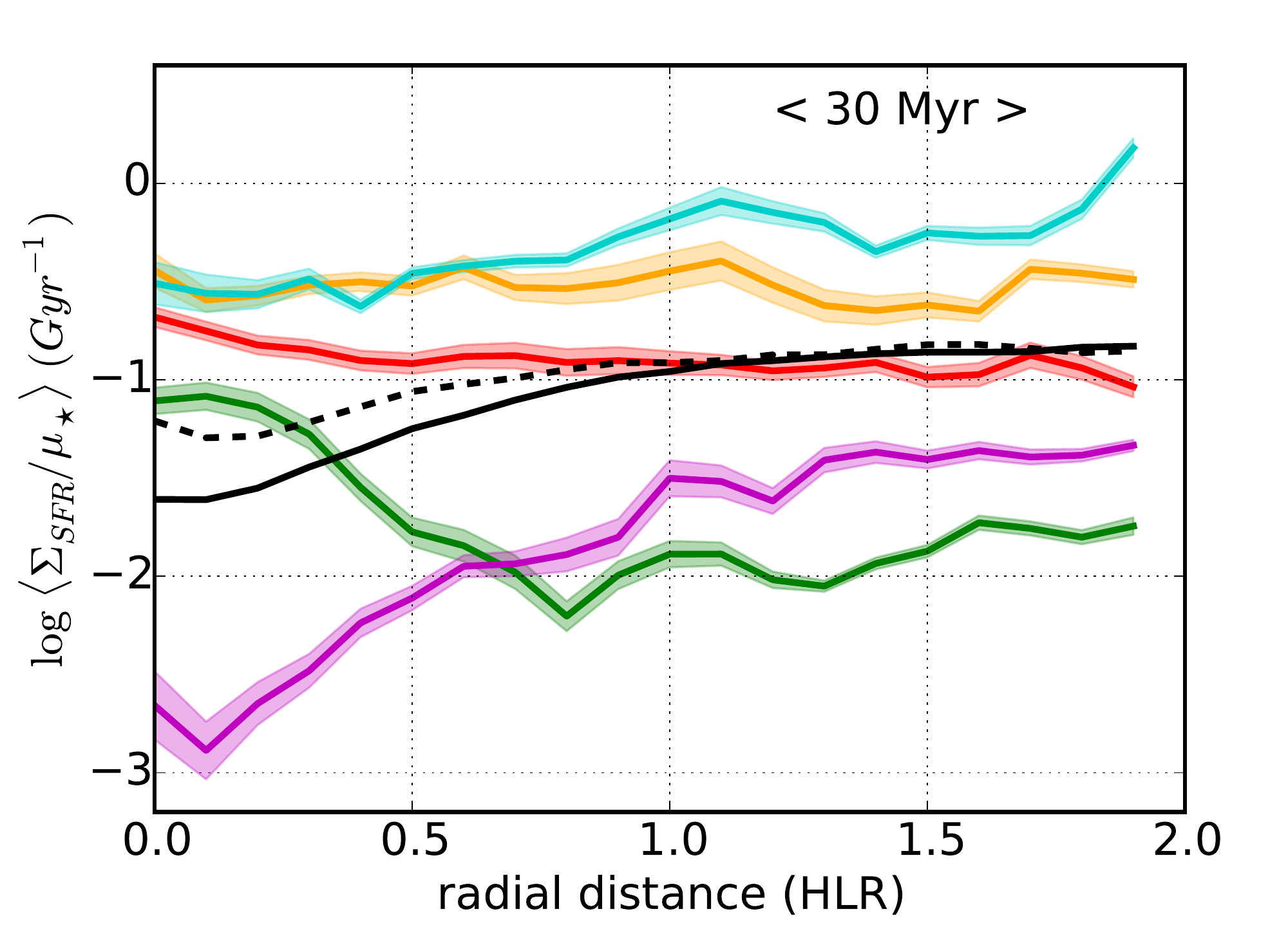}
\includegraphics[width=0.5\textwidth]{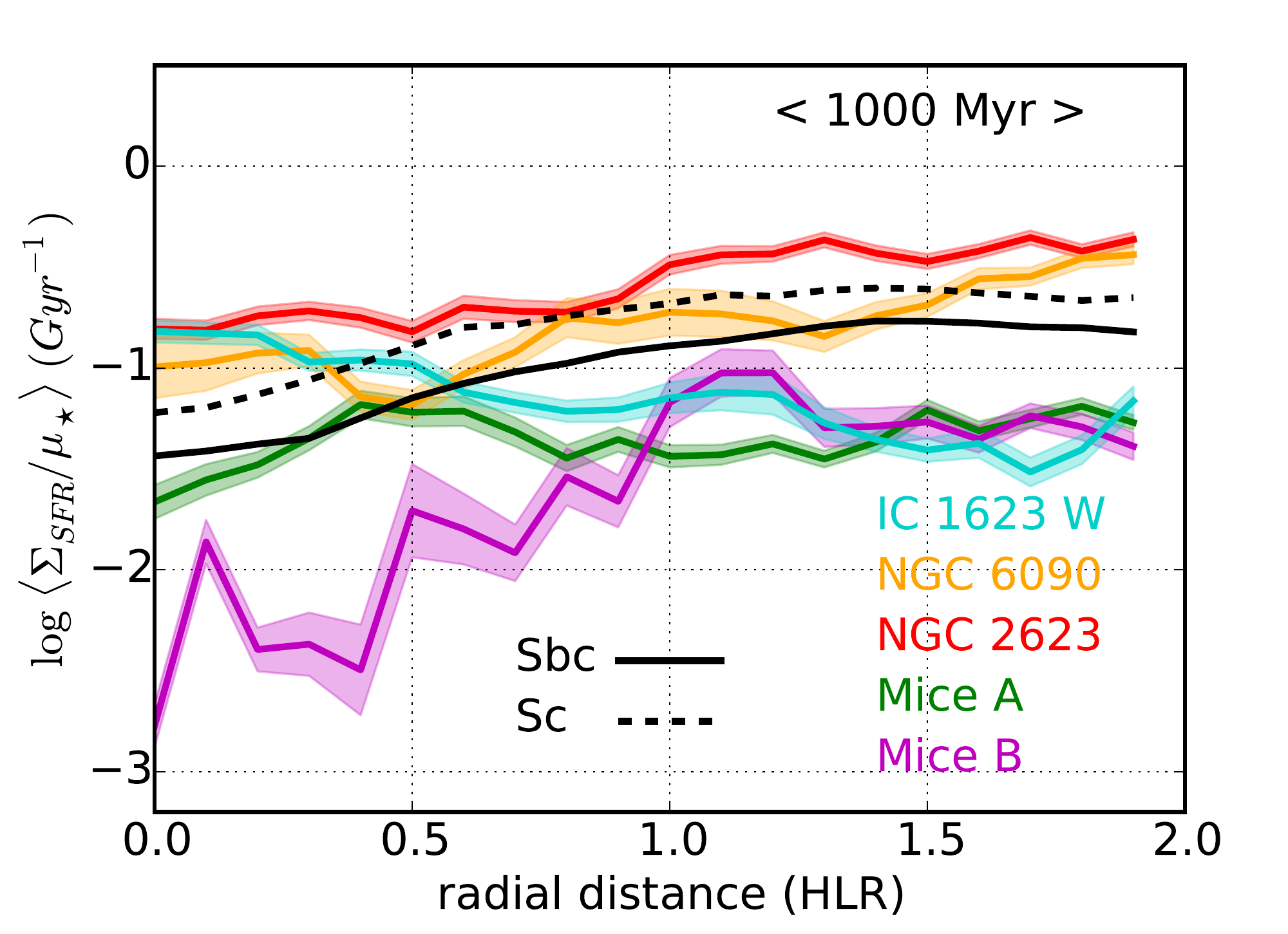}
\caption{The radial profiles of the specific star formation 
rate ($\Sigma_{SFR}$/$\mu_{\star}$) obtained for two different 
time scales: the last 30 Myr, and 1.0 Gyr. 
The uncertainties, calculated as the standard 
error of the mean, are shaded in light colors.} 
\label{fig:RadialProfiles_sSFR}
\end{figure}

The local sSFR allows to compare the relative star formation 
between different regions, as it is normalized by the amount of 
stellar mass. Fig.\ \ref{fig:RadialProfiles_sSFR} shows the radial 
profiles of sSFR of the mergers and control spirals; 
the comparisons are done using different 
star forming time scales as defined above. The first result 
to note is that in the time scale of 30 Myr, the radial profiles 
$\Sigma_{SFR}(R)$/$\mu_\star(R)$ of LIRG-mergers are flatter 
than in spirals, and above or equal to 0.1 Gyr$^{-1}$. 
This is an interesting result 
because sSFR = 0.1 Gyr$^{-1}$ is a threshold to separate 
star-forming galaxies from quiescent systems \citep{peng10}. 
Thus,  the disk regions (R$ > $ 1 HLR) 
of our spiral control sample 
and the LIRG-mergers of the sample are  above this 
threshold. But Mice are significantly below, indicating 
that at the present epoch this system is forming stars with 
an efficiency that is significantly below to what they did in the past.

Specifying the results in more detail:

\begin{itemize}
\item{{\em $t_{SF}$ = 30 Myr}: 
The sSFR($t_{SF}$ = 30 Myr) in the central 0.5 HLR of IC 1623 W, 
NGC 6090, and NGC 2623 is respectively 0.93 dex (0.67 dex), 
0.94 dex (0.68 dex), and 0.65 dex (0.38 dex) higher than in Sbc (Sc) 
galaxies. In Mice A is similar to Sbc (Sc) galaxies, 
by 0.15 dex ($-$0.12 dex), and in Mice B significantly decreased, 
$-$1.03 dex ($-$1.30 dex).
In the outer parts (1--1.5 HLR), we find that sSFR($t_{SF}$ = 30 Myr) 
is enhanced with respect Sbc (Sc) galaxies in the early-stage merger 
LIRGs IC 1623 W, 0.69 dex (0.68 dex), 
and NGC 6090, 0.36 dex (0.33 dex), similar to them in NGC 2623, 
$-$0.04 dex ($-$0.07 dex), and decreased in Mice A and Mice B by 
$-$1.04 dex ($-$1.07 dex) and $-$0.57 dex ($-$0.60 dex), respectively.  
With respect to the gradient in the inner 1 HLR, it is 
almost flat in NGC 6090 (0.00 dex/HLR), and 
NGC 2623 ($-$0.23 dex/HLR) in comparison to the clearly 
positive gradient in Sbc (0.65 dex/HLR) 
and Sc (0.30 dex/HLR) galaxies.
The positive gradient in the control spirals indicates 
a central shut down of the star formation in comparison 
with that at 1 HLR, and that the quenching is progressing 
inside-out \citep{gonzalezdelgado16}. However, the flattening of 
the local sSFR(R) in the mergers indicates that these galaxies 
are forming stars more actively than in the past.
In IC 1623 W (0.33 dex/HLR) the gradient is positive and more similar to 
Sc galaxies, pointing to less recent star formation in the center than 
at 1 HLR. 
By contrast, in Mice A ($-$0.78 dex/HLR) the gradient is negative but 
steeper than in the LIRGs, indicating that the increase of the 
star formation in the last 30 Myr is more important in the nucleus that 
in the disk. In Mice B (1.16 dex/HLR) the gradient is positive, 
but steeper than in the control spirals. 

Again, when considering the young-intermediate 
time-scale ($t_{SF} \lesssim$ 300 Myr) the 
results are approximately consistent with the 
$t_{SF} \lesssim$ 30 Myr ones. The star formation is 
enhanced in the central regions of IC 1623 W (by a factor 29), 
NGC 6090 (factor 6), 
and NGC 2623 (by a factor 7) in comparison with the control spirals.
In the outer parts, only IC 1623 W presents a significant
enhancement (by a factor 4) of the star formation with respect 
to control spirals. On the contrary, in the outer parts of NGC 6090, 
and NGC 2623, the star formation is not so significantly 
increased ($\lesssim$ factor 1.7) with respect to Sbc--Sc galaxies,
when considering the 300 Myr period.}

\item{{\em $t_{SF}$ = 1 Gyr}:
The sSFR($t_{SF}$ = 1 Gyr) in the center of the mergers
differ with respect to Sbc (Sc) galaxies by 
0.55 dex (0.30 dex) in NGC 2623, 
0.43 dex (0.18 dex) for IC 1623 W, 
0.31 dex (0.06 dex) for NGC 6090, 
$-$0.08 dex ($-$0.33 dex) for Mice A, 
and $-$0.94 dex ($-$1.19 dex) for Mice B.
Analogously, in the outer parts we find differences 
between our mergers and Sbc (Sc) galaxies of 
0.38 dex (0.19 dex) in NGC 2623, 
0.07 dex ($-$0.12 dex) in NGC 6090, 
$-$0.36 dex ($-$0.55 dex) in Mice B,
$-$0.42 dex ($-$0.61 dex) in IC 1623 W, 
and $-$0.56 dex ($-$0.75 dex) in Mice A.
With respect to the gradient in the inner 1 HLR, 
the sSFR($t_{SF}$ = 1 Gyr) gradient is from more positive 
to more negative; 
1.60 dex/HLR in Mice B, 0.55 dex/HLR in Sbc galaxies, 
0.54 dex/HLR in Sc galaxies, 0.32 dex/HLR in NGC 2623, 
0.27 dex/HLR in NGC 6090, 
0.23 dex/HLR in Mice A, and $-$0.33 dex/HLR in IC 1623 W.
The sSFR($t_{SF}$ = 1 Gyr) gradient is positive for 
the control spirals and all the mergers except IC 1623 W. 
This points to a 
quenching in the central parts of the galaxies in comparison with 
the outer parts in the last 1 Gyr. IC 1623 W, however, presents 
the opposite behavior.}

\end{itemize}

The sSFR can be interpreted as the inverse of a time scale 
for the SF ($\tau$). These results in the last 30 Myr indicate 
that $\tau$ is much shorter in LIRG-mergers than in spiral galaxies 
of similar mass. In particular, this time is $\sim$2.7 Gyr for IC 1623 W,
$\sim$3.2 Gyr for NGC 6090, $\sim$7.0 Gyr for NGC 2623, while for the 
Mice and control spirals $\tau$ is above 
the Hubble time \citep{gonzalezdelgado16}. 
This means that the three LIRG mergers can double 
their actual stellar mass faster than Sbc-Sc galaxies, if the SFR 
proceeds at the same rate than in the last 32 Myr. 

In contrast, for $t_{SF} = 1$ Gyr, $\tau$ is quite similar for the 
early-stage LIRG
mergers and the Sbc-Sc galaxies ($\sim$ 10 Gyr). This reinforces 
the result that the enhancement 
of star formation in these objects occurs in time scales of few 100 Myr. 
On the contrary, for the more advanced merger NGC 2623, 
$\tau (< 1 Gyr) \sim$ 5 Gyr, pointing to a star formation history 
more extended in time.

\section{Discussion}
\label{sec:Discussion}

\subsection{Global enhancement of the star formation in mergers}

It is well established that star-forming galaxies of the local Universe 
show a correlation between their recent star formation rate and galaxy 
stellar mass known as the main sequence of star forming 
galaxies (MSSF) \citep{brinchmann04}. The logarithmic slope of this 
relation is sub-lineal and close to 1, and the dispersion is small, 
of 0.2--0.3 dex \citep{renzinipeng15}. High redshift galaxy surveys 
have also proved that  this relation persists at least to 
redshift $\sim$ 4 \citep{peng10,wuyts11,speagle14}. 

Recently, a bimodal distribution 
in the SFR--M$_{\star}$ plane is found in z = 4--5 galaxies; 
probing the existence of a starburst sequence, with a much 
more effective star formation mode (i.e. mergers, perturbations) than 
the secular halo gas accretion occuring in main-sequence 
star forming galaxies. Previous studies failed in recognising 
the starburst sequence, as they only focuse in 
the most massive systems (M$_{\star}$ $>$ 10$^{10}$ M$_{\odot}$). 
When low and intermediate mass galaxies are taken into account, 
the starburst sequence is clearly evident, and contributes to $>$ 50$\%$ 
of the cosmic SFR density at z = 4--5 \citep{caputi17}.
\begin{figure*}
\includegraphics[width=0.85\textwidth]{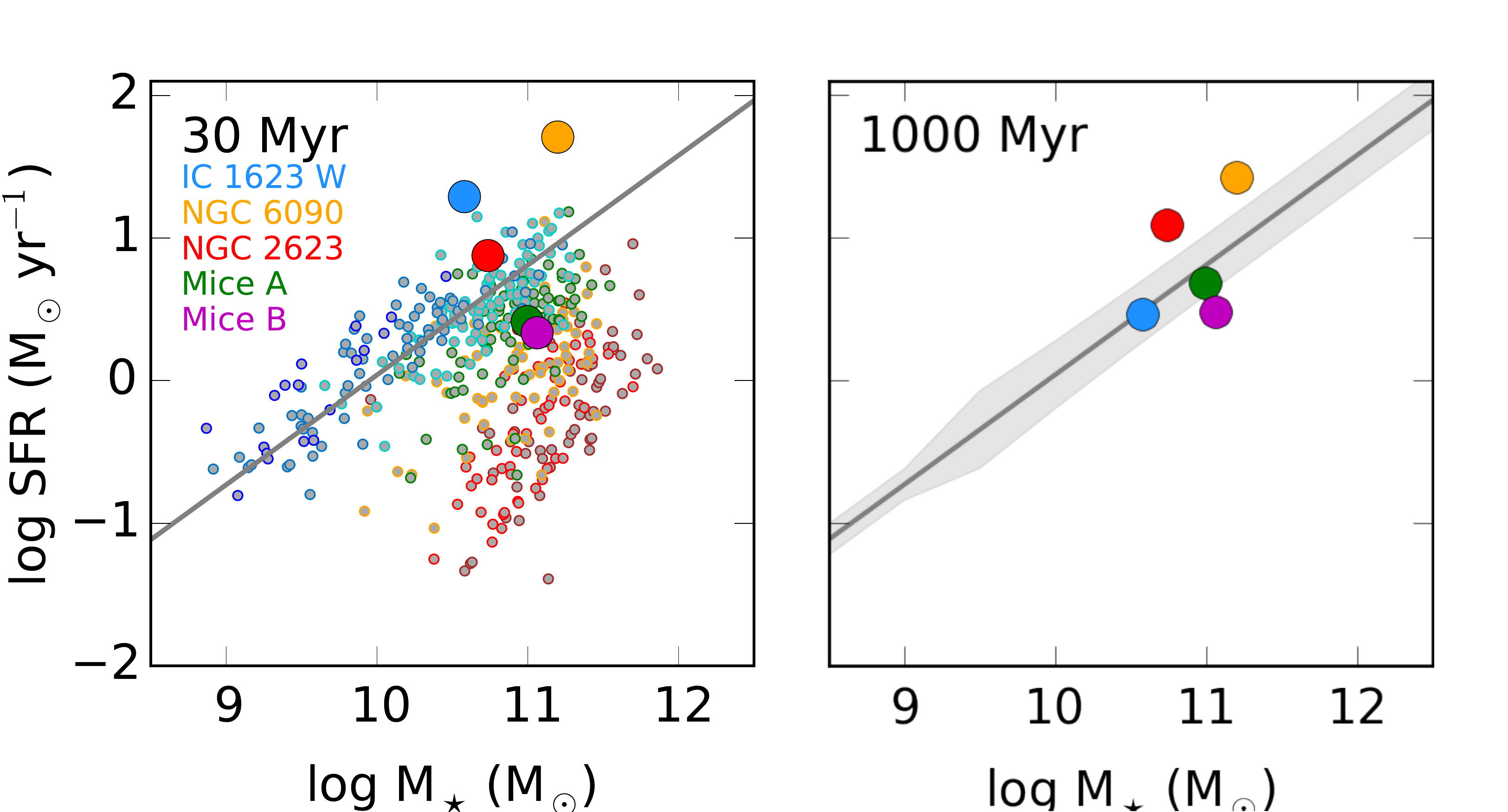}
\centering
\caption{Global average SFR versus the 
galaxy stellar mass for the four mergers of our sample at the 
two different time scales ($t_{SF} = $ 30 Myr (left) 
and 1 Gyr (right)). 
For comparison, we show the MSSF (grey line) as derived 
by \protect\citet{gonzalezdelgado16} using late type Sc 
spirals of CALIFA and using a $t_{SF} =$ 30 Myr. 
In the left panel, all the galaxies are shown with small dots,
colour-coded by Hubble type following 
\protect\cite{gonzalezdelgado16}, with a brown to blue 
color palette 
that represents Hubble types from ellipticals to late spirals.
In the right panel the light grey shaded regions indicate the 1$\sigma$ 
dispersion of the sample. 
The positions of the Mice, the pre-merger LIRGs IC 1623 W and NGC 6090, 
and NGC 2623, are shown with solid circles following the same colour 
coding as in previous figures.} 
\label{fig:MSSF}
\end{figure*}
 
Local galaxies with recent intense starbursts, 
as (U)LIRGs, can also be significantly above of the MSSF. 
In a recent study 
\citet{guo16} found that roughly three quarters of the advanced 
ULIRGs of their sample are located above 1$\sigma$ of the MSSF. 
However, this is not always true for LIRGs; as the enhancement of 
the star formation with respect to the MSSF depends on the time 
scale and merger phase. 
For example, \citet{pereira15} in their sample of local LIRGs, 
find that the global SFH of these galaxies can be grouped in 
three categories: 1) almost half of the objects with a recent starburst 
in the last 10--100 Myr; 2) a similar fraction of galaxies with 
constant SFR; and 3) only a few objects with a declining SFR. 
In the first group, all of them are significantly above 
the MSSF in time scales of less than 30 Myr; but at 1 Gyr, 
the objects are on the 
MSSF or below. Most of the LIRGs that are above the MSSF 
are mergers and have a high sSFR$> $ 0.8 Gyr$^{-1}$.

Fig.\ \ref{fig:MSSF} shows the global average SFR vs 
galaxy stellar mass for the four mergers of our sample 
at $t_{SF} = $ 30 Myr and 1 Gyr time scales.
For comparison, the MSSF as derived 
by \citet{gonzalezdelgado16} using late type Sc spirals of CALIFA 
and $t_{SF} =$ 30 Myr is drawn as a greyline. 
With the exception of the Mice, 
the global SFR in the other systems are enhanced with respect to 
the MSSF galaxies, in the $t_{SF} \leq 30$ Myr time-scale. 
In particular, the SFR is enhanced by a factor $\sim$6 in IC 1623 W, 
$\sim$6 in NGC 6090, and $\sim$2 in NGC 2623. 
The Mice is seen after first passage \citep{wild14}, 
and it is probably the least advanced merger, so the star formation 
is still not enhanced. This result was expected given that the 
Mice have FIR luminosities below 10$^{11}$ L$_\odot$, 
the limit to be classified as a LIRG. 

It is also interesting that although the LIRG-mergers are above 
the MSSF at $t_{SF} \leq 30$ Myr, at $t_{SF} = 1$ Gyr 
IC 1623 W is on the MSSF, while 
in NGC 6090 the SFR is enhanced by 
a factor $\sim$3 with respect to Sc MSSF galaxies, but 
is half the one measured at $t_{SF} \leq 30$ Myr. 
On the contrary, in NGC 2623 there is a slight increase of 
the SFR in this longest timescale, which is a 
factor $\sim$3 larger than in Sc MSSF galaxies. This is 
expected given the known presence and relevance of intermediate-age 
stellar populations in this system \citep{cortijo17b}. 

This points out that a major phase 
of star formation in the merger LIRGs occurs in time 
scales of 10$^7$ yr to few 10$^8$ yr.
Our results are more in agreement 
with \citet{rodriguezzaurin09, rodriguezzaurin10} where the optical 
light of local ULIRGs (mostly mergers in different evolutionary 
stages) have important contribution of stellar populations 
of 300--500 Myr. They also agree with \citet{pereira15} where the 
duration of the bursts is on time scales of $\leq$ 10 to few 100 Myr 
for the LIRGs that show merger signatures. Star formation time 
scales of 40--260 Myr have been also found by \citet{marcillac06} 
in their LIRGs at redshift  $z\sim$ 0.7, but none of their galaxies show
time scales of less than a few 10 Myr as we detect in LIRG-mergers. 
In the more advanced merger NGC 2623 an 
even more intense phase of star formation 
occurred in a longer 
time-scale of $\sim$1 Gyr. These results are in agreement with those 
of \citet{alonsoherrero10} that studied a small sample of LIRGs, 
and found that the dominant stellar population is at least 1--2 Gyr old. 
However, we note that most of their galaxies are isolated or weakly 
interacting local LIRGs.

\subsection{The extent and relevance of the different phases of star formation }

It is interesting now to discuss the relevance of the different phases 
of enhancement of the star formation and their spatial extent. 
Using the specific SFR intensity as a mass-independent SFR tracer, 
what we have found is:

\begin{itemize}
\item{{\em $t_{SF}$ = 30 Myr}:
In the early-stage merger LIRG IC 1623 W the star formation in the last 30 Myr 
is enhanced with respect to the control spirals both in the central 
regions (by a factor  7) and in the "disk" (by a factor 5), 
but in the early-stage merger NGC 6090 the enhancement is important 
in the central region (by a factor 7) while in the "disk" it is
much more moderate (factor 2) in comparison to control spirals.
For the merger NGC 2623, the young star formation 
in the last 30 Myr is enhanced in the central 
region (by a factor 3), while at the outer parts it is similar 
to the control spirals. On the contrary, in Mice A, the young star formation 
in the last 30 Myr is similar to the control spirals in the centre, 
but decreased by a factor 11 in the disk, and in Mice B it is decreased 
both in the centre (by a factor 14), and in the disk (by a factor 4).

The results are qualitatively the same when considering 
a longer young-intermediate time scale of $\lesssim$300 Myr, 
with some differences only in the numbers.}

\item{{\em $t_{SF}$ = 1 Gyr}:
However, the star formation in the last 1 Gyr is only 
enhanced (by a factor 3) in the central regions of NGC 2623, 
with respect to control spirals, while in the pre-mergers 
the enhancement is very low (by factors of $<$2). In the centre 
of Mice A it is comparable to the control spirals, 
and in Mice B it is decreased by a factor 11. 
In the outer regions, the star 
formation in the last 1 Gyr is enhanced with respect to 
the control spirals by a factor 2 in NGC 2623, while for 
the pre-merger LIRGs it is comparable to the 
control spirals (NGC 6090), or a factor 3--4 lower 
(IC 1623 W, Mice A and B), 
as if the star formation has been inhibited in this time period.}

\end{itemize}

We can also quantify the relative impact of these different 
periods of star formation by comparing the amount of mass formed 
in each period. Using the mass fractions reported in 
Table \ref{tab:natbib} 
we obtain very similar numbers compared to those previously reported, 
and therefore, reaching totally consistent conclusions when considering 
the amounts of mass formed relative to the non-interacting spirals.
Basically, the two pre-merger LIRGs 
formed 3--7 times more mass, in the 
last 30 Myr, than the non-interacting control spirals, both in the centres, 
and in the disks, while the merger NGC 2623 
formed 4 times more mass 
in its centre than Sbc-Sc spirals, but comparable in the disk. 
In the 300 Myr timescale, IC 1623 W is the system 
which formed more mass in comparison to the control spirals, 
by a factor $\sim$21 in the centre, and 4 in the disks. 
The two other LIRGs, NGC 6090 and NGC 2623, formed 
significantly more mass than the spirals in their 
centres (a factor 5--6), and less in their disk 
(but still a factor $\sim$2 more). 
In the 1 Gyr time-scale, only NGC 2623 formed significantly more 
mass than the non-interacting spirals, by a factor 3 in the centre, 
and 2 in the disk.

Hence, we conclude that the recent star formation in the three LIRG mergers 
is enhanced with respect to the Sbc-Sc galaxies. However, the amount and the 
spatial extension of the enhancement of the star formation depends on the 
time scale and the galaxy. Thus, 
on short ($t_{SF} = 30 Myr$) time scales the 
enhancement is quite significant in the three systems, 
but on longer time-scales ($t_{SF} = 1 Gyr$) it is only relevant in the 
advanced merger NGC 2623. In the Mice, the star formation is mostly 
inhibited with respect to the control 
spirals, specially in Mice B. 
We think this is due to a lack of gas in this progenitor, because it 
has a significantly lower gas mass fraction ($\sim$6$\%$) 
in comparison to the Mice A ($\sim$13$\%$) \citep{wild14}, and 
typical Sbc (Sc) spirals $\sim$13 (16)$\%$ \citep{rubin85}. 
This gas deficiency could be caused by negative 
feedback from the AGN which is present in the centre 
of Mice B \citep{gonzalezmartin2009,haan2011,masegosa2011}, 
as Mice B has a total stellar mass 
of $\sim$1.5 $\times$ 10$^{11}$ M$_{\odot}$, compatible with the regime 
where feedback is dominated by AGN outflows \citep{shankar2006}.

\subsection{Evolutionary scheme}
High spatial resolution hydrodynamic simulations of 
disk galaxies mergers predict that 
extended starbursts 
arise spontaneously after the first 
pericenter passage, due to fragmentation of the gas clouds 
produced by the increase of the supersonic turbulence of ISM. 
According to these models, a merger-induced nuclear starburst 
is also present, but it occurs later in the merger sequence, 
after the second pericenter 
passage \citep{teyssier2010,powell13}.

Although a systematic spatially resolved analysis of a 
large and complete observational sample of mergers is necessary 
to properly test the validity of these simulations and 
attain more general conclusions, our sample provides some 
clues from the observational point of view. 
The most advanced merger in our sample, NGC 2623, 
which has already passed coalescence, seems to 
reproduce the predictions from simulations, 
showing both an extended SFR enhancement 
about $\lesssim$1 Gyr ago, relic of the first pericenter 
passage epoch, and a current SFR enhancement in 
the last 30 Myr, located in the central 0.5 HLR 
but not in the outer parts, consistent with the second 
pericenter passage and final coalescence. 
This is consistent to what we already 
found through the stellar population analysis in 
this system \citep{cortijo17b}.
We note that advanced post-coalescence systems are ideal places 
where to apply the fossil method to unveil past star formation 
epochs, as both the extended and nuclear starbursts should be 
identifiable in the spectra of most of these systems 
if the current merger scenario is correct. 

Analogously, in the two pre-merger LIRGs IC 1623 W and NGC 6090, 
which are somewhere located between first pericenter passage and 
prior to coalescence, the most remarkable increase of the SFR with 
respect to non-interacting spirals occurred in the last 30 Myr, and 
is spatially extended, in agreement with current simulations.
We note that IC 1623 W could represent a more evolved state as the 
extended SFR enhancement was already traced by $\sim$300 Myr populations. 
This is in agreement to what we found in \citep{cortijo17}. 
The $\sim$ 1Gyr populations do not show significantly 
enhanced mass/SFR contributions with respect to non-interacting 
spirals in neither the outer parts nor the centres.

Contrary to expected, although Mice passed 
the first pericenter passage about 170 Myr ago \citep{barnes04}, 
where a starburst should have occurred, there exists no evidence 
of an enhanced SFR in the last 30 Myr or 300 Myr in this 
system: on the contrary, the star formation is mainly inhibited.
The Mice are probably the least evolved 
merger of the sample, and the fact that 
the gas fraction of Mice B is smaller than 
in most non-interacting spirals shows that there are 
many factors that determine when, where, and 
with which intensity the starbursts will occur.
Also, the Mice are close to a prograde orbit \cite{barnes04}, 
while simulations by \cite{dimatteo2007} show that retrograde 
encounters (when the galaxies spin is antiparallel) have 
larger star formation efficiencies, even if prograde 
encounters develop more pronounced asymmetries than retrograde 
ones, whose remnants are more compact.

What simulations are showing us is that the merger timescale, 
SF enhancement, and remnant properties of 
galaxy mergers, depend on several factors such as: 
the morphologies of the progenitors, 
the availability of gas (gas-rich or gas-poor), and the 
orbital characteristics. 
It is therefore necessary to make a detailed spatially 
resolved study of mergers covering all the merger stages, 
with information about the total gas content and the kinematics, 
to shed more light on the merger evolutionary sequence.

\section{Summary and conclusions}
\label{sec:Summary}
Using IFS data from the CALIFA survey and PMAS in LArr mode, 
we have analyzed the spatially resolved star formation history 
of a small sample of local mergers: the early-stage 
mergers (Mice, IC 1623, and NGC 6090) 
and a more advanced merger, NGC 2623, 
to determine/quantify if there is an enhancement of the star 
formation and trace its time scale and spatial extent.
A full spectral fitting analysis was performed using the 
\starlight\ code and a combination of the single stellar populations models 
by \citet{vazdekis10} and \citet{gonzalezdelgado05}, using 156 spectra 
of 39 different ages from 1 Myr to 14 Gyr, and four 
metallicities ($Z =$ 0.2, 0.4, 1, and 1.6 $Z_\odot$). The spectral 
fitting results are processed 
through \pycasso\ pipeline to derive the 2D ($R \times t$) maps, from which 
we obtain the spatially resolved star formation rate (SFR), 
specific sSFR, and the intensity of the SFR ($\Sigma_{SFR}$), 
over three different timescales (30 Myr, 300 Myr, and 1 Gyr).

Our main results are:

\begin{enumerate}
\item A major phase of star 
formation in the merger LIRGs is occurring in time 
scales of 10$^7$ yr to few 10$^8$ yr, with global SFR enhancements 
of $\sim$2--6 with respect to the MSSF galaxies.
In the more advanced merger NGC 2623 a previous 
phase of star formation occurred in a longer 
time-scale of $\sim$1 Gyr.

\item From our spatially resolved analysis, we have quantified 
the extension and relative impact of the different periods 
of star formation in the last 30, 300, and 1000 Myr.  
Using the sSFR as a mass-independent SFR tracer, we find 
that the two pre-merger LIRGs present a sSFR 
enhancement in the last 30 Myr of 2--7 times that in 
non-interacting control spirals, both in the centres, 
and in the disks, while in the merger NGC 2623 the 
enhancement is a factor 3 in the centre, but comparable to Sbc-Sc 
galaxies in the outer regions. 
In the 300 Myr timescale, IC 1623 W is the system which presents 
a greater sSFR enhancement in comparison to the control spirals. 
Finally, in the 1 Gyr time-scale, only NGC 2623 formed significantly more 
mass than the non-interacting spirals, by a factor 3 in the centre, 
and 2 in the disk.

\item The spatially resolved study of the three LIRG-mergers 
reveals that their SFH is consistent with the predictions 
from high spatial resolution simulations. 
Extended starbursts arise after the first 
pericenter passage (consistent with the enhancement of 
the extended SF in the last $\lesssim$30--300 Myr in IC 1623 W 
and NGC 6090, and $\lesssim$1 Gyr in NGC 2623), while 
a nuclear starburst occurs later in more advanced mergers, 
after the second pericenter 
passage/coalescence (i.e. as reflected by the enhancement 
of the central SF in NGC 2623 in the last 30 Myr).
A systematic analysis of a larger and more complete 
observational sample of mergers is necessary 
to further test the validity of these simulations.

\item  The Mice is the only system not 
showing a significant enhancement of the spatially extended 
star formation, but rather it is inhibited, that is more 
severe in Mice B. 
The Mice are not classified as LIRG, and it is probably 
the less advanced merger in our sample.
The fact that the gas fraction of Mice B 
is smaller than in most non-interacting spirals, and that 
the Mice are close to a prograde orbit, presents a new 
challenge for the models, which must cover a larger 
space of parameters in terms of the availability of gas 
and the orbital characteristics.

\end{enumerate}

\begin{acknowledgements} 
CALIFA is the first legacy survey carried out at Calar Alto. The CALIFA collaboration would like to thank the IAA-CSIC and MPIA-MPG as major partners of the observatory, and CAHA itself, for the unique access to telescope time and support in manpower and infrastructures.  We also thank the CAHA staff for the dedication to this project.
Support from the Spanish Ministerio de Econom\'\i a y Competitividad, through projects 
AYA2016-77846-P, AYA2014-57490-P, AYA2010-15081, and Junta de Andaluc\'\i a FQ1580, 
AYA2010-22111-C03-03, 
AYA2010-10904E, AYA2013-42227P, RyC-2011-09461, AYA2013-47742-C4-3-P, EU SELGIFS exchange 
programme FP7-PEOPLE-2013-IRSES-612701, and CONACYT-125180 and DGAPA-IA100815.
We also thank the Viabilidad, Dise\~no, Acceso y Mejora funding program, ICTS-2009-10, 
for funding the data acquisition of this project. 
ALdA, EADL and RCF thanks the hospitality of the IAA and the support of CAPES 
and CNPq. RGD acknowledges the support of CNPq (Brazil) through Programa 
Ci\^encia sem Fronteiras (401452/2012-3). We thank the support of the 
IAA Computing group. 
This research made use of Python (http://www.python.org); 
Numpy \citep{vanderwalt2011}, and Matplotlib \citep{hunter2007}.
\end{acknowledgements}



\bibliographystyle{aa}
\bibliography{2DSFH_LIRG}

\end{document}